\documentclass[aps,amsmath,onecolumn, showpacs,superscriptaddress,pre,10.5pt,nofootinbib]{revtex4-1}
\usepackage{amssymb}
\usepackage{amsmath}
\usepackage{graphicx} 
\usepackage{array}
\usepackage{physics}
\usepackage{mathtools}
\usepackage{amssymb}
\usepackage{verbatim}
\usepackage{color}
\usepackage{bm}
\usepackage{amsthm} 
\usepackage[normalem]{ulem}

\usepackage{color}
\usepackage[dvipsnames]{xcolor}
\usepackage[latin1]{inputenc}
\definecolor{Blue}{rgb}{0.00, 0.00, 1.00}
\definecolor{Red}{rgb}{1.00, 0.00, 0.00}

\usepackage[pdfpagemode=UseNone,
		bookmarksopen=false,
		colorlinks=true,
		urlcolor=blue,
		citecolor=red,
		citebordercolor=blue,
		linkcolor=blue, 
		urlcolor=black, 
		filecolor=red]{hyperref}
		
\newcommand{\bea}{\begin{eqnarray}}
\newcommand{\eea}{\end{eqnarray}}
\newcommand{\be}{\begin{equation}}
\newcommand{\ee}{\end{equation}}
\newcommand{\bee}{\begin{equation*}}
\newcommand{\eee}{\end{equation*}}

\newcommand{\InverseW}{\mathbin{\rotatebox[origin=c]{180}{$W$}}}

\colorlet{Mycolor1}{green!10!orange!90!}
\newcommand{\isEquivTo}[1]{\underset{#1}{\sim}} 
\newcommand{\nocontentsline}[3]{}
\newcommand{\tocless}[2]{\bgroup\let\addcontentsline=\nocontentsline#1{#2}\egroup}

\begin{document}

\title{Matrix Kesten Recursion, Inverse-Wishart Ensemble and Fermions in a Morse Potential}

\author{Tristan Gauti\'e}
\affiliation{Laboratoire de Physique de l'Ecole Normale Sup\'erieure,
PSL University, CNRS, Sorbonne Universit\'e, Universit\'e de Paris,
24 rue Lhomond, 75231 Paris, France.}
\author{Jean-Philippe Bouchaud}
\affiliation{Capital Fund Management, 23 Rue de l'Universit\'e, 75007 Paris, France.}
\author{Pierre Le Doussal}
\affiliation{Laboratoire de Physique de l'Ecole Normale Sup\'erieure,
PSL University, CNRS, Sorbonne Universit\'e, Universit\'e de Paris,
24 rue Lhomond, 75231 Paris, France.} 

\begin{abstract} 
The random variable $1+z_1+z_1z_2+\dots$ appears in many contexts and was shown by Kesten to exhibit a heavy tail
distribution. We consider natural extensions of this variable and its associated recursion to $N \times N$ matrices either real symmetric $\beta=1$
or complex Hermitian $\beta=2$. In the continuum limit of this recursion, we show that the matrix distribution converges to the inverse-Wishart ensemble of random matrices.
The full dynamics is solved using a mapping to $N$ fermions in a Morse potential, which are non-interacting for $\beta=2$.
At finite $N$ the distribution of eigenvalues exhibits heavy tails, generalizing Kesten's results in the scalar case.
The density of fermions in this potential is studied for large $N$, and the power-law tail of the eigenvalue distribution is
related to the properties of the so-called determinantal Bessel process which describes the hard edge universality of random matrices.
For the discrete matrix recursion, using free probability in the large $N$ limit, we obtain a self-consistent equation for the
stationary distribution. The relation of our results to recent works of Rider and Valk\'o, Grabsch and Texier, as well as Ossipov, is discussed.
\end{abstract}


\maketitle

{
	\hypersetup{linkcolor=black}
	\tableofcontents
}

\bigskip 

\section{Introduction} 
\label{sec:Introduction}

The dynamical generation of random variables with heavy tails is a fascinating and
ubiquitous phenomenon. In physics it occurs, among many others, in problems
such as diffusion in random media \cite{BouchaudGeorges}, directed polymers in random media \cite{DerridaSpohn}, Anderson localization in an external field
\cite{SouillardPRL1984,Progodin1980,GeorgesAspect2018},  growing networks \cite{Dorogovtsev}, avalanches in
driven elastic systems \cite{ABBM} or in self-organized critical systems such as sandpiles
\cite{Bak}. Power-law tails are also ubiquitous in economics and finance \cite{Bouchaud2001,Gabaix}: wealth distribution in models of economic agents \cite{BouchaudMezard}, returns of financial markets \cite{Gabaix}, city and firm sizes \cite{Axtell, GabaixZipf, SornetteZipf}, etc.  

One of the simplest example giving rise to power-law tails is a stochastic recursion equation studied in a seminal paper by Kesten
\cite{Kesten1973}, which we will call here the Kesten recursion. It was studied in a scalar and a vector form. It occurs naturally in its scalar form in models of random walks in random environments \cite{Solomon1975,Kesten1975}
and in its vector form to describe the mean population in multi-specie branching processes in random environments.
Its continuum limit appears in the problem of Arrhenius diffusion of a particle in a one-dimensional random force field, where it predicts the generation of broad distributions of waiting times, which can be seen as effective traps for the particle \cite{BouchaudPLD1990,BouchaudGeorges}.

In this paper our main aim is to introduce and study a matrix realization of the Kesten recursion, which we call
the matrix Kesten recursion. There are several ways to generalize the scalar recursion ($N=1$) to the case
of a $N \times N$ matrix. Here we will study two such generalizations, one with real symmetric matrices (denoted $\beta=1$)
and the other with Hermitian matrices ($\beta=2$). We will obtain two sets of exact results. 
First we will provide a solution for the continuum limit of this recursion
for arbitrary size $N$ of the matrix. It shows that the marginal distribution of the few largest eigenvalues of this matrix
exhibit a power-law distribution with a similar exponent as for the scalar case. These eigenvalues however also exhibit strong correlations, as is typical in random matrix theory
\cite{MehtaBook,Forrester,RMTBook,AllezBouchaud2012}. We will also provide some exact results in the limit of a large size of the matrix both for
the discrete recursion and its continuum limit. These results unveil a connection to a classical random matrix ensemble, the 
inverse-Wishart ensemble, well studied in various other contexts
\cite{BunBouchaudPotters2017,RMTBook,AllezBouchaud2013}.

Here we also unveil a remarkable connection between the 
continuum limit of the matrix Kesten recursion and the quantum mechanics of $N$ fermions in
a Morse potential. The eigenvalues of the matrix are related to the positions of the fermions. 
The large time limit is obtained by considering the ground state of these fermions which 
can be obtained exactly in both cases. For Hermitian matrices $\beta=2$ the
fermions are non-interacting and the full dynamics at finite time can be obtained in a simple way. Such fermion problems are
important in the context of cold atoms in traps, and the present results thus add the Morse potential to the list of potentials in one dimension 
for which the quantum correlations can be computed using their connection to random matrices \cite{DeanLeDoussal19,LACTLeDoussal17,LACTLeDoussal18,NadalMaj09,CundenMezzOconnell18,Forrester,MehtaBook}. 

Our study also has connections with previous works on matrix Brownian functionals by Rider and Valk\'o in probability theory and by
Grabsch and Texier in the study of mesoscopic quantum transport in disordered wires. Our model, in its 
continuum version, leads to a different matrix dynamics compared to the ones considered in the aforementioned works. Remarkably however, we show that the evolution of the eigenvalues can be made to coincide by choosing the appropriate
discretization (It\^o versus Stratonovich). Hence we find that two different models with two different noise
terms and two different discretization prescriptions lead to the exact same dynamics for the eigenvalues. 

Let us now introduce the main 
objects of interest and summarize the known results.

\subsection{Overview of the Kesten recursion}

The scalar Kesten recursion is the following system of equations on the sequence $(Z_n)_{n\geqslant 0}$:
\begin{equation}
\label{eq:def}
 \left\{
    \begin{array}{ll}
        Z_{0} = 0 \\[5pt]
Z_{n} = \xi_{n}\left(1+Z_{n-1}\right)
    \end{array}
\right.
\end{equation}
where $(\xi_n)_{n\geqslant 0}$ is a sequence of independent, identically distributed positive random variables. The solution of this recursion is formally:
\begin{equation}
\label{eq:FormalSum}
Z_{n} \  = \ \sum_{j=0}^{n-1} \xi_{n} \cdots \xi_{n-j} \ = \ \sum_{j=0}^{n-1} \prod_{k=n-j}^n \xi_k\; .
\end{equation}
{Note that for a given integer $n$, the {\sc{iid}} hypothesis gives, by permuting the $\xi_k$ variables as $(0,1,\cdots\!, n) \to (n, \cdots\!, 1,0)$, the following equality in law:
}
\begin{equation}
\label{eq:FormalSumReordered}
Z_n \ \overset{\mathrm{law}}{=}  \   \sum_{j=0}^{n-1} \prod_{k=0}^j \xi_k 
\end{equation}
Harry Kesten studied the limit as $n \to \infty$ of such recursions in \cite{Kesten1973,Kesten1975,Kesten1984}. 
Under certain conditions, depending on the sign of $\mathbb{E}[ \log \xi ]$,
the sum in \eqref{eq:FormalSumReordered} may either grow unboundedly with $n$, if $\mathbb{E}[ \log \xi ] \geqslant 0$, or, 
 if $\mathbb{E}[ \log \xi ] < 0$, converge to a positive random variable denoted $Z_\infty$.
Kesten showed that in that case the tail of the distribution of $Z_\infty$ exhibits a power-law behaviour with exponent $1+\nu$ such that
\begin{equation} 
\label{eq:conditionKappa}
\mathbb{E} \left[  \xi^\nu \right] =1 \; .
\end{equation}
It is easy to see why this condition arises. From the definition \eqref{eq:def} one must have that in law $Z_\infty \overset{\mathrm{law}}{=} \xi(1+ Z_\infty)$, where in the r.h.s $\xi$ and $Z_\infty$ are taken as independent random variables. Denoting $p(\xi)$ the probability density function (PDF) of $\xi$, we see that the PDF 
$P(Z)$ of $Z=Z_\infty$ must obey the integral equation 
\be
P(Z) = \int dZ' P(Z') \int d\xi p(\xi) \delta(Z- \xi(1+ Z')) = \int \frac{d\xi}{\xi} p(\xi) P(\frac{Z}{\xi}-1) 
\ee 
If one assumes that $P(Z) \sim Z^{-(1+\nu)}$ at large $Z$, the relation \eqref{eq:conditionKappa}
follows. 
This and related integral equations admit a variety of possible behaviors 
which were studied in details in physics 
\cite{Calan} and in mathematics \cite{goldie1991}. 
Large deviations and rate of convergence were also studied in \cite{Buraczewski2013,Buraczewski2015}.

The Kesten recursion has a close connection to the problem of a Random Walk in a Random Environment (RWRE) \cite{Solomon1975,Kesten1975, Sinai1982, DerridaPomeau1982, Derrida1983a,Kesten1986,Calan,BouchaudPLD1987,PLD1989,ComtetDean,MonthusPLD1999}. The RWRE setting is the study of a random walk where the hopping probabilities are random variables that constitute a random environment, that the random walk explores. This problem has been widely studied on the lattice, and in the continuum.  Let us explain how the Kesten recursion occurs from RWRE in one dimension, following an argument by Solomon \cite{Solomon1975}. We denote by $\alpha_n = \mathbb{P}_{n,n+1}$ and $\beta_n =\mathbb{P}_{n,n-1}=1- \alpha_n$ the transition probabilities of a random walk on the $\mathbb{Z}$ lattice, from site $n$ to its two neighbours. In the random environment setting, these are random variables. We suppose the sequence $(\alpha_n,\beta_n)_{n\in \mathbb{Z}}$ to be fixed, such that a random environment has been chosen beforehand. Let $\mu_n$ be the mean first passage time from site $n$ to site $n+1$ for a random walk in this environment: the first passage time is 1 with probability $\alpha_n$, if the first step from site $n$ is taken to the right, or $(1+\mu_{n-1} + \mu_n)$ in mean, with probability $\beta_n$, if the first step is taken to the left:
\begin{equation}
\mu_{n}=\alpha_{n}+\beta_{n}\left(1+\mu_{n-1}+\mu_{n}\right)
\end{equation}
Defining $\xi_n = \frac{\beta_n}{\alpha_n}$, a prominent random variable to understand the behaviour of the RWRE, and $Z_{n}=\frac{1}{2}\left(\mu_{n}-1\right)$, we obtain directly the Kesten recursion $Z_n = \xi_n (1+ Z_{n-1})$, with the correct initialization $Z_0= 0$ if we choose $\alpha_0 =1$ such that the random walk is constrained to the positive sites. The mean first passage time $\mu_n$ from site $n$ to site $n+1$ is thus distributed as the Kesten random variable. The quantity $- \mathbb{E}[ \log \xi ]$ thus representes the effective bias of the environment. If it is positive, the random walker typically moves to the right, and $\mu_\infty$ is
a finite random variable, which can be large with a power-law distribution corresponding to all possible explorations backward from site $n$ before reaching site $n+1$.

The Kesten recursion also appears in a simplified model of Directed Polymers in Random Media \cite{DerridaSpohn}, where the polymer lives on a complete graph \cite{BouchaudMezard}. The time evolution of the partition function of polymers ending on site $a = 1, \cdots , N$ at time $n$ is given by:
\begin{equation}
    z_{n+1}^a = \eta_n^a \left((1-\varepsilon) z_n^a + \frac{\varepsilon}{N} \sum_{b=1}^N z_n^b \right),
\end{equation}
where $N$ is the total number of sites and $\varepsilon$ the hopping rate of the polymer. On the complete graph, the Laplacian takes a mean-field form that allows one to simplify the analysis considerably. Indeed, introducing $\overline{z}_n = \sum_b z_n^b/N$, one can then show that for large $N$, the rescaled partition function $Z^a_n:=(1 - \varepsilon) z^a_n/\varepsilon \overline{z}_n$ obeys a Kesten recursion:
\begin{equation}
    Z_{n+1}^a = \xi_n^a (1 + Z_n^a), 
\end{equation}
with $\xi_n := (1- \varepsilon) \eta_n/\mathbb{E}[\eta]$, see \cite{BouchaudMezard} for details. 

The Kesten recursion is again useful in the study of the random-field Ising chain at low temperature, as shown by Derrida and Hilhorst in \cite{Derrida1983}. In this setting, the free energy of the chain involves infinite products of random 2 x 2 matrices and can be expressed as $F(\varepsilon)=\lim _{N \rightarrow \infty}(1 / N) \log \left(a_{N} / a_{0}\right)$, where the sequence of random variable $(\frac{a_{i+1}}{a_i})_i$ satisfies a recursion, which can be mapped to the Kesten recursion in the low temperature limit. A similar connection to quantum models, such as random Dirac operators in 1D, 
was studied in e.g. \cite{Texier1997,Texier1999,Steiner1999}.

In \cite{Dufresne1990}, Dufresne exhibits the Kesten recursion in an investment management problem: consider that a savings account is credited every year by a unit currency. Define the random variable $\xi_n = 1 + R_n$, with $R_n$ the rate of return during the $n$th year. The accumulated value in the account $S_n$ is then given as a function of the preceding value $S_{n-1}$, through the Kesten recursion:
\begin{equation}
S_n = \xi_n (1+ S_{n-1})\; .
\end{equation}

Finally, these recursions also appeared early on in biosciences \cite{Paulson1972,Feldman1973}.

\subsection{Scalar, vector and matrix models} 
\label{sec:Models}

The discrete recursion defined in \eqref{eq:def} was studied for scalars and vectors by Kesten in \cite{Kesten1973}, in the slightly more general form $Z_n = \xi_n Z_{n-1} + q_n$. We will be interested throughout this work only in the special case where $q_n =\xi_n$. In the scalar case, as mentioned above, the variable $Z_\infty$ is in the domain of attraction of a stable law and there exists an exponent $\nu$ such that the following limit exists and is nonzero: 
\begin{equation}
0 < \lim _{t \rightarrow \infty}  t^\nu  \  \mathbb{P} \left( Z_\infty >t \right) <\infty
\end{equation}
Furthermore, $\nu$ is characterized by \eqref{eq:conditionKappa} (Th. 5 of \cite{Kesten1973}).
De Calan et al.\ give in \cite{Calan} an argument for this characterization by studying poles of the Mellin transforms associated with the relevant distributions in this problem.

The vector case for this discrete recursion was also studied by Kesten, promoting $Z_n$ and $q_n$ to $d$-dimensional vectors and $\xi_n$ to a $d \times d$ matrix. It is shown, under suitable hypothesis, that in this case also there exists $\nu$ such that for any unit $d$-dimensional vector $x$ the following limit exists and is nonzero: 
\begin{equation}
0 < \lim _{t \rightarrow \infty} t^{\nu} \, \mathbb{P}\left( x^T . \ Z_\infty >  t \right) < \infty
\end{equation}
In this case however $\nu$ is not characterized by a simple moment equation. 
 
A continuous version of the scalar recursion was studied in physics as a Langevin equation describing the diffusion of a particle in a one-dimensional white noise Gaussian random field \cite{BouchaudPLD1990} (see section 4.3 there), the interpretation of the Kesten variable being the mean sojourn time of a particle in some region of space. It was studied in mathematics at about the same time by Dufresne in \cite{Dufresne1990,Dufresne2001}. In these works the exact distribution was obtained as an inverse Gamma law (see below). Note that, by contrast, in the discrete case, apart from special cases \cite{Calan}, only the power-law behaviour is accessible. One way to obtain a meaningful continuous version is to consider the variable $U_t = Z_{n=\frac{t}{{\rm d}t}} \, {\rm d}t$, i.e.\ the Kesten variable multiplied by the vanishing time interval. In the context of RWRE it corresponds to going from number of steps to continuous time duration. Writing the random variable $\xi$ in exponential form as $\xi_n = e^{\eta_n}$, the reordered equivalent form \eqref{eq:FormalSumReordered} gives for the variable $U$ at time $t = n {\rm d}t$ :
\begin{equation}
\label{eq:ContinuumReordered}
U_{t}   \ \overset{\mathrm{law}}{=} \ \sum_{j=0}^{n-1} \ e^{\; \sum\limits_{k = 0}^j \eta_k}  {\rm d}t   \quad \overset{{\rm d}t \to 0 }{\longrightarrow}  \quad \int_0^t e^{A_s} \ \mathrm{d}s
\end{equation}
where the process $A_s$ is the continuous limit of the random walk $\sum\limits_{k = 0}^j \eta_k$. In order to ensure a  well-defined limit for the random walk, the {\sc{iid}} $\eta$ random variables must scale as:
\begin{equation}
\eta = \gamma {\rm d}t + \alpha \sqrt{{\rm d}t} X, \quad \quad X \sim \mathcal{N}(0,1)
\end{equation}
where $X$ follows the standard centered normal law. In that case, 
the limit process is a Brownian motion with a drift $ A_t = \gamma t + \alpha W_t$, where we denote throughout this work $W_t$ as the (real scalar) standard Brownian motion. The infinite-time limit of the $U$ variable can then be written as the following Brownian exponential functional:
\begin{equation}
\label{eq:Uinfini}
U_\infty \overset{\mathrm{law}}{=} \int_{0}^{\infty} e^{\gamma t+\alpha W_{t}} \mathrm{d} t
\end{equation}
where $\gamma= \mathbb{E} \left[ \log \xi \right]/{\rm d}t$ and $\alpha^{2}=\mathbb{V}\left[ \log \xi \right]/{\rm d}t$ have well defined limits. Such objects have been widely studied in the mathematical literature \cite{MatYor1,MatYor2}. For some applications in physics see e.g. \cite{Comtet1998,Comtet2005}.

It is also possible to derive the stochastic evolution of $U$ from the random evolution of $Z$. Note that the scaling for $\eta$ shows that one must choose the i.i.d random variables $\xi$ of the following form in the continuous limit: 
\begin{equation} \label{choicexi} 
\xi = 1 + (\gamma + \frac{\alpha^2}{2}) {\rm d}t + \alpha \sqrt{{\rm d}t} X, \quad \quad X \sim \mathcal{N}(0,1)
\end{equation}
The random recursion $Z_{n} = \xi_{n}\left(1+Z_{n-1}\right)$ becomes
$U_{t+{\rm d}t} = (1 + (\gamma + \frac{\alpha^2}{2}) {\rm d}t + \alpha {\rm d}W_t) ({\rm d}t + U_t)$ which,
expanded up to order ${\rm d}t$ using ${\rm d}W_t^2={\rm d}t$, gives the following It\^o stochastic differential equation (SDE) for $U_t$ in the continuous limit, denoting $m = \gamma +\frac{\alpha^2}{2}$ and $\sigma = \frac{\alpha}{\sqrt{2}}$
\footnote{Note that $U_t$ is the continuum analogue of Eq. \eqref{eq:FormalSum} and can be written as $\int_0^t e^{A_t - A_s} \mathrm{d}s$ 
which has the same law (at fixed $t$) as $\int_0^t e^{A_s} \mathrm{d}s$ in Eq. \eqref{eq:ContinuumReordered} by a reordering of variables analogous in the continuum to the discrete reordering employed from Eq. \eqref{eq:FormalSum} to \eqref{eq:FormalSumReordered}. However, note that these processes are not equivalent as can be seen by differing stochastic evolutions. In particular, the evolution of $U_t$ is closed in $U_t$ while it is not the case for the quantity on the r.h.s. of Eq. \eqref{eq:ContinuumReordered}.
\label{footnoteReordering}} 
:
\begin{equation}
\label{eq:StochasticEvolution1D}
{\rm d}U_t = \left( 1+ m \; U_t \right) {\rm d}t + \sqrt{2}\sigma \; U_t {\rm d}W_t \quad \quad \quad \quad \text{(It\^o)}
\end{equation}  
It is possible to obtain the exact stationary large time limit, e.g. using the associated 
Fokker Planck equation, i.e. the PDF of the variable $U_\infty$ which exists
whenever $\gamma<0$ (or equivalently  $m < \sigma^2$). One finds
\cite{BouchaudPLD1990,Dufresne1990,BouchaudMezard} that the inverse of $U_\infty$ follows a gamma law 
\begin{equation}
U_\infty^{-1} \ \sim  \ \Gamma\left( 1 - \frac{m}{\sigma^2}, \sigma^2 \right)
\end{equation}
such that $U_\infty$ itself follows an inverse-gamma law with the following PDF denoted by $P$
\begin{equation}
\label{eq:1DInverseGamma}
P(U_\infty)=\frac{ \sigma ^{-2 + 2\frac{m}{\sigma^2}}}{\Gamma\left(1 - \frac{m}{\sigma^2} \right) } \ U_\infty^{-2 + \frac{m}{\sigma^2}}  \ e^{-\frac{1}{\sigma^{2} U_\infty}} \; .
\end{equation} 
with the heavy tail exponent $1+\nu=2 - \frac{m}{\sigma^2}$.\\

The aim of the present work is to propose a random matrix extension of the Kesten recursion, and to study it both in the discrete and continuous settings. In the continuous setting, the main result will be the convergence to the inverse-Wishart distribution which is the natural matrix extension of the inverse-gamma law, see Appendix \ref{app:WishartInverseWishart} for a precise 
definition. Before  
introducing the model, let us mention results from relevant random matrix models. Until now
only continuum models have been considered, while here we extend the discrete Kesten recursion
to the matrix world. Rider and Valk\'o have extended some identities related to the Dufresne inverse-gamma law
\eqref{eq:1DInverseGamma}
 to matrices in \cite{RiderValko}. In particular they show that, defining the matrix geometric Brownian motion $M_t$ such that $d M_{t}=M_{t} d H_{t}+\left(\frac{1}{2}-\mu\right) M_{t} d t$, where ${\rm d}H_t$ are a matrix of $N^2$ independent standard Brownian motions (see Section \ref{sec:RD}), the random matrix defined by the following integral
\begin{equation}
\label{eq:IntegralRiderValko}
\int_{0}^{\infty} M_{s} M_{s}^{T} \mathrm{d} s
\end{equation}
follows an inverse-Wishart law. This model is a direct matrix extension of the integral expression of the continuous Kesten problem in Eq. \eqref{eq:Uinfini} in which the integrand is a scalar geometric Brownian motion with drift.  
We note that Rider and Valk\'o further extend the geometric \textit{M - X } L\'evy and  \textit{2M - X } Pitman theorems, by considering the $M_{t}^{-1}\left(\int_{0}^{t} M_{s} M_{s}^{T} d s\right) M_{t}^{-T}$ and $M_{t}^{-1} \int_{0}^{t} M_{s} M_{s}^{T} d s$ matrix processes. The matrix Dufresne property and the \textit{2M - X} theorems were also covered and extended by O'Connell in \cite{OConnell2019}.

Another important matrix model in relation to this paper is the one introduced by Grabsch and Texier in \cite{GrabschTexier2016} to model the stochastic evolution of the Riccati matrix associated to a random-mass Dirac equation on the half-line, as a model for a multichannel disordered wire. Furthermore, the same authors have recently recovered in \cite{GrabschTexier2020} a similar stochastic evolution for a different model, namely the Wigner-Smith time delay matrix of a disordered multichannel model obeying a Schr\"odinger equation with random potential (see also \cite{Ossipov}).
In this last setting, their model has very close connections with the integral representation of Rider and Valk\'o. In fact, the general problem of the Wigner-Smith time delay in chaotic cavities has close connections with the present work, and power law tails have also been obtained in that problem \cite{TexierMajumdar13,FyodorovSommers97}. For a general review on the Wigner-Smith time delay see \cite{Texier2016}.

A precise explanation of the difference between the matrix stochastic model introduced in this work with the ones from Rider-Valk\'o and Grabsch-Texier will be adressed in section \ref{sec:LinksRVGT}.
 
\subsection{Definition of the matrix Kesten recursion}

For $N \in \mathbb{N}^*$, let us define the Kesten recursion for a sequence of symmetric positive semidefinite $N \times N$ matrices $(Z_n)_{n \geqslant 0}$,  with  $(\xi_n)_{n \geqslant 0}$ a sequence of independent identically distributed ({\sc{iid}}) symmetric positive semidefinite $N \times N$ matrices and $I$ the $N \times N$ identity matrix, as: 
\begin{equation}
\label{eq:defmodel}
Z_{n+1}=\sqrt{\epsilon I +Z_{n}}  \ \xi_{n} \ \sqrt{ \epsilon I +Z_{n}} 
\end{equation} 
We have introduced a real parameter $\epsilon >0$ in the model, which will be given a value of order $1$ in the discrete setting, and taken to $0$ in the continuous limit. We notice that the symmetric positive semidefinite property of the sequence $\xi$ implies by recursion the same property for the sequence $(Z_n)_{n \geqslant 0}$, as we denote for each $n$ by $\sqrt{\epsilon I +Z_{n}}$ the principal square-root of $\epsilon I +Z_{n}$. The properties of $\xi_n$ and $Z_n$ can be summed up by stating that they are symmetric matrices with only nonnegative eigenvalues.
One expects that for large $n$ the Kesten matrix $Z_n$ tends to a random matrix $Z_\infty$ and one of the aim of this paper
is to study the properties of the distribution of $Z_\infty$.

A generalization where all matrices are Hermitian positive semidefinite will be considered as well. We define the usual Dyson index to be $\beta =1$ in the real symmetric case and $\beta = 2$ in the complex Hermitian case.

The initial motivation for studying such a matrix generalization of the Kesten problem arose from an attempt to define a matrix analogue of the Directed Polymer problem, for which the Kesten recursion is a mean-field version (as we explained in the previous section). 

\subsection{Outline}

The outline of this work is as follows.
In section \ref{sec:ContinuousTime}, we study the continuous limit of this model for a fixed arbitrary integer 
$N$. We obtain the stationary distribution for the Kesten matrix in section \ref{sec:stat}
and provide a mapping 
to a quantum problem of interacting fermions in the Morse potential in section \ref{sec:finite}. 
The case $\beta=2$ corresponding to non-interacting fermions is studied in details in section \ref{sec:NonInteractingFermionsbeta2}
both for finite $N$ and in the large $N$ limit. In addition to their relation to the matrix Kesten recursion, the results in that section are of interest for the physics of
trapped fermions. 
The section \ref{sec:largeN} is devoted to the large $N$ limit, where free probability results can be obtained for the discrete model, as well as for the continuous limit.  
We characterize the stationary random matrix $Z_\infty$ by its $\mathcal{S}$-transform, which obeys the 
self-consistent equation \eqref{scS}. In the continuum limit a resolvent analysis is performed which 
shows agreement with the finite $N$ results.
In section \ref{sec:LinksRVGT} we discuss the links with the Rider-Valk\'o and Grabsch-Texier models.
In particular we show in section \ref{sec:LinksRVGTConnectionsPresentWork} that the flow of the eigenvalues coincide provided appropriate 
correspondence is performed between the discretization prescriptions of the stochastic equations.
Finally the six appendices contain additional details and information.

\section{Finite \texorpdfstring{$N$}{N} Matrix Kesten recursion, in continuous time}
\label{sec:ContinuousTime}
 
In this section we study the continuous limit of the Kesten recursion for $N \times N$ matrices, with $N$ a fixed arbitrary integer. We consider here only Gaussian noise and restrict to strictly positive definite matrices 
$\xi_n$ and $Z_n$ (including the initial condition $Z_0$). 
Remembering the picture laid out in section \ref{sec:Models} for the scalar case, we choose the following structure for the matrices $\xi_n$ which ensures their positive definite nature 
\begin{equation}
\xi_n = \exp \left(  \gamma \epsilon I + \sigma \sqrt{\epsilon } B_n \right)
\end{equation}
where $\epsilon$ is the parameter introduced in \eqref{eq:defmodel}.

The definition of the {\sc{iid}} noise matrices $B_n$ will be different in the real and complex cases. If $\beta =1$, the  matrices $B_n$ are independent and sampled from the Gaussian Orthogonal Ensemble (GOE), i.e. $B_n = ({H_n+H_n^T})/{\sqrt{2}}$ where the entries of $H_n$ are $N^2$ independent real variables following $\mathcal{N}(0,1)$. If $\beta =2$, the  matrices $B_n$ are independent and sampled from the Gaussian Unitary Ensemble (GUE), i.e. $B_n = ({H_n+H_n^T)/2 + i (\tilde{H}_n - \tilde{H}_n^T )}/{2}$ where $H_n$ and $\tilde{H}_n$ are independent samples with the same distribution as above.  
The linear combinations are chosen such that the off-diagonal entries of $B_n$ are real (if $\beta =1$) or complex (if $\beta=2$) Gaussian variables with variance $1$ in both cases. The PDF of the matrix $B_n$ is rotationally invariant and proportional to $\exp({- {\beta} \ {\rm Tr} B_n B_n^\dagger}/4)$. With this choice the spectrum of $B_n/\sqrt{N}$ converges in density at large $N$ to a semi-circle of support 
$[- 2,2]$ for any $\beta$. 
In summary:
\begin{equation} 
\label{eq:DefinitionBn} 
B_n = \frac{H_n + H_n^T}{\sqrt{2\beta}} + i \frac{\tilde{H}_n - \tilde{H}_n^T}{2}\delta_{\beta=2}
\end{equation}

Using the above definitions for $B_n$, the small-$\epsilon$ expansion of the matrix exponential definition for $\xi_n$ gives to order $\epsilon$:
\begin{equation}
\label{eq:structuresigma}
\xi_{n}=(1+m \, \epsilon) I +\sigma \, \sqrt{\epsilon} B_{n}
\end{equation}
where parameters $m=\gamma + \frac{1}{2} \sigma^2(N+\delta_{\beta=1})$ and $\sigma$ 
respectively tune the mean and the noise in $\xi_n$. Indeed for the sake of the
continuum limit obtained below we can replace $\epsilon B_n^2 \to \epsilon I (N+\delta_{\beta=1})$. Note that for $N=1$, \eqref{eq:structuresigma} with $\epsilon={\rm d}t$ identifies with \eqref{choicexi}. With the definition \eqref{eq:structuresigma}, the Kesten recursion for $Z$, Eq. \eqref{eq:defmodel}, becomes:
\begin{equation}
Z_{n+1} = (1+ m \epsilon) (\epsilon I + Z_n) +\sigma \sqrt{\epsilon} \sqrt{\epsilon I +Z_{n}} B_{n} \sqrt{\epsilon I +Z_{n}}
\end{equation}
To first order in $\epsilon$, the square-root of the sum of two commuting matrices can be expanded as:
\begin{equation}
\sqrt{\epsilon I+Z_{n}}=\sqrt{Z_{n}} \sqrt{I+\epsilon Z_{n}^{-1}}=\sqrt{Z_{n}}\left(I+\frac{\epsilon}{2} Z_{n}^{-1} + \mathcal{O}(\epsilon^2) \right)  =\sqrt{Z_{n}}+\frac{\epsilon}{2} \sqrt{Z_{n}^{-1}}+ \mathcal{O}(\epsilon^2) 
\end{equation}
since we have restricted to the case where $Z_n$ has only strictly positive eigenvalues {
Finally, the evolution of $Z$ reads:
\begin{equation}
Z_{n+1}=Z_{n}+\epsilon I+m \epsilon Z_{n}+ \sigma \sqrt{\epsilon} \sqrt{Z_{n}} B_{n} \sqrt{Z_{n}} + \mathcal{O}(\epsilon^{3/2})
\end{equation} 

In the $\epsilon \to 0$ limit, we define a continuous time matrix process $U_t = Z_{t/\epsilon}$, defining time as $t = n \epsilon$. Notice that the insertion of $\epsilon$ in the definition of the recursion avoids the need for a rescaling between $U$ and $Z$. We obtain the time evolution for $U_t$ as the following It\^o matrix SDE: 
 
\begin{equation}
\label{eq:EvolutionU}
{\rm d}U_t = \left( I +m U_t\right) {\rm d}t +\sigma \sqrt{U_t} {\rm d}B_{t} \sqrt{U_t}   \quad  \quad  \quad  \quad \text{(It\^o)}
\end{equation}
with $B_t = ({H_t + H_t^T})/{\sqrt{2\beta}} + i ({\tilde{H}_t- \tilde{H}_t^T})/2 \delta_{\beta=2}  $ where the entries of $H_t$ and $\tilde{H}_t$ are $2 N^2$ independent standard Brownian motions. We remark that taking $N=1$ in the real case $\beta=1$ recovers exactly the continuous Kesten evolution \eqref{eq:StochasticEvolution1D} as expected, because the diagonal terms in the matrix $B_t$ are $\sqrt{2}$ times a standard Brownian. Fixing $N=1$ in the complex case $\beta=2$ requires a rescaling of $\sigma$ by $\sqrt{2}$ to fit the scalar case because a diagonal term of $B_t$ is then a standard Brownian motion.

One method to study the matrix stochastic equation \eqref{eq:EvolutionU} is to write the evolution of
the eigenvalues of $U_t$. The flow that we study is such that $U_t$ is non-degenerate (see below)
and its eigenvalues, denoted $\{\lambda_i (t) \}_{1 \leqslant i \leqslant N}$, are distinct and positive. 
One can then use non-degenerate perturbation theory, as detailed in Appendix \ref{app:PT}. The evolution of the set $\{\lambda_i (t) \}_{1 \leqslant i \leqslant N}$ is obtained as the joint stochastic equations, for 
all $1 \leqslant i \leqslant N$, see \eqref{eq:PTmodelsqUdBsqU}, 
\begin{equation}
\label{eq:EvolutionLambdai}
{\rm d}\lambda_{i}=\left(1+m \lambda_{i}\right) {\rm d}t+  \sigma^2\sum\limits_{\substack{1 \leqslant j \leqslant N \\j\neq i}} \frac{\lambda_{i} \lambda_{j}}{\lambda_{i}-\lambda_{j}} {\rm d}t+\sqrt{\frac{2}{\beta}} \sigma \lambda_{i} {\rm d}W_{i}
  \quad  \quad \quad \quad \text{(It\^o)}
\end{equation} 
where the $(W_i(t))_{1 \leqslant i \leqslant N}$ are $N$ independent standard Brownian motions. 
Note that the evolution of the eigenvalues is obtained with no a priori knowledge of the evolution of
the eigenvectors. As compared to the case $N=1$, which recovers the scalar continuous Kesten evolution,
we note the interaction between the eigenvalues in the form ${\lambda_{i} \lambda_{j}}/{\lambda_{i}-\lambda_{j}}$ for each pair. Note the similarity with the usual Dyson Brownian motion (DBM) \cite{DysonBM65}. However the
difference is that here, because of the multiplicative noise, the repulsion depends also multiplicatively on the eigenvalues.

From the stochastic equations \eqref{eq:EvolutionLambdai} one can obtain the evolution of the
joint probability distribution function (JPDF) $P(\vec{\lambda},t )$ of the set of eigenvalues $\{ \lambda_i\}_{1 \leqslant i \leqslant N}$ denoted by the vector $\vec{\lambda}$. One finds that the Fokker-Planck equation associated with the eigenvalue evolution of the It\^o process $U_t$ is:
\begin{equation}
\label{eq:FPEigenval}
\frac{\partial P(\vec{\lambda},t)}{\partial t}=\frac{1}{2} \sum_{i=1}^N \frac{\partial^{2}}{\partial \lambda_{i}^{2}}\left(\left(\sqrt{\frac{2}{\beta}} \sigma \lambda_{i}\right)^{2} P(\vec{\lambda},t)\right)-\sum_{i=1}^N \frac{\partial}{\partial \lambda_{i}}\left(\left(1+m \lambda_{i}+  \sigma^2\sum\limits_{\substack{1 \leqslant j \leqslant N \\j\neq i}}\frac{\lambda_{i} \lambda_{j}}{\lambda_{i}-\lambda_{j}}\right) P(\vec{\lambda},t)\right)
\end{equation}
where the initial condition is specified by the eigenvalues of $U_0$ (which we keep arbitrary at this stage).

 \subsection{Large-time limit: stationary solution}
 \label{sec:stat} 

In this section we obtain the stationary solution of the \eqref{eq:FPEigenval}. The equation 
\eqref{eq:FPEigenval} can be written as $\frac{\partial P(\vec{\lambda},t)}{\partial t}= - \sum_i \partial_{\lambda_i} J_i$, which defines the components of the current (or flux) $J_i$. The stationary solution
$P_\mathrm{stat}(\vec{\lambda})$ is such that $\sum_i \partial_{\lambda_i} J_i=0$, and here it will be found as the normalizable solution of the zero-flux condition $J_i=0$ for all $i$, namely
\be \label{Ji} 
0 = J_i = \frac{\partial}{\partial \lambda_{i}}\left( \frac{1}{\beta}\left(  \sigma \lambda_{i}\right)^{2} P_\mathrm{stat} \right)-   \left(1+m \lambda_{i}+ \sigma^{2} \sum\limits_{\substack{1\leqslant j\leqslant N\\ j\neq i}}    \frac{\lambda_{i} \lambda_{j}}{\lambda_{i}-\lambda_{j}}\right) P_\mathrm{stat}
\ee
Inspired by the DBM we will look for a solution of the form
\be
P_\mathrm{stat}(\vec{\lambda}) = F(\vec{\lambda}) \; \big| \Delta(\vec{\lambda}) \big| ^\beta 
\ee 
in terms of $\Delta(\vec{\lambda})= \prod_{1 \leqslant i < j \leqslant N} (\lambda_j - \lambda_i) $ the determinant of the Vandermonde matrix $(\lambda_i^{j-1})_{1\leqslant i,j \leqslant N}$. Using the following identities:
\begin{equation}
   \partial_{\lambda_i} \log |\Delta(\vec{\lambda})| = \sum_{j \neq i} \frac{1}{\lambda_i-\lambda_j}; \qquad  \sum_{j \neq i} \frac{\lambda_i^2}{\lambda_i-\lambda_j} = (N-1) \lambda_i + \sum_{j \neq i} \frac{\lambda_i \lambda_j}{\lambda_i-\lambda_j},
\end{equation} 
where the second one is used when computing the derivative in the first term
in \eqref{Ji}, one finds that $F(\vec{\lambda})$ must obey the set of equations for $1 \leqslant i \leqslant N$
\be
\frac{\sigma^2}{\beta} \lambda_i^2 \partial_{\lambda_i} F = 
(1 + \lambda_i (m - \frac{2 \sigma^2}{\beta} - \sigma^2 (N-1))) F 
\ee 
where the interaction term has cancelled. Integrating this equation one finally obtains the stationary 
measure for any $N$ as 
\begin{equation}
\label{eq:EigenJoint}
P_\mathrm{stat}(\vec{\lambda})=K \ \big| \Delta(\vec{\lambda}) \big| ^\beta  \ \prod_{i=1}^N \lambda_{i}^{ \beta \left( \frac{m}{\sigma^{2}} - N +1 \right) -2 } \  e^{-\frac{\beta}{\sigma^{2}} \sum_{i=1}^N \frac{1}{\lambda_{i}}}
\end{equation}
with $K$ a normalization constant. We note that for $N=1$ the Vandermonde factor is absent 
and setting $\beta=1$ one recovers exactly the inverse gamma law for $\lambda_1 \equiv U_\infty$, i.e. Eq.
\eqref{eq:1DInverseGamma}, 
$\beta=1$ (and also for $\beta=2$ setting in \eqref{eq:EigenJoint} $\sigma \to \sigma \sqrt{2}$).

One recognizes in \eqref{eq:EigenJoint} the joint PDF of the eigenvalues of the
inverse-Wishart ensemble. We recall the (white) inverse-Wishart matrix distribution
over $N \times N$ real ($\beta=1$) or Hermitian ($\beta=2$) positive matrices, 
with parameters $T,C$
  \begin{equation}
\mathcal{P}_{\mathrm{iw}}\left( M \right) = Q_{T, N, C, \beta}   \
\operatorname{det}(M)^{ -\frac{\beta}{2} (T + N -1 ) -1 }
  \ e^{- C \frac{\beta}{2} \operatorname{tr} M^{-1}}
 \end{equation} 
 and its joint eigenvalue distribution: 
\begin{equation}
P_{\mathrm{iw}}(\vec{\lambda} ) =K_{T, N, C, \beta}  \ \left|\Delta(\vec{\lambda})\right|^{\beta}  \ \prod_{i=1}^{N} \lambda_{i}^{-\frac{\beta}{2} (T + N -1 ) -1 }
 \ e^{-C \frac{\beta}{2} \sum_{i=1}^N \frac{1}{\lambda_{i}}  } 
\end{equation}
where the definition and normalization constants are detailed in Appendix \ref{app:WishartInverseWishart}. We denote $q=\frac{N}{T} $. We thus see in \eqref{eq:EigenJoint} that $P_\mathrm{stat}$ is the Inverse-Wishart eigenvalue joint distribution with the following correspondence of parameters:
\begin{equation}
\label{eq:parameters}
\left\{
    \begin{array}{ll}
        T = N  - \frac{2 m }{\sigma^2 } + \frac{2}{\beta }  -1  \\[10pt]
        \frac{1}{q} = \frac{T}{N} = 1- \frac{1}{N} \left( \frac{2 m }{\sigma^2 } - \frac{2}{\beta }  +1  \right)  \\[10pt]
        C = \frac{2}{ \sigma^2}  \\[10pt]
        K= K_{T, N, C, \beta}   
        =  \frac{1}{N!} \ \sigma ^{- N \left( \beta (N-1-\frac{2m}{\sigma^2}) +2  \right)} 
\frac{ \Gamma\left(\frac{\beta}{2}\right)^N  \ \pi^{\frac{\beta}{2}N(N-1) }}{\Gamma_{N,\beta}( \frac{\beta}{2}N) \ \Gamma_{N,\beta}( \frac{\beta}{2} (N- \frac{2m}{\sigma^2}-1) +1) }
    \end{array}
\right. 
\end{equation}  
The eigenvector-basis rotation invariance of ${\rm d}B_t$ in \eqref{eq:EvolutionU} then allows us to conclude that the full matrix distribution $\mathcal{P}_{\mathrm{stat}}\left( U \right) $ is a stationary distribution for this model, with:
\begin{equation}
\label{eq:StationaryMatrixDist}
\mathcal{P}_{\mathrm{stat}}\left( U \right) = Q  \ \operatorname{det}(U)^{ \beta \left( \frac{m}{\sigma^{2}} - N +1 \right) -2 }
 \ e^{- \frac{\beta}{\sigma^2} \operatorname{tr} U^{-1}}
\end{equation}
where $Q =Q_{T, N, C, \beta}   =  \frac{ \sigma ^{- N \left( \beta (N-1-\frac{2m}{\sigma^2}) +2  \right)}  }{ \Gamma_{N,\beta}( \frac{\beta}{2} (N- \frac{2m}{\sigma^2}-1) +1) }  $.

We note that this distribution is well-defined only when the arguments of the $\Gamma$ functions in the denominator of $K_{T, N, C, \beta}$ are positive in order to avoid the poles of $\Gamma$ at negative integers: $1- \beta \frac{m}{\sigma^2} + \frac{\beta}{2}(j-1) >0 \ , \ \forall j \in [1, N]$. As a consequence, the normalizability condition for the stationary distribution does not depend on $N$:
\begin{equation}
\label{eq:ConstraintNormalizability}
m < \frac{\sigma^2}{\beta}
\end{equation}
It is indeed similar to the $N=1$ case obtained in continuous scalar Kesten evolution \eqref{eq:StochasticEvolution1D}, as can be seen in the real case $\beta=1$.

Another way to interpret this constraint is that the Inverse-Wishart distribution is well-defined only if the corresponding Wishart samples are almost surely invertible, i.e. they have full-rank. Since Wishart samples are obtained as $X^\dagger X$ where $X$ is a $T\times N$ matrix with independent Gaussian entries, they have full-rank only if $T > N-1$. Again, this yields \eqref{eq:ConstraintNormalizability}. 

In order to shed light on the stationary solution, we detail the marginal density for the largest eigenvalue $\lambda_1$. In the limit where $\lambda_1$ is large, for a fixed value of $N$, its distribution reads
\begin{equation}
\label{eq:QueueMarginalLargestEval}
P(\lambda_1) \quad 
\stackrel{\lambda_1 \to \infty}{\approx} 
\quad
\frac{N \; (2/\sigma^2)^{1-\frac{\beta m} {\sigma^2}} \; \Gamma(1+\frac{\beta}{2}) \; \Gamma(1+\frac{\beta}{2} (N-\frac{2m}{\sigma^2}))}{\Gamma(1+\frac{\beta}{2}N) \; \Gamma(1-\frac{\beta m}{\sigma^2}) \; \Gamma(1-\frac{\beta m}{\sigma^2} + \frac{\beta}{2})}
\ \lambda_1^{-2 + \beta \frac{m}{\sigma^2}} \; ,
\end{equation}
see Appendix \ref{app:WishartInverseWishart} for the derivation. We see that the tail exponent of the maximal eigenvalue distribution is independent of $N$ and coincides with the $N=1$ scalar Kesten result \eqref{eq:1DInverseGamma} as can be seen simply by injecting $\beta =1$. We will show below (see section \ref{sec:ConstantSigmaParagraph}) that, in the limit of large $N$ and fixed $\sigma$, this tail is related to the universal statistics at the hard-edge of the Wishart random matrix ensemble, which is given by the Bessel determinantal point process.

Note here some connections with the statistics of the Wigner-Smith time delay studied in the 
context of chaotic cavities \cite{Brouwer97,Brouwer99,Cunden16}.
In Ref. \cite{TexierMajumdar13}, the proper time delay $\tau$ was studied. It has the same distribution as the following sum over the eigenvalues
$\lambda_i$ studied here, i.e. $\tau= \frac{1}{N} \sum_{i=1}^N \lambda_i$. The parameters used in \cite{TexierMajumdar13}
correspond here to $\sigma^2=2$ and $m=-N$. With these parameters we see that our result 
\eqref{eq:QueueMarginalLargestEval} exhibits the same power law as the one 
conjectured in \cite{FyodorovSommers97} and obtained in \cite{TexierMajumdar13} in the large $N$ limit from a Coulomb gas calculation. This suggests that
the sum is dominated by the largest eigenvalue. We will return to the large $N$ limit below.

\subsection{Finite-time solution: Morse-Sutherland quantum mapping}
 \label{sec:finite}

In the case of the usual Dyson Brownian motion, there exists a mapping onto a quantum system, the Calogero-Sutherland model \cite{RMTBook}.
In the present case, as we now show, there also exists such a mapping, albeit onto a different quantum model. To elicit this
mapping we  
first perform a change of variables by defining 
\be 
\lambda_i = e^{x_i} \; .
\label{eq:Definitionx}
\ee
From It\^o rules, the stochastic evolution of the $x$ variables has additive noise following:
\begin{equation}
\label{eq:LangevinMu}
{\rm d}x_i = \mathcal{F}_i {\rm d}t + \sqrt{\frac{2}{\beta}} \sigma {\rm d}W_i \qquad \mathcal{F}_i := e^{-x_i} + m - \frac{\sigma^2}{\beta} +  \sigma^2 \sum\limits_{\substack{1\leqslant j\leqslant N\\ j\neq i}}  \frac{1}{e^{ x_i - x_j} -1},
\end{equation}
with an initial condition $x_i = x_0  \ \; \forall i \in [1,N]$ at $t=0$ with $x_0$ a large negative constant, and $\mathcal{F}_i$ the force felt by particle $x_i$. Let us denote $\tilde P(\vec{x},t)$ the probability distribution at time $t$ for the set of $x$ variables, obtained from $P(\vec{\lambda},t)$ through the change of variables. The Fokker-Planck equation verified by
$\tilde P(\vec{x},t)$ 
is:
\begin{equation}
\label{eq:FokkerPlanckQuantumMapping}
\frac{\partial \tilde P}{\partial t } =  \sum_{i=1}^N   \left( \frac{\sigma^2}{\beta} \frac{\partial^2 \tilde P }{\partial x_i^2} - \frac{\partial}{\partial x_i} ( \mathcal{F}_i \tilde P)  
\right)
= - \hat H_{FP} \tilde P 
\end{equation}
where we have introduced the generator $ \hat H_{FP} $. One can map this generator to a Schr\"odinger problem in the fashion of \cite{Gautie1} as follows.
Denoting $\tilde P_\mathrm{stat}(\vec{x}) $ the distribution obtained from $P_\mathrm{stat}(\vec{\lambda}) $ under this change of variables, we define $\psi(\vec{x},t)$ such that:
\begin{equation}
\tilde P(\vec{x},t) = \left( \tilde P_\mathrm{stat}(\vec{x}) \right)^{\frac{1}{2}} \ \psi(\vec{x},t)
\end{equation}
Injecting $P$ into \eqref{eq:FokkerPlanckQuantumMapping} yields the following Schr\"odinger equation on $\psi(\vec{x},t)$:
\begin{equation}
\label{eq:FokkerPlanckQuantumMapping2}
\frac{\partial \psi }{\partial t} =  \sum_{i=1}^N \left( \frac{\sigma^2}{\beta} \frac{\partial^2 \psi}{\partial x_i^2} 
- \left(  \beta\frac{\mathcal{F}_i^2}{4 \sigma^2} + \frac{1}{2}\frac{\partial \mathcal{F}_i}{\partial x_i}  \right) \ \psi
\right)
= - \frac{\sigma^2}{\beta} (  \hat{H} - E_0) \psi 
\end{equation} 
where the constant $E_0$ is given in \eqref{eq:DefEground}. The stochastic system is hereby mapped to a quantum system with the following Hamiltonian:
\begin{eqnarray} 
\hat{H} &=& \sum_i  \left( - \frac{\partial^2}{\partial x_i^2} + V(x_i) +\sum\limits_{\substack{1\leqslant j\leqslant N\\ j\neq i}}  V_\mathrm{int}(x_i,x_j )   \right)
\end{eqnarray} 
with a 1-particle potential $V$ of the Morse form $A e^{-2 x} - B e^{-x}$ \cite{Morse29} and with Sutherland $\sinh^{-2}$ interaction potential $V_\mathrm{int}$ \cite{SutherlandBook}: 
\be 
\label{eq:QuantumPotential}
{\large
\begin{array}{ll}
V(x_i) = \beta^2 \frac{e^{-2 x_i}}{4 \sigma^4}  - \frac{\beta}{\sigma^2}
 \left(  \frac{\beta}{2}(N-1-  \frac{m}{ \sigma^2} )+1\right) e^{- x_i}  
  \\[5pt]
  V_\mathrm{int}(x_i,x_j ) = 
   \frac{\beta (\beta -2) }{16} 
  \frac{1 }{\sinh(\frac{x_i-x_j}{2})^2} 
\end{array}
}
\ee

We thus see that the stochastic generator $\hat H_{FP}$ and the quantum hamiltonian $\hat H$ are related by the operator transformation
\be 
\label{eq:OperatorRelation}
\frac{\sigma^2}{\beta} (\hat{H} - E_0) = \tilde P_\mathrm{stat}(\vec{x})^{-1/2} \hat H_{FP} \tilde P_\mathrm{stat}(\vec{x})^{1/2} 
\ee 
The eigenenergies ${\cal E}$ of $\hat H_{FP}$ and $E$ of $\hat H$ are also related through
\be
\mathcal{E} = \frac{\sigma^2}{\beta}  ( E - E_0)
\ee

The lowest Fokker-Planck eigenvalue $\mathcal{E}$ is of course equal to $0$, the eigenvalue of the stationary state, such that the constant $E_0$ introduced in \eqref{eq:FokkerPlanckQuantumMapping2} is indeed the ground-state energy of $\hat{H}$ and is computed from the correspondence to be
  \begin{equation}
  \label{eq:DefEground}
E_0 =
- N   \frac{( \beta m- \sigma^2 )^2}{4\sigma^4} 
 +   \frac{N(N-1)}{4} (\frac{\beta^2}{\sigma^2} m - \beta  \frac{\beta+2}{2}  )  
 - N(N-1)(N-2)\frac{\beta^2}{12} 
  \end{equation}
It turns out that the ground state of the quantum problem is $N!$ degenerate \cite{Calogero69}. Indeed one can 
choose the sign of the wavefunction for each ordering of the $x_i$. This can be seen from the 
common feature in Random Matrix Theory that the eigenvalues of matrix processes do not intersect. It can be verified in particular in \eqref{eq:LangevinMu}: fixing two indices $i$ and $j$, let us assume that $x_i > x_j$ at some time $t$; the evolution of the difference follows, when $x_i - x_j \to 0$, with $\tilde{W}$ a standard Brownian Motion:
\begin{eqnarray}
{\rm d}( x_i - x_j) &=& ( e^{- x_i} - e^{-x_j} ) \left(  1 + \beta \sigma^2 \sum_{k \neq i,j} \frac{e^{x_k} }{(e^{x_i} - e^{x_k}) ( e^{x_j} - e^{x_k}) } \right) {\rm d}t +  \sigma^2 \frac{e^{ x_i - x_j} + 1}{e^{ x_i - x_j} -1}{\rm d}t + 2 \frac{\sigma}{\sqrt{\beta}}  {\rm d}\tilde{W} \\
&\sim&  \frac{  2  \sigma^2 }{x_i- x_j} {\rm d}t + 2 \frac{\sigma}{\sqrt{\beta}} {\rm d}\tilde{W}, \qquad \qquad (x_i - x_j \to 0)
\end{eqnarray}
Scaling the time variable by an adequate factor, we see that this equation becomes the SDE $2X dX = {\beta} {\rm d}t + 2 X {\rm d}\tilde{W}$, i.e. a $(\beta + 1)$-Bessel process describing the norm of a Brownian motion in $\beta +1$ dimensions. For any $\beta \geqslant 1$ this process never hits zero as a consequence of the recurrence properties of Brownian motion, and we can conclude that $x_i \neq x_j$ for all times. As a consequence the wave-function $\psi$ can be 
chosen to have the fermionic symmetry (i.e antisymmetric in all its arguments). Hence we are led to study
a a system of $N$ fermions in the Hamiltonian described above.

\subsection{ The case \texorpdfstring{$\beta=2$}{b=2}: non-interacting fermions in the Morse potential}
\label{sec:NonInteractingFermionsbeta2}

In the special case $\beta =2$, the mapping is simplified as the interaction term vanishes, leading to a system of $N$ non-interacting spinless fermions in the Morse potential. From Eq. \eqref{eq:Definitionx} the position of the fermions $x_i$ are related to the eigenvalues $\lambda_i$ of the Kesten matrix via $x_i=  \log \lambda_i$. Using the known spectrum of the single-particle model, we detail in section \ref{spectrum}
the ground-state wavefunction for the system of $N$ fermions. It is constructed as usual as a Slater determinant where the $N$
single-particle states of energy below the Fermi energy $\epsilon_F$ are occupied.
This provides an interpretation of the stationary measure of the model studied in this paper,
which is related to the inverse-Wishart ensemble. It also unveils a correspondence between the inverse-Wishart random
matrix ensemble and non-interacting fermions in a Morse potential. 

The present study thus adds the Morse potential to the list of potentials in one dimension which
are related to random matrices \cite{DeanLeDoussal19,LACTLeDoussal17,LACTLeDoussal18,NadalMaj09,CundenMezzOconnell18,Forrester,MehtaBook}. It includes the harmonic oscillator (related to the GUE), the inverse square potential (related to
the Wishart-Laguerre ensemble) and the Jacobi box potential (related to the Jacobi ensemble). 
Note however that a distinct feature of the Morse potential is that it can only accommodate a finite number of bound states, $N_b$. We say that the Morse potential is full when all the bound states are occupied by the $N$ fermions.

We study in \ref{sec:largeNMorse} this correspondence in the large $N$ limit. There are two interesting scaling regimes depending how
the parameter $\sigma$ is chosen in that limit. 
If $\sigma$ is scaled as $\mathcal{O}(1/\sqrt{N})$ the support of the density of the Fermi gas is a finite interval.
If $\sigma$ remains of $\mathcal{O}(1)$ in that limit, the Morse potential is almost full and the Fermi gas extends very far to the right.
In that regime, we find that the fluctuations of the largest eigenvalues of the Kesten matrix, equivalently of the rightmost fermions in the Morse potential, are described by the so-called Bessel determinantal point process up to a change of variable.

Finally we write an exact formula for the full dynamics of the model in section \ref{sec:finitetimeresults}. 
It is interesting to note that the integral functional of the geometric Brownian motion, which is equivalent to the continuous scalar Kesten recursion as we recalled in Section \ref{sec:Introduction}, has a prominent interest in the field of quantitative finance \cite{yor}. In the specific setting of the Asian option, the quantum dynamics in the Morse potential has proved useful to compute the fair price of the financial product in \cite{Zhang10}. Our present study can thus be viewed as an $N$-particle generalization of this analytical correspondence, although there is no clear financial applications of this generalization.
 
\subsubsection{Energy spectrum and ground state of \texorpdfstring{$N$}{N} fermions} 
 
\label{spectrum} 
 
The Morse potential is described by the following single-particle Hamiltonian (in units where the mass is set to $1/2$ and $\hbar=1$)
\begin{equation}
\label{eq:1particleHamiltonian}
\hat{H}_1 = - \frac{\partial^2}{\partial x^2} + V(x) \quad , \quad V(x)=  g^2( e^{- 2 (x - x_0) } - 2 e^{-(x-x_0)})
\end{equation} 
where the parameters $g$ and $x_0$ denote respectively the strength of the potential and the position of the minimum. The value at the minimum $V(x_0)=-g^2$
is negative and the potential tends to zero at $x=+\infty$, see Fig \ref{fig:Morse}. For the Kesten matrix model studied in this paper the values of these parameters are fixed as
\be
\label{eq:linkgxzero} 
g = N-\frac{m}{\sigma^2} \quad \text{and} \quad e^{-x_0} = \sigma^2 \left(N-\frac{m}{\sigma^2}\right) \; ,
\ee
and depend explicitly on $N$, the number of fermions, and this has important consequences
discussed below. However the study of the fermion problem that we perform below is valid for general $g$ and $x_0$.

\begin{figure}[!t]
\centering
\includegraphics[width= 0.6 \textwidth]{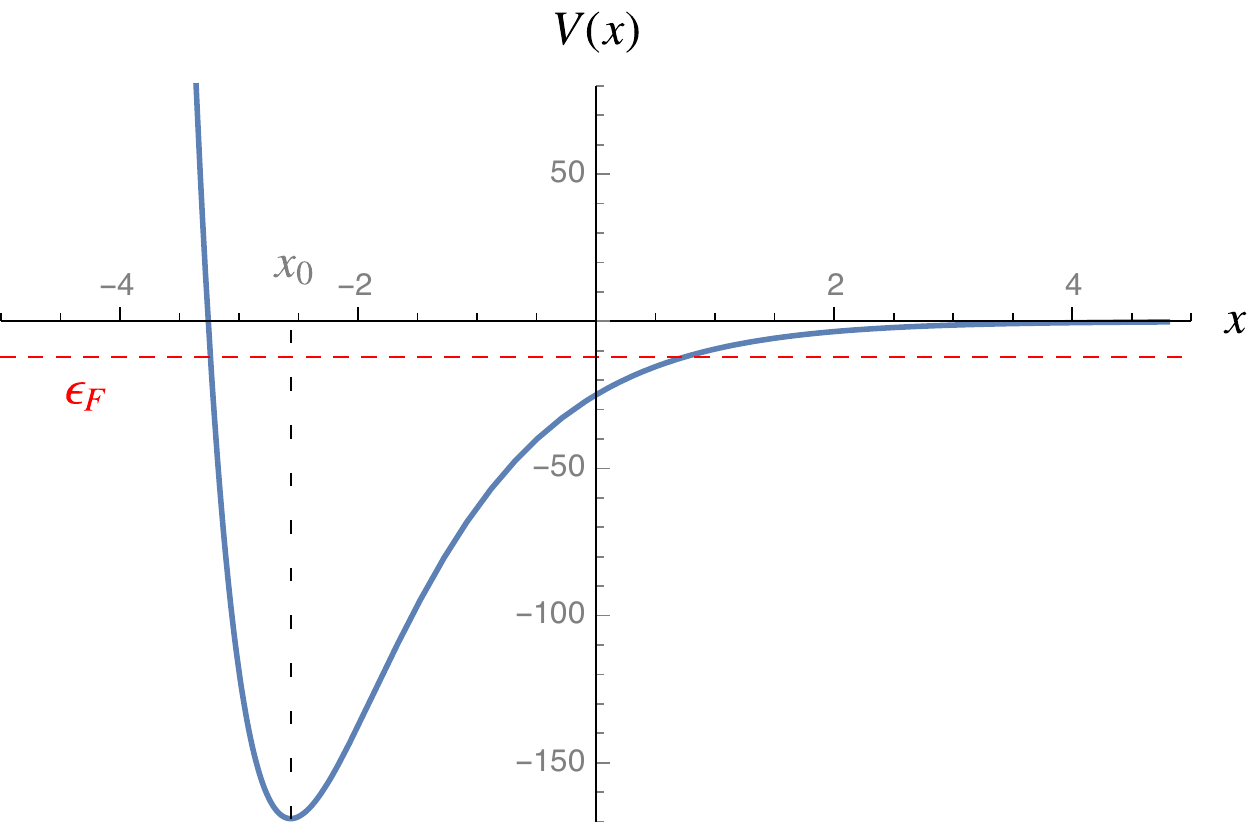}
\caption{The Morse potential $V(x)$ in \eqref{eq:1particleHamiltonian}, plotted with parameters $x_0\simeq -2.565$ and $g=13$, which correspond to 
$\sigma=1$, $m=-3$ and $N=10$ in the Kesten model. The energy of the highest occupied state by the $N=10$ fermions (i.e. the Fermi energy $\epsilon_F$ given in Eq.
\eqref{eq:LastOccupiedLevelGroundState}) is indicated by the dashed line.
For these parameters the Morse potential contains $N_b=13$ bound states, it is thus almost full.}
\label{fig:Morse}
\end{figure}

It is shown in Appendix \ref{app:MorsePresentation} that this potential exhibits a finite number $N_b= \lfloor \left(g + \frac{1}{2}\right)^- \rfloor$ of bound states, the largest integer strictly smaller than $g + \frac{1}{2}$, as well as a continuous branch of diffusion states with positive energies. The bound states in the discrete branch of the spectrum have the following eigenenergies $\epsilon_k$ and eigenfunctions $\Psi_k$ with $0\leqslant k < g - \frac{1}{2}$:}
\begin{eqnarray}
    \label{eq:EigenergiesEigenfunctionsbeta2}
    \epsilon_k &=& - \left(g - k - \frac{1}{2}\right)^2  = - \left( N-\frac{m}{\sigma^2} - k - \frac{1}{2}\right)^2 \\
    \psi_{k}(x) &=& N_{k} \left( \frac{2}{\sigma^2} e^{-x}\right)^{N-\frac{m}{\sigma^2}-k-\frac{1}{2}} e^{-  \frac{1}{\sigma^2} e^{- x }} L_{k}^{(2N-\frac{2m}{\sigma^2}-2 k-1)}\left(\frac{2}{\sigma^2} e^{- x } \right) 
\end{eqnarray}  
where $L_k^{(\alpha)}$ is the generalized Laguerre polynomial and $N_k = \left[\frac{k !(2 g-2 k-1)}{\Gamma(2 g-k)}\right]^{\frac{1}{2}} $.

The ground-state wavefunction of the $N$-fermion model, denoted $\vec{0}$, is then simply the Slater determinant obtained from the $N$ lowest 1-particle eigenstates $k=0,1, \cdots, N-1$:
\begin{eqnarray}
          \Psi_{\vec{0}} ( \vec{x} ) &=&  \frac{1}{\sqrt{N!}} \det_{1\leqslant i,j \leqslant N} \left( \psi_{i-1}(x_j)\right) \\
        E_{\vec{0}} &=&  \sum_{k=0}^{N-1} \epsilon_k  
\end{eqnarray}  
It can be verified as expected that $E_{\vec{0}}$ is exactly the ground-state energy obtained by inserting $\beta =2$ in \eqref{eq:DefEground}. It can also be verified that the stationary solution of the matrix Kesten problem corresponds to the ground state through  $\tilde P_\mathrm{stat}(\vec{x}) = \big| \Psi_{\vec{0}}(\vec{x} ) \big|^2 $. Indeed, we can develop the modulus square expression:
\be 
 \big| \Psi_{\vec{0}}(\vec{x} ) \big|^2 = \frac{1}{N!} \prod_{k=1}^{N}   N_{k-1}^2 \left( \frac{2}{\sigma^2} e^{-x_k}\right)^{2(N-\frac{m}{\sigma^2}-\frac{1}{2})} e^{-  \frac{2}{\sigma^2} e^{- x_k }} 
 \left( \det_{1\leqslant i,j \leqslant N}
 (\frac{2}{\sigma^2} e^{-x_j})^{-i+1}
 L_{i-1}^{(2N-\frac{2m}{\sigma^2}-2 i+1)}\left(\frac{2}{\sigma^2} e^{- x_j } \right) \right)^2
\ee
The polynomials  $X^{i-1}
 L_{i-1}^{(2N-\frac{2m}{\sigma^2}-2 i+1)}\left( {X}^{-1}   \right)$ forming a family of increasing degree $i-1$ in $X$, the determinant in this expression is of the Vandermonde form. More precisely, the leading term in these polynomials have coefficient
  $\binom{2N-\frac{2m}{\sigma^2}- i}{i-1}$, the constant term of $L_{i-1}^{(2N-\frac{2m}{\sigma^2}-2 i+1)}$, where the binomial is extended to non-integers by a standard analytic continuation. This gives, extracting constants from the Vandermonde:
\be 
 \big| \Psi_{\vec{0}}(\vec{x} ) \big|^2 = \frac{\left( \sigma^2/2 \right)^{N(N-1)}}{N!} \prod_{k=1}^{N}   N_{k-1}^2 
 \binom{2N-\frac{2m}{\sigma^2}- k }{k-1}^2  \left( \frac{2}{\sigma^2} e^{-x_k}\right)^{2N-2\frac{m}{\sigma^2}-1} e^{-  \frac{2}{\sigma^2} e^{- x_k }} 
\big|\Delta \left( \{  e^{x} \}\right) \big|^2
\ee
where the set $\{e^{x_i}\}_{1 \leqslant i \leqslant N}$ is denoted here as $\{  e^{x} \}$ instead of a vector, for ease of notation. Rearranging the constant terms, we obtain in this case $\beta =2$:
\be 
\label{eq:GroundStateJPDF}
 \big| \Psi_{\vec{0}}(\vec{x} ) \big|^2 = K  \
 \big|\Delta \left( \{e^{x}\}\right)\big|^\beta \
 \prod_{k=1}^{N}    \left(  e^{x_k}\right)^{\beta ( \frac{m}{\sigma^2}-N+1)-1} e^{-  \frac{\beta}{\sigma^2} e^{- x_k }}  \quad , \quad   \beta = 2
\ee 
where $K$ is the constant given earlier in \eqref{eq:parameters}. Changing variables to $\lambda = e^x$ then gives exactly the stationary distribution obtained earlier in \eqref{eq:EigenJoint},  for the case $\beta=2$:
\be 
P_\mathrm{stat}(\{ \lambda \} ) = K \ \big|\Delta \left( \vec{\lambda}\right)\big|^2 \
\prod_{k=1}^{N}     \lambda_k  ^{-2N+2\frac{m}{\sigma^2}} e^{-  \frac{2}{\sigma^2}   \frac{1}{\lambda_k}} 
\ee 
This shows that the modulus square of the ground-state wavefunction is indeed the stationary distribution on the $x$ variables: $\tilde P_\mathrm{stat}(\vec{x}) = \big| \Psi_{\vec{0}}(\vec{x} ) \big|^2 $. 

As we mentioned above the Morse potential has a finite number of bound states $N_b= \lfloor \left(g + \frac{1}{2}\right)^- \rfloor$.
Hence for a generic value of $g$ it can only accommodate $N \leq N_b$ fermions in their ground state. Since for the matrix
Kesten problem $g=N- \frac{m}{\sigma^2}$ depends itself on $N$, we must check for consistency and for the existence of the solution. The ground-state wavefunction, as a Slater determinant of quantized states, exists if all states are well-defined i.e. $N-1 < g - \frac{1}{2}$, or equivalently $ m < \frac{\sigma^2}{2}$. We see that, in this case $\beta=2$, this condition agrees perfectly with the condition obtained in the stochastic study for the normalizability of the stationary solution \eqref{eq:ConstraintNormalizability}. In the ground state, we further note that the Fermi energy $\epsilon_F$, the last occupied energy level, is independent of the number of particles:
\begin{equation}
\label{eq:LastOccupiedLevelGroundState}
     \epsilon_F   =  \epsilon_{N-1} = - \left( \frac{1}{2} - \frac{m}{\sigma^2} \right)^2
\end{equation} 
which holds only for $ m < \frac{\sigma^2}{2}$, and is strictly negative. Note that
when $- \frac{1}{2} \leqslant \frac{m}{\sigma^2} <\frac{1}{2} $ the potential is full with $N=N_b$,
while for $\frac{m}{\sigma^2} < - \frac{1}{2} $ one has $N<N_b$.
The region of parameters $m \geqslant \frac{\sigma^2}{2}$
corresponds in the fermion problem to a positive Fermi energy, in which case there are continuum eigenstates extending to infinity,
and no finite-$N$ ground state. It can be given a sense for finite $N$ at finite time (see below) and in the Kesten problem
it corresponds to runaway behavior of the Kesten matrix variable to infinity, with no stationary state.

The fermionic setting suggests a determinantal rewriting of the stationary solution, as in \cite{DeanLeDoussal19,Gautie2}:
\begin{eqnarray}
\tilde P_\mathrm{stat}(\vec{x}) &=& \frac{1}{N!} \left( \det_{1 \leqslant i,j \leqslant N}(\psi_{i-1}(x_j)) \right)^2
= \frac{1}{N!} \det_{1 \leqslant i,j \leqslant N} \left( \sum_{k=0}^{N-1} \psi_{k}(x_i) \psi_k(x_j)
\right) = 
 \frac{1}{N!}
\det_{1 \leqslant i,j \leqslant N}  K_{N}(x_i,x_j) 
\end{eqnarray}
where the kernel $K_{N}$ is:
\begin{equation}
\label{eq:KernelFermions}
K_{N}(x,y) = (2/\sigma^2)^{2(N-\frac{m}{\sigma^2}-\frac{1}{2} ) }  
   \sum_{k=0}^{N-1}  \frac{ \sigma^{4k} N_k^2}{4^{k}}     e^{-(x+y)(N-\frac{m}{\sigma^2}-k-\frac{1}{2})} e^{-  \frac{1}{\sigma^2}(  e^{- x }+e^{-y})}     L_k^{(2N-\frac{2m}{\sigma^2}-2k-1)}(\frac{2}{\sigma^2}e^{-x})
    L_k^{(2N-\frac{2m}{\sigma^2}-2k-1)}(\frac{2}{\sigma^2}e^{-y})
\end{equation}
By the orthonormalization of the eigenfunctions $\int \psi_{k}^{*}(x) \psi_{k^{\prime}}(x) \mathrm{d}x=\delta_{k, k^{\prime}}$, this kernel satisfies the reproducibility property, which can be written as:
\begin{equation}
    \int K_{N}(x, z) K_{N}(z, y) \mathrm{d} z=K_{N}(x, y)
\end{equation} 
This property implies that the joint distribution $\tilde P_\mathrm{stat}$ describes a determinantal point process (DPP) with kernel $K_N$. 
From standard results on DPP's \cite{DeanLeDoussal19,Johansson05,Borodin11}, this implies that all $n$-point correlations can also be written as determinants in terms of $K_N$.
In particular, the 1-particle density in the ground state
$P_1(x)$ is given as
\bea
\label{eq:P1exactdensity}
    P_1(x) &:=& \frac{1}{N} \mathbb{E}\left[ \sum_{k=1}^N \delta(x-x_k) \right] = \frac{1}{N} K_N(x,x) \\
    &=&\frac{(2/\sigma^2)^{2(N-\frac{m}{\sigma^2}-\frac{1}{2} ) } }{N} 
   \sum_{k=0}^{N-1}  \frac{ \sigma^{4k} N_k^2}{4^{k}}     e^{-2x (N-\frac{m}{\sigma^2}-k-\frac{1}{2})} e^{-  \frac{2}{\sigma^2} e^{- x } }     \left(L_k^{(2N-\frac{2m}{\sigma^2}-2k-1)}(\frac{2}{\sigma^2}e^{-x})\right)^2 
\eea
where in the definition the average is taken over the ground-state JPDF. This formula has been used to plot the fermion density in Figs. \ref{fig:densitysigmatilde} and \ref{fig:densityconstantsigma}.

The above results were obtained for $\beta=2$. In the case $\beta=1$, which corresponds to fermions in the Morse potential
in presence of Sutherland like interactions given in \eqref{eq:QuantumPotential}, the ground-state wavefunction is also
obtained from \eqref{eq:GroundStateJPDF} substituting $\beta=1$.

\subsubsection{Large \texorpdfstring{$N$}{N} behavior} 
\label{sec:largeNMorse}

When $N$ is large, the density of the Fermi gas in the potential $V(x)$ \eqref{eq:1particleHamiltonian} is well described in the bulk by the 
semi-classical limit, also called local density approximation (LDA) \cite{DeanLeDoussal19}, $P_1(x) \simeq P_\mathrm{bulk}(x)$. The LDA density $P_\mathrm{bulk}$ is supported on an interval $[x_-,x_+]$ and vanishes at the edges:
\be
\label{eq:LDADensity}
P_\mathrm{bulk}(x)=\frac{1}{\pi N} \sqrt{E_\mathrm{eff} - V(x)} = \frac{ge^{x_0}}{\pi N } \sqrt{( e^{-x}- e^{-x_+})(e^{-x_-} - e^{-x})  }
\ee
where the edges $x_\pm$ verify $V(x_\pm) =E_\mathrm{eff}$. Here $E_\mathrm{eff}$ is an effective Fermi energy defined such that the normalization condition $1= \int_{x_-}^{x^+} P_\mathrm{bulk}(x) \mathrm{d}x$ holds. One obtains the relations $e^{-x_-}+e^{-x_+}=2e^{-x_0}$ and $E_\mathrm{eff}=-g^2e^{-(x_+ + x_-)+2x_0}$
which leads to $e^{- x_\pm}= e^{-x_0} (1 \mp \sqrt{1+ \frac{E_\mathrm{eff}}{g^2}})$, hence the positions of the edges are given by
\be \label{log} 
x_\pm = x_0 - \ln (1 \mp \sqrt{1 + \frac{E_\mathrm{eff}}{g^2} }  ) 
\; .
\ee

Now we note that one can rewrite \eqref{eq:LDADensity} upon the change of variable $\Lambda= 2 g e^{-(x-x_0)}$ and using 
the above result for $x_\pm$ as
\be 
P_\mathrm{bulk}(x) \mathrm{d}x
= \frac{g - \sqrt{- E_\mathrm{eff}}}{N} \rho_{\rm MP}(\Lambda) \mathrm{d} \Lambda 
\quad , \quad \Lambda_\pm = 2 g e^{-(x_\mp-x_0)}
\ee
where 
\be
\rho_{\rm MP}(\Lambda)= \frac{2}{\pi \Lambda (\sqrt{\Lambda_+} - \sqrt{\Lambda_-})^2} \sqrt{ (\Lambda - \Lambda_-) (\Lambda_+ - \Lambda) }
\ee 
is the Marcenko-Pastur distribution, which is normalized to unity $\int_{\Lambda_-}^{\Lambda^+} {\rm d}\lambda \rho_{\rm MP}(\Lambda)=1$, see Appendix \ref{app:WishartInverseWishart}. Hence the normalization condition for $P_\mathrm{bulk}(x)$ immediately implies that the the effective Fermi energy
is related to the number of fermions as 
\be \label{effFermi} 
E_\mathrm{eff} =  - (g-N)^2
\ee  
The appearance of the Marcenko-Pastur distribution originates from the fact that the variables $\Lambda_i= 2 g e^{-(x_i-x_0)}$
obey, in the fermionic ground state, the statistics of the eigenvalues of a Wishart matrix. This fact will be used
again below. Inserting the result for $E_\mathrm{eff}$ in Eq. \eqref{log} we find that the positions of the edges become  
\be \label{xpm} 
x_\pm  =  
x_0 - \ln (1 \mp 
\sqrt{\frac{N}{g}} \sqrt{
2-\frac{N}{g}
}  )
\; .
\ee
for $N<g$. One sees that as $N/g \to 1^-$, the variable $x_+-x_0$ diverges which means that, relative to the minimum of the well, the right edge moves to $+\infty$. Hence one 
recovers the condition on the number of particles discussed above, which in the large $N$ limit reads $N<g$. Thus to study the large $N$ limit we need to choose the strength of the Morse potential $g$ to
scale as $N$. 

We now apply these results to our Kesten matrix problem and use the parameter values $g= N - \frac{m}{\sigma^2}$ and $e^{-x_0}=\sigma^2 g$ given in \eqref{eq:linkgxzero}.
In that case the effective Fermi energy \eqref{effFermi}  becomes
\be 
\label{eq:EeffParamKesten}
E_\mathrm{eff} = -\frac{m^2}{\sigma^4}
\ee

At this point we should make an important distinction: depending on the scaling of $\sigma$ with $N$, the asymptotic behavior of the system will be drastically different. \\

\paragraph{Scaling $\sigma \sim 1/\sqrt{N}$.} 

\begin{figure}[!ht]
\centering
\includegraphics[width=0.45\textwidth]{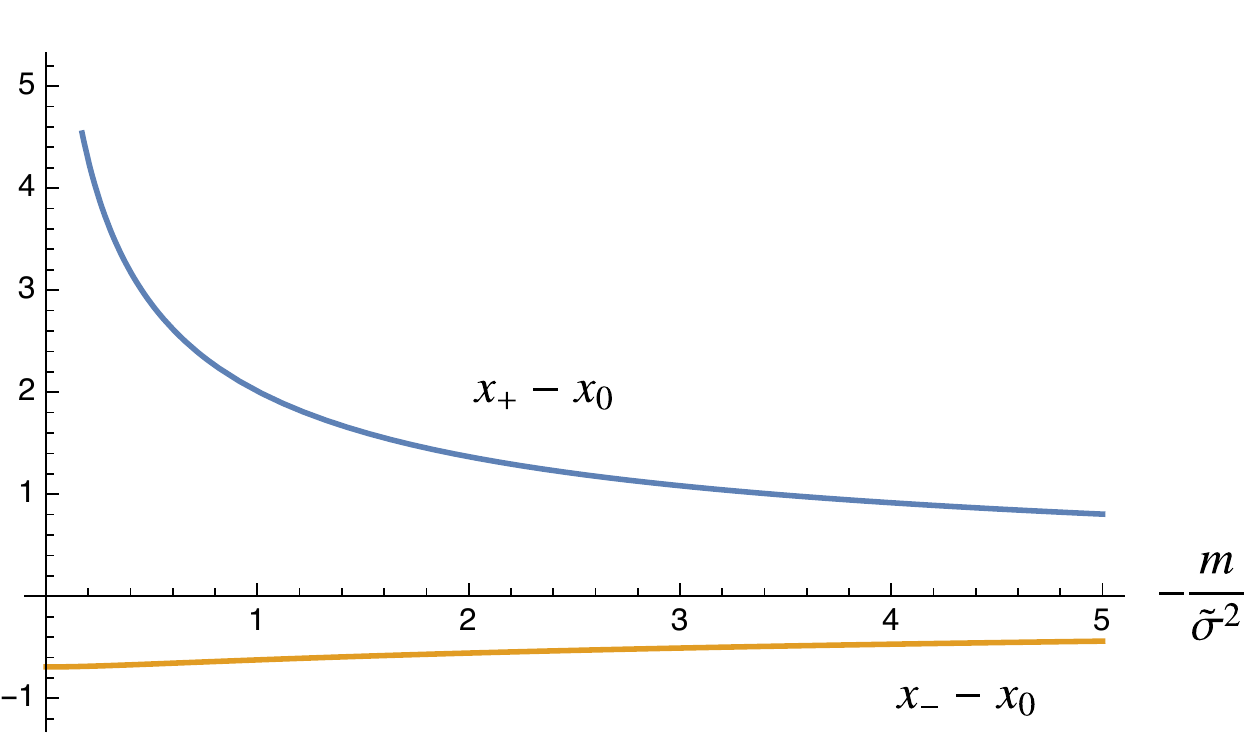}
\hspace{.5cm}
\includegraphics[width=0.45\textwidth]{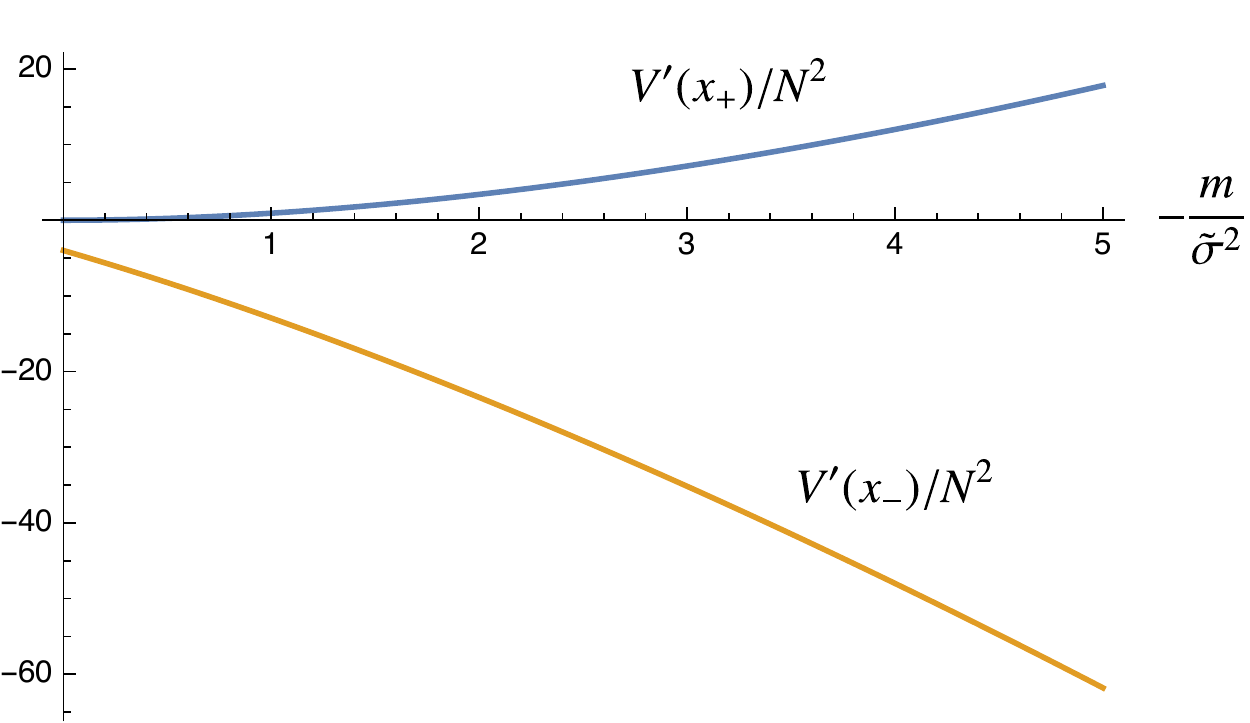}  
\caption{In the regime where $\sigma = \tilde \sigma /\sqrt{N}$, we plot the relative positions of the two edges $x_\pm -x_0$ (left) and the derivative of the potential at the edges $V'(x_\pm)/N^2$ (right) as a function of the positive variable $-\frac{m}{\tilde \sigma^2}$. Note that $x_+ - x_0$ diverges when $-\frac{m}{\tilde \sigma^2}$ tends to $0$, while $x_- - x_0$ tends to $-\ln 2$.}
\label{fig:xplusmoins}
\end{figure}

In the case where  $ \sigma=\tilde \sigma/\sqrt{N}$ with $\tilde \sigma$ kept constant, one has $g=N(1-\frac{m}{\tilde\sigma^2})$
and $e^{-x_0}=\tilde \sigma^2 (1-\frac{m}{\tilde\sigma^2})$.
Because of the exact condition $m <\frac{\sigma^2}{2}\sim 1/N$, here we must choose $m < 0$ which ensures that $N<g$ is satisfied. We see that the effective Fermi energy \eqref{eq:EeffParamKesten} coincides in this small $\sigma$ limit with the last occupied level \eqref{eq:LastOccupiedLevelGroundState}. In that regime we see from \eqref{xpm} that the asymptotic support of the fermion density is $[x_-,x_+]$ such that the positions of the two 
edges $x_\pm = x_0   - \ln (1 \mp 
\frac{ \sqrt{ 1 - 2\frac{m}{\tilde\sigma^2}}}{1- \frac{m}{\tilde\sigma^2}}
  )$ remain fixed and of order $\mathcal{O}(1)$, see Fig. \ref{fig:xplusmoins}.
  
  In the bulk, the density is given by the LDA in \eqref{eq:LDADensity}:
\be 
P_\mathrm{bulk}(x) = \frac{1}{\pi \tilde\sigma^2}
 \sqrt{( e^{-x}- e^{-x_+})(e^{-x_-} - e^{-x})  }
\ee
  
  At these points, the potential has the following derivative:
  \be 
  V'(x_\pm) = 2 N^2   \sqrt{ 1 - 2\frac{m}{\tilde\sigma^2}}  \left( \pm \left(1- \frac{m}{\tilde\sigma^2} \right) - \sqrt{ 1 - 2\frac{m}{\tilde\sigma^2}} \right) 
  \ee

\begin{figure}[!ht]
\centering
\includegraphics[width= 0.75 \textwidth]{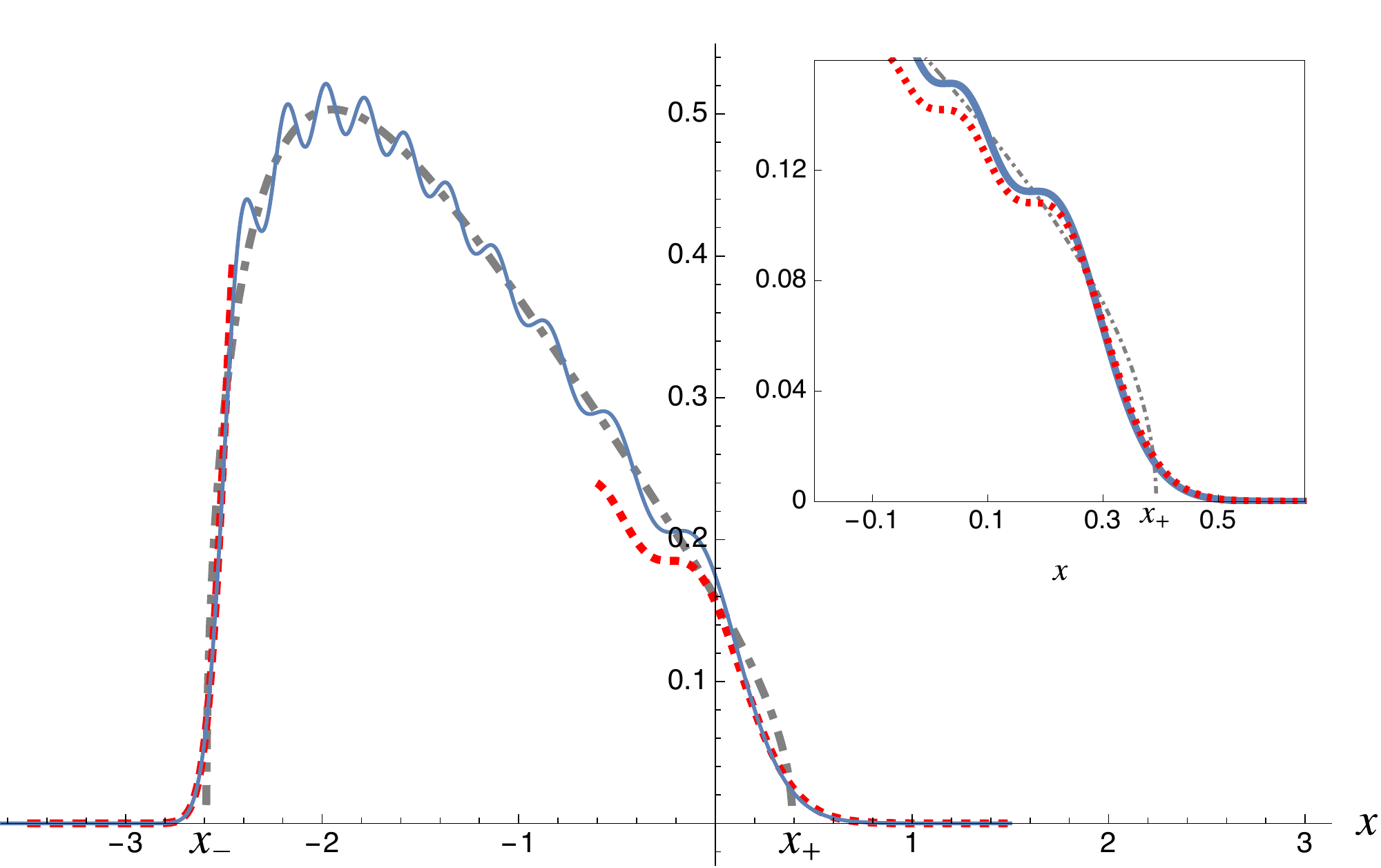}
\caption{\emph{Scaling $\sigma \sim 1/\sqrt{N}$ :\ } Plot of the 1-particle density $P_1(x)$ given in \eqref{eq:P1exactdensity} for $N=10$ particles with $\tilde \sigma=2$, such that $\sigma=\frac{2}{\sqrt{10}}$, and $m=-3$ (solid blue line). For these parameters, we have $x_-\simeq -2.59$, $x_0\simeq -1.95$ and $x_+ \simeq 0.39$. The bulk density $P_{\rm bulk}(x)$ given in \eqref{eq:LDADensity} obtained from the LDA is the dashed gray line. The left and right edge densities $P^\mathrm{edge}_\pm(x)$ obtained from the Airy kernel in \eqref{eq:PedgesAiryplusminus} are the dashed red lines. The inset shows the region near the right edge for $N=50$, with $\tilde \sigma=2, m=-3$ as above. We see that $P^\mathrm{edge}_+(x)$ is closer to the exact density for this higher number of particles.}
\label{fig:densitysigmatilde}
\end{figure}

Near the two edges, denoting $
 w_\pm =  V'(x_\pm) ^ {-1/3} = \mathcal{O} \left( N^{-2/3}\right) $ the length scale which characterizes the width of the edge regions, the density is correctly described by the one obtained from the Airy kernel such that \cite{DeanLeDoussal19}:
\be  
\label{eq:PedgesAiryplusminus}
P^\mathrm{edge}_\pm(x) \simeq \frac{1}{N \abs{w_\pm}} F_{1}\left(\frac{x-x_\pm}{w_\pm}\right)
, \quad
 F_{1}(z)=\left[\mathrm{Ai}^{\prime}(z)\right]^{2}-z[\mathrm{Ai}(z)]^{2}
\ee 
The agreement of the 1-particle density with the approximations in the bulk and at the edges, in the case of the $\sigma \sim 1/\sqrt{N}$ scaling, is illustrated in Fig. \ref{fig:densitysigmatilde}.

\paragraph{Constant $\sigma$.} 
\label{sec:ConstantSigmaParagraph}
 In the case where $\sigma$ is a constant of order 1, the position of the minimum of the potential is asymptotically ${x_0 = - \ln(\sigma^2 N)}+ \mathcal{O}(\frac{1}{N})$. $g$ scales as $N$, such that the left edge is at a fixed distance of the minimum $x_- = x_0 - \ln 2 + \mathcal{O}(\frac{1}{N^2})$, while the right edge is at diverging distance from the minimum ${x_+ = x_0 + 2 \ln (\sigma^2 N) - \ln (\frac{m^2}{2}) + \mathcal{O}(\frac{1}{N})}$. In the translated frame where the position of the minimum is fixed, the Fermi gas only has one edge to the left, while extending infinitely to the right for $N = + \infty$.  
  
In the vicinity of the left edge, the Airy kernel characterizes the behavior of the Fermi gas. With $V'(x_-)= -4 g^2 \simeq -4 N^2$ and $w =V'(x_-)^{-1/3} = \mathcal{O}(N^{-2/3})$, the density is then described as above by:
\be  
\label{eq:PleftedgeAiry}
P^\mathrm{left \; edge}(x) \simeq \frac{1}{N \abs{w}} F_{1}\left(\frac{x-x_-}{w}\right)
, \quad
 F_{1}(z)=\left[\mathrm{Ai}^{\prime}(z)\right]^{2}-z[\mathrm{Ai}(z)]^{2}
\ee  

In order to obtain the behavior of the Fermi gas to the far right, we will use again the mapping to the Wishart matrix ensemble. As mentioned above, it is easily seen that the ground-state JPDF \eqref{eq:GroundStateJPDF} can be mapped, under the change of variables $\Lambda = 2 g e^{-(x-x_0)} = \frac{2}{\sigma^2}e^{-x}$, to the following Wishart eigenvalue JPDF of (fixed) parameter $-\frac{2m}{\sigma^2}$:
\be 
P_\mathrm{Wishart} ( \vec{\Lambda} ) = 
 K  \
 \big|\Delta \left( \vec{\Lambda}\right)\big|^2 \
 \prod_{k=1}^{N} \Lambda_k^{-\frac{2m}{\sigma^2}} e^{- \Lambda_k} 
\ee 
Under this mapping, the large $x$ behavior can be accessed by studying the left edge of the Wishart model $\Lambda \simeq 0$. It is well-known that this hard-edge behavior in the Wishart ensemble is characterized by the Bessel kernel \cite{DeanLeDoussal19,TracyWidom94}, such that the kernel 
$K_W$
of the DPP of the $\Lambda_i$'s is, in the large $N$ limit for $\Lambda = \mathcal{O}(1/N)$:
\be 
K_W(\Lambda_1,\Lambda_2) \simeq 4N K_{\mathrm{Be}, -\frac{2m}{\sigma^2}}( 4N \Lambda_1, 4N \Lambda_2), \quad 
K_{\mathrm{Be}, \nu}(u, v)=\frac{\sqrt{v} \mathrm{~J}_{\nu}^{\prime}(\sqrt{v}) \mathrm{J}_{\nu}(\sqrt{u})-\sqrt{u} \mathrm{~J}_{\nu}^{\prime}(\sqrt{u}) \mathrm{J}_{\nu}(\sqrt{v})}{2(u-v)}
\ee
where $J_\nu$ is the Bessel function of order $\nu$.
This is equivalent to the fact that in the large $N$ limit the eigenvalues near the edges can be written as $\Lambda_i \simeq \frac{b_i}{4 N}$ 
where the $b_i$'s form the Bessel DPP of Kernel $K_{\mathrm{Be}, \nu}$. The Bessel DPP is defined as a limit process for $N=+\infty$ 
with $b \in [0,+\infty[$. Its mean density $\rho(b) = \mathbb{E}[ \sum_{i=1}^{+\infty} \delta(b-b_i) ]$ 
decays at large $b$ as $\sim 1/\sqrt{b}$. Many of its properties are known, such as the PDF of the minimum
$b_{\rm min}= \min_i b_i$ \cite{TracyWidom94}. 

We can now translate these properties to describe the fermions which are {\it very far to the right}:
their positions can be written as 
\be
e^{-x_i} = \frac{\sigma^2}{8 N} b_i \quad , \quad x_i = \ln N - \ln(\sigma^2 b_i/8) \quad , \quad x_i = x_+ + \ln( \frac{4m^2}{\sigma^4 b_i})
\ee

 We deduce the fermion kernel $K_N$ \eqref{eq:KernelFermions} in the large $x,y$ regime, ie $x-x_+ = \mathcal{O}(1)$ and $y-x_+ = \mathcal{O}(1)$ :
 
\be 
K_N (x,y) \simeq \frac{8N}{\sigma^2}e^{-(x+y)/2} 
K_{\mathrm{Be}, -\frac{2m}{\sigma^2}}
\left( \frac{8N}{\sigma^2}e^{-x} , \frac{8N}{\sigma^2}e^{-y} \right)
\ee  
and in particular the density:
\bea
\label{eq:PfarrightBessel}
P^\mathrm{far \ right}(x) &\simeq& \frac{8}{\sigma^2}e^{-x} K_{\mathrm{Be}, -\frac{2m}{\sigma^2}}
\left( \frac{8N}{\sigma^2}e^{-x} , \frac{8N}{\sigma^2}e^{-x} \right)  \\ &=& 
\frac{2}{\sigma^2}e^{-x} \left(
J_{-\frac{2m}{\sigma^2}} \left( \frac{2\abs{m}}{\sigma^2}e^{-\frac{x-x_+}{2}} \right)^2
-
J_{-\frac{2m}{\sigma^2}+1} \left( \frac{2\abs{m}}{\sigma^2}e^{-\frac{x-x_+}{2}}\right)
J_{-\frac{2m}{\sigma^2}-1} \left( \frac{2\abs{m}}{\sigma^2}e^{-\frac{x-x_+}{2}}\right)
\right)  
\eea

The fluctuations of the position of the rightmost fermion $x_\mathrm{max} = x_+ + \ln( \frac{4m^2}{\sigma^4 b_\mathrm{min}})$ are of order 1 and can be obtained from the PDF of $b_\mathrm{min}$:
\be 
P_{b_\mathrm{min}}(s) = \frac{\sigma(s)}{s} e^{-\int_0^s \frac{\sigma(x)}{x} \mathrm{d}x}
\ee 
 where $\sigma$ satisfies a Painlev\'e III equation (Eq. (1.21) in \cite{TracyWidom94}). The PDF of $b_\mathrm{min}$ vanishes at $s=0$ as $s^{-\frac{2m}{\sigma^2}}$.
Taking into account the prefactor (Eq. (1.22) in \cite{TracyWidom94}), we deduce the tail of the distribution of the rightmost fermion as:
\be 
P_{\mathrm{max}}(x)
\ \stackrel{x \to \infty}{\simeq} \ \frac{(m^2/\sigma^4)^{1-\frac{2m}{\sigma^2}}}{ \Gamma(1-\frac{2m}{\sigma^2}) \Gamma(2-\frac{2m}{\sigma^2} )}e^{-(1-\frac{2m}{\sigma^2})(x-x_+)}
\ee 
Finally, we note that changing variables to $\lambda=e^x$ and injecting the value of $x_+$ given at the beginning of the paragraph, we recover exactly the tail of the marginal distribution of the largest eigenvalue $\lambda_1$ in \eqref{eq:QueueMarginalLargestEval} for the present case $\beta=2$, with matching prefactor in the large $N$ limit:
\be 
P(\lambda_1) \ \stackrel{\lambda_1 \to \infty}{\simeq} \
 \frac{(2 N/\sigma^2)^{1-\frac{2m}{\sigma^2}}}{ \Gamma(1-\frac{2m}{\sigma^2}) \Gamma(2-\frac{2m}{\sigma^2} )}
 \lambda_1^{-2 +\frac{2m}{\sigma^2}}
\ee  
 
\begin{figure}[!ht]
\centering
\includegraphics[width= 0.75 \textwidth]{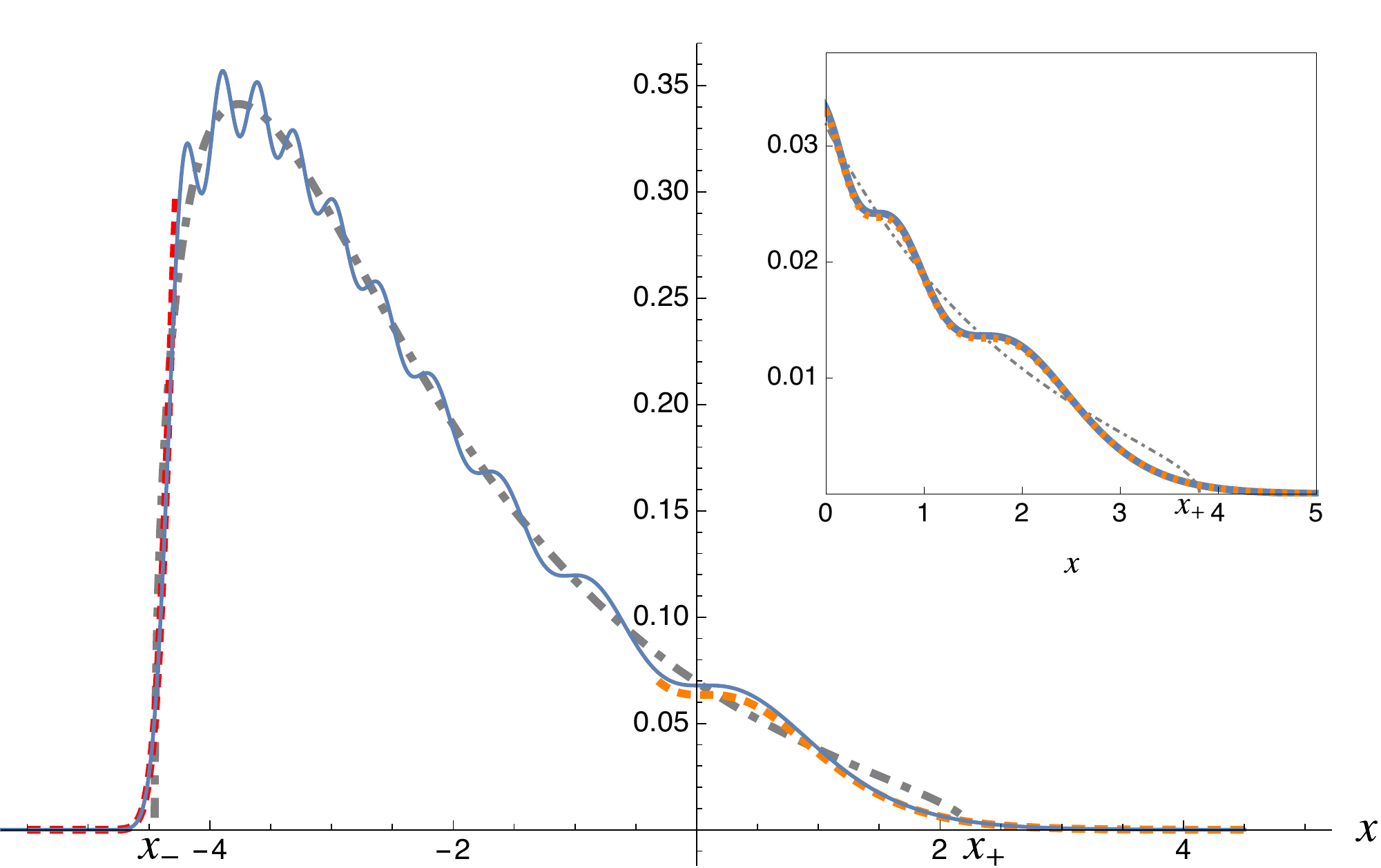}
\caption{\emph{Constant $\sigma$ :\ } Plot of the 1-particle density $P_1(x)$ given in \eqref{eq:P1exactdensity} for $N=10$ particles with $\sigma=2,m=-3$ (solid blue line). For these parameters, we have $x_-\simeq -4.45$, $x_0\simeq -3.76$ and $x_+ \simeq 2.26$. The bulk density $P_{\rm bulk}(x)$ given in \eqref{eq:LDADensity} obtained from the LDA is the dashed gray line. The density
at the left edge $P^\mathrm{left \; edge}(x)$ (red dashed line) is given by the Airy kernel, see Eq. \eqref{eq:PleftedgeAiry}. The density $P^\mathrm{far \ right}(x)$ on the far right
near $x_+ \sim \log N$ (orange dashed line) is obtained from 
the Bessel kernel, and given in Eq. \eqref{eq:PfarrightBessel}. The inset shows the region near the right edge for $N=50$, with $\sigma=2,m=-3$ as above. For these parameters, we have $x_+ \simeq 3.81$. We see that the exact density is very closely approximated by $P^\mathrm{far \ right}(x)$ in that region.}
\label{fig:densityconstantsigma}
\end{figure}

The agreement of the 1-particle density with the approximations in the bulk and at the edges, in the case of constant $\sigma$, is illustrated in Fig. \ref{fig:densityconstantsigma}.

Note again that in the interacting case $\beta=1$, the ground-state wavefunction, \eqref{eq:GroundStateJPDF} with $\beta=1$, can also
be analyzed in the large $N$ limit. It corresponds to an inverse Wishart matrix model with $\beta=1$. For the case $\sigma=\mathcal{O}(1)$, the left edge
is now described, upon the same change of variable, by the universality of the Gaussian orthogonal ensemble. The fermions on the
far right, corresponding to the largest eigenvalues of the continuous matrix Kesten recurrence, are now described,
upon this change of variable, by the $\beta=1$ Bessel point process (which is not determinantal, see
\cite{Rider09} for its description). 

\subsubsection{Finite time results}
\label{sec:finitetimeresults}

The mapping of the matrix Kesten recursion onto the quantum problem allows us to study its time evolution. In the case $\beta=2$ one 
can obtain an exact formula for this time evolution. 

Taking into account the bound states and the continuous branch of the spectrum, the Euclidean propagator, or Green's function in imaginary time, of a 1-particle system in the Morse potential \eqref{eq:1particleHamiltonian} is, see App. \ref{app:MorsePresentation}:
\bea
G(x,x',t)  &=&  \sum_{k=0}^{\lfloor \left(g - \frac{1}{2} \right)^- \rfloor} \psi_k(x) \psi_k(x') e^{- (g - k - \frac{1}{2})^2 t} \nonumber \\
&& + \int_0^{+\infty} \frac{dp}{2 g \pi^2} p \sinh(2 \pi p) |\Gamma(i p - g + \frac{1}{2})|^2   e^{\frac{x+x'}{2} - x_0} W_{g, i p}(2 g e^{-(x-x_0)}) W_{g, i p}(2 g e^{-(x'-x_0)}) e^{- p^2 t} 
\eea
in terms of the Whittaker $W$ function. From completeness, this propagator satisfies $G(x,x',0) = \delta(x-x')$. The propagator of the $N$-particle system of non-interacting fermions is then simply:
\be
G_N(\vec{x},\vec{x}',t)= \bra{\vec{x}} e^{-t \hat{H} } \ket{\vec{x}'}= \det_{1 \leqslant i,j \leqslant N } G(x_i,x_j',t)
\ee
which satisfies $G_N(\vec{x},\vec{x}',0) = \delta(\vec{x}-\vec{x}')$. Assuming a known initial position for the fermions $\vec{x}_0$, the $N$-particle wavefunction at a finite time $t$ is $ \psi(\vec{x},t) =G_N(\vec{x},\vec{x}_0,t) $. From this quantum result, we obtain using Eq. \eqref{eq:OperatorRelation} the finite time solution of the stochastic equation \eqref{eq:LangevinMu}:
\begin{equation}
\tilde P( \vec{x} ,t)   
=    \langle x | e^{- t H_{FP} } | x_0 \rangle =  \left(\frac{\tilde P_\mathrm{stat}( \vec{x} )}{\tilde P_\mathrm{stat}( \vec{x}_0 )} \right)^{\frac{1}{2} }   G_N(\vec{x},\vec{x}_0,t) e^{E_0 t}, \quad \quad \tilde P( \vec{x} ,0) = \delta( \vec{x} - \vec{x}_0)   
\end{equation}
Upon change of variable this gives directly the following finite-time solution for the stochastic evolution of the $\lambda_i = e^{x_i}$ variables:
\be 
P(\vec{\lambda},t) = \tilde  P( \vec{x} = \ln \vec{\lambda},t)  \; \prod_{k=1}^N \frac{1}{\lambda_i}
\ee
 where the vector $\ln \vec{ \lambda}$ denotes the set $\{\ln \lambda_i\}_{i\leqslant i \leqslant N}$.
 
\section{Discrete recursion in the large \texorpdfstring{$N$}{N} limit}
\label{sec:largeN} 

In the large $N$ limit, we study the discrete matrix Kesten recursion \eqref{eq:defmodel} with tools from free probability theory and through the analysis of the resolvent evolution. We treat first the case of the discrete recursion, and then study the continuum version. In the continuum case,
the results obtained by these two methods are compared with the $N \to \infty$ limit of the finite $N$ results obtained in the previous section
(noting that here we scale the parameter $\sigma=\mathcal{O}(1/\sqrt{N})$ in that limit). We note that the Dyson index $\beta$ does not have the same significance in the free case, since all the information is contained in the eigenvalue density: results apply to both $\beta=1,2$ cases. 
In the last subsection, we complete the study with the analysis of the expected characteristic polynomial for arbitrary $N$.

\subsection{Free probability approach}

The property of freeness is a generalization of the concept of independence to non-commuting random variables, such as random matrices \cite{Voiculescu85,Voiculescu91,Voiculescu95, MingoSpeicher2017,TulinoVerdu,Novak14,BunBouchaudPotters2017,RMTBook}. The field of free probability theory has seen the development of numerous results and tools relevant to the study of free variables, see Appendix \ref{app:FreeProba} for a brief summary of the main transforms used in this work and \cite{RMTBook} for a precise introduction to freeness and detailed consequences for random matrices. The general result that we apply in this section to establish a connection to free probability is the following: two large symmetric matrices whose eigenbasis are randomly rotated with respect to one another are free, see \cite{BunBouchaudPotters2017}.

\subsubsection{Discrete matrix recursion}

With help from free probability tools, applicable in the large $N$ limit, we study the discrete matrix recursion of Eq. \eqref{eq:defmodel}:
\begin{equation}
\label{eq:defmodelpart3}
Z_{n+1}=\sqrt{\epsilon I +Z_{n}}  \ \xi_{n} \ \sqrt{ \epsilon I +Z_{n}} 
\end{equation}
We make two assumptions on the distribution of $\xi$: i) its support is of order 1 in the large $N$ limit, ii) it is invariant under rotations. By the rotational invariance of $\xi$, $\epsilon I + Z_n$ and $\xi_n$ in the Kesten recursion \eqref{eq:defmodel} become free random matrices in the $N \to \infty$ limit. The model under study then defines $Z_{n+1}$ as the free (symmetrized) multiplication of $\epsilon I + Z_n$ and $\xi_n$. Free multiplication is endowed with a very powerful property \cite{BunBouchaudPotters2017,RMTBook}: the $\mathcal{S}$-transform -- defined below and in App. \ref{app:FreeProba} -- of the product matrix $Z_{n+1}$ factorizes to the product of the $\mathcal{S}$-transforms of $\epsilon I+Z_n$ and $\xi_n$:
\begin{equation}
\label{eq:FactorizationStransforms}
\mathcal{S}_{Z_{n+1}} = \mathcal{S}_{\epsilon I + Z_n} \times \mathcal{S}_{\xi}
\end{equation} 
where the dependence on $n$ in $\mathcal{S}_{\xi}$ has been dropped as the $\xi_n$ matrices are {\sc{iid}}. In the following, we use this property to write a recursion on $\mathcal{S}_{Z_n}$, in order to characterize the evolution of the spectrum through the evolution of the $\mathcal{S}$-transform. See App. \ref{app:FreeProba} for more details on the free probability transforms used in this section.

We introduce $Y_n =  \epsilon I + Z_n$ for ease of notation. Let us express the $\mathcal{S}$-transform of this variable in terms of $\mathcal{S}_{Z_n}$. 
For a matrix $M$, the eigenvalue density is defined as $\rho :=  \mathbb{E} \left[ \frac{1}{N} \sum_{\lambda \in \mathrm{Sp}(M)} \delta_\lambda \right] $.
For $u \in \mathbb{R}$, the spectrum densities of $Z_n$ and $Y_n$ are related as:
\begin{equation}
\rho_{Y_{n}}(u)=\rho_{Z_{n}}(u-\epsilon)
\end{equation}
For a matrix $M$, the Stieltjes transform is defined as $\mathfrak{g}_M(z):=\int \frac{\rho_M(u)}{z-u} \mathrm{d} u$. For $z \in \mathbb{C}$ away from the support of $\rho_{Y_n}$, the Stieltjes transforms of $Y_n$ and $Z_n$ are related as:
\begin{equation}
\label{eq:RelationStieltjesVZ}
\mathfrak{g}_{Y_{n}}(z)=\int \frac{\rho_{Y_{n}}(u)}{z-u} \mathrm{d} u=
\int \frac{\rho_{Z_{n}}(u)}{z-\epsilon - u} \mathrm{d} u =\mathfrak{g}_{Z_{n}}(z-\epsilon)
\end{equation}
The $\mathcal{T}$-transform is defined from the Stieltjes transform as $\mathcal{T}(z) := z\mathfrak{g}(z) -1 $. As a consequence of \eqref{eq:RelationStieltjesVZ}: 
\begin{equation}
\mathcal{T}_{Y_n}(z) =z \, \mathfrak{g}_{Y_{n}}(z) - 1 = z \, \mathfrak{g}_{Z_{n}}(z-\epsilon) -1
= \mathcal{T}_{Z_n}(z - \epsilon) + \epsilon \, \mathfrak{g}_{Z_{n}}(z-\epsilon) 
\end{equation}
We then have the following relationship between the $\mathcal{T}$-transforms of $Z_n$ and $Y_n$:
\begin{equation}
\label{eq:TVfromTU}
\mathcal{T}_{Y_n}(z) = \frac{z \, \mathcal{T}_{Z_n}(z- \epsilon) + \epsilon }{z- \epsilon}
\end{equation} 
Defining finally the $\mathcal{S}$-transform as $ \mathcal{S}(\omega):=\frac{\omega+1}{\omega \mathcal{T}^{-1}(\omega)}$, we have $ \mathcal{S} (\mathcal{T}( z ) ) = \frac{\mathcal{T}( z ) +1 }{z \mathcal{T}( z )}$. Applying this to $Y_n$ and injecting \eqref{eq:TVfromTU} yields:
\begin{equation}
\mathcal{S}_{Y_n} (\mathcal{T}_{Y_n}( z ) ) =  \frac{\mathcal{T}_{Z_n}(z - \epsilon) +1 }{z \, \mathcal{T}_{Z_n}(z- \epsilon) + \epsilon}
\end{equation}
such that:
\begin{equation}
\frac{1}{\mathcal{S}_{Y_n} (\mathcal{T}_{Y_n}( z ) )} =  \frac{(z-\epsilon)\mathcal{T}_{Z_n}(z - \epsilon) }{\mathcal{T}_{Z_n}(z- \epsilon) +1} + \epsilon
=
\frac{1}{\mathcal{S}_{Z_n} (\mathcal{T}_{Z_n}( z-\epsilon ) )} + \epsilon 
\end{equation}
Applying this to $z = \mathcal{T}_{Y_n}^{-1}( z ) $ and applying \eqref{eq:TVfromTU}, we finally obtain the desired relation between the $\mathcal{S}$-transforms of $Z_n$ and $Y_n$:
\begin{equation}
\frac{1}{\mathcal{S}_{Y_n} (\omega) } = \frac{1}{\mathcal{S}_{Z_n} \left( \, \omega(1- \epsilon \,\mathcal{S}_{Y_n} (\omega) )   \, \right) } + \epsilon
\end{equation}
Applying the factorization of the $\mathcal{S}$-transforms in the free product $\mathcal{S}_{Z_{n+1}} = \mathcal{S}_{Y_n} \times \mathcal{S}_{\xi_n}$, we obtain the main result of this section in the form of the following recursion for $\mathcal{S}_{Z_n}$, the $\mathcal{S}$-transform of the matrix evolving under Kesten recursion:
\begin{equation}
\label{eq:StransformRecursion}
\frac{\mathcal{S}_{\xi}(\omega)}{\mathcal{S}_{Z_{n+1} } (\omega) } = \frac{1}{\mathcal{S}_{Z_{n} }\left( \, \omega(1- \epsilon \,\frac{\mathcal{S}_{Z_{n+1} }(\omega)}{\mathcal{S}_{\xi}(\omega)} )   \, \right) } + \epsilon
\end{equation} 
where we recall that $\mathcal{S}_\xi$ is the $\mathcal{S}$-transform of the noise matrix that enters the recursion. 

If we assume the existence of a stationary solution $Z_\infty$ to the Kesten recursion, then the $\mathcal{S}$-transform of $Z_\infty$ must verify the following self-consistent relation:
\begin{equation} \label{scS} 
\frac{\mathcal{S}_{\xi}(\omega)}{\mathcal{S}_\infty (\omega) } = \frac{1}{\mathcal{S}_\infty \left( \, \omega(1- \epsilon \,\frac{\mathcal{S}_\infty(\omega)}{\mathcal{S}_{\xi}(\omega)} )   \, \right) } + \epsilon
 \end{equation}  
 
\subsubsection{Continuum limit}

We now consider again the continuum limit of the Kesten recursion which is obtained by taking the noise matrix as $\xi_{n}=(1+m \, \epsilon) I +\sigma \, \sqrt{\epsilon} B_{n} $, in the same way as in \eqref{eq:structuresigma}, and then taking the limit $\epsilon \to 0$. Here, we study the large $N$ limit with a different method from the previous section \ref{sec:largeNMorse}, and we use the recursion relation for the $\mathcal{S}$-transform which we derived above in Eq. \eqref{eq:StransformRecursion}.

We study the case where $\tilde{\sigma}= \sqrt{N}\sigma$ remains fixed as $N \to +\infty$, such that the spectrum of $\sigma B_n$ has semi-circle density supported on $[ -2 \tilde{\sigma}, 2 \tilde{\sigma}]$, for both $\beta = 1,2$ as defined in \eqref{eq:DefinitionBn}. As discussed in Sec \ref{sec:largeNMorse}, we need to assume $m<0$. Note that we also assume $1 + m \epsilon > 2 \tilde \sigma \sqrt{\epsilon}$, i.e. 
$\epsilon < \epsilon_c= (\frac{\tilde\sigma -\sqrt{\tilde\sigma ^2-m}}{m})^2$
so that $\xi_n$ is positive definite as required.
It is shown in Appendix \ref{app:FreeProba} that the $\mathcal{S}$-transform corresponding to the resulting distribution for $\xi$ is then \eqref{eq:StransformSigma}:
\begin{equation} \label{Sxichoice} 
\mathcal{S}_\xi (\omega) =  \frac{-1-m \epsilon + \sqrt{(1+m\epsilon)^2 +4 \tilde{\sigma}^2 \epsilon \omega}}{2 \tilde{\sigma}^2 \epsilon \omega} 
\end{equation}

Let us define the following $\epsilon$-expansion for $\mathcal{S}_{Z_n}$ and $\mathcal{S}_{Z_{n+1}}$:
\begin{equation}
\mathcal{S}_{Z_n}(\omega) = \mathcal{S}_{n}^{(0)} (\omega) + \epsilon \mathcal{S}_{n}^{(1)} (\omega) + \epsilon^2 \mathcal{S}_{n}^{(2)} (\omega) + \cdots 
\end{equation}
Inserting these expressions in the $\mathcal{S}$-transform recursion relation \eqref{eq:StransformRecursion} yields the following relations at zeroth and first orders in $\epsilon$:
\begin{eqnarray}
 \mathcal{S}_{n+1}^{(0)}(\omega) &=& \mathcal{S}_{n}^{(0)}(\omega)  \\
\mathcal{S}_{n+1}^{(1)}(\omega)  -  \mathcal{S}_{n}^{(1)}(\omega) &=& - \mathcal{S}_{n}^{(0)} (\omega) \left( m +  \tilde{\sigma}^2 \omega + \mathcal{S}_{n}^{(0)}( \omega) + \omega {\mathcal{S}_{n}^{(0)}}'(\omega) \right)
\end{eqnarray}
As a consequence, the following evolution equation for $\mathcal{S}_{Z_{n}}$ holds at first order in $\epsilon$ :
\begin{equation}
\mathcal{S}_{Z_{n+1}}(\omega)  - \mathcal{S}_{Z_{n}}(\omega) = - \epsilon \mathcal{S}_{Z_{n}}(\omega) \left( m + \tilde{\sigma}^2 \omega + \mathcal{S}_{Z_{n}}(\omega) + \omega \mathcal{S}_{Z_{n}}'(\omega) \right)
\end{equation}

In the continuous limit, we denote as in section \ref{sec:ContinuousTime} the continuous-time matrix process $U_t = Z_{t/\epsilon}$, defining time as $t = n \epsilon$. The $\mathcal{S}$-transform of $U_t$ is then a function of two variables $\omega$ and $t$:
\begin{equation}
\label{eq:EvolSU}
\frac{\partial}{\partial t } \mathcal{S}(\omega,t) = - \mathcal{S}(\omega,t) \left( m +  \tilde{\sigma}^2 \omega + \mathcal{S}(\omega,t) + \omega \frac{\partial}{\partial \omega } \mathcal{S}(\omega,t)  \right)
\end{equation}

The stationary solution is the solution of $0=m +  \tilde{\sigma}^2 \omega + \mathcal{S}( \omega) + \omega \frac{\partial}{\partial \omega } \mathcal{S}(\omega) $, namely:
\begin{equation}
\mathcal{S}_\infty(\omega) = -m - \frac{\tilde{\sigma}^2}{2} \omega = \abs{m} \left( 1 - \frac{\tilde{\sigma}^2}{2 \abs{m} }\omega \right)  
\end{equation}
From the scaling property $\mathcal{S}_{a M} (\omega)  = \frac{1}{a}  \mathcal{S}_{ M} (\omega)  $ detailed in Appendix \ref{app:FreeProba}, $\mathcal{S}_\infty$ is the transform of:
\begin{equation}
\label{eq:UinfinityInverseMP}
U_{\infty} \   \stackrel{\mathrm{law}}{=} \   \frac{1}{\abs{m}} \InverseW
\end{equation}
where $\InverseW$ follows an inverse-Wishart matrix probability distribution with parameter $\kappa = \frac{\abs{m}}{\tilde{\sigma}^2}$, see Appendix \ref{app:WishartInverseWishart}. Indeed, for such a distribution, the $\mathcal{S}$-transform is linear and equal to $\mathcal{S}_{\small{\InverseW}}(\omega) = 1 -\frac{\omega}{2 \kappa} = 1 - \frac{\tilde{\sigma}^2}{2 \abs{m} }\omega $ \cite{BunBouchaudPotters2017}. For completeness, we note that $\InverseW$ is distributed as the inverse of a Wishart matrix with parameter $q$ given by $\frac{1}{q} = 2 \kappa +1  = 1 + \frac{2 \abs{m}}{\tilde{\sigma}^2}$, up to a scaling factor of $1-q$, see App. \ref{app:WishartInverseWishart} for details. The eigenvalue density, or spectral distribution, of $U_\infty$ has thus a finite support $[\lambda_+,\lambda_-]$ and is given by:
\begin{equation}
\label{eq:DensityUinfini}
\rho_{U_\infty}(\lambda) =    \frac{\abs{m}}{\pi \lambda^{2}\tilde{\sigma}^2 } \sqrt{\left(\lambda_{+}-\lambda\right)\left(\lambda-\lambda_{-} \right) } , \quad  \lambda \in\left[\lambda_{-}, \lambda_{+}\right], \quad \text{with} \quad \lambda_\pm= \frac{1}{\abs{m}}+ \frac{\tilde{\sigma}^2}{m^2}\left[ 1 \pm \sqrt{2 \frac{\abs{m}}{\tilde{\sigma}^2} +1}\right]
\end{equation}
where we recall that $\abs{m} =-m$. One can check that this agrees, upon the change of variables $\lambda=e^x$, with the density \eqref{eq:LDADensity}. This equivalence shows a perfect matching between the stationary solution in the continuous limit for the discrete-time/large-$N$ setting and the one obtained 
at finite $N$ in Eq. \eqref{eq:StationaryMatrixDist} as $N$ becomes large. Indeed, as $N \to \infty$, all the information about the matrix distribution is contained in the eigenvalue density. 

The matching with the previous situation can also be proved directly from the finite-$N$ stationary distribution as follows: 
let us denote $X$ a matrix sampled from the stationary distribution in Eq. \eqref{eq:StationaryMatrixDist}. We see by a rescaling that $
X  \ \overset{\mathrm{law}}{=} \ 
{2V  }/{\sigma^2}$
where $V$ is distributed proportionally to $\operatorname{det}(V)^{ \beta \left( \frac{m}{\sigma^{2}} - N +1 \right) -2 }  \exp({- {\beta} \operatorname{tr} V^{-1}/2}) $. The factor in the exponential is now correct to apply the convergence of the eigenvalue density of $V$ to the inverse Marcenko-Pastur density for this inverse-Wishart matrix, see App. \ref{app:WishartInverseWishart}.  The parameter $q=N/T$, and thus the related parameter $\kappa =( \frac{1}{q} -1)/2$, are asymptotically, from \eqref{eq:parameters} by injecting $\sigma = \frac{\tilde{\sigma}}{\sqrt{N}}$:
 $
 \frac{1}{q} = 1 + \frac{2 \abs{m}}{ \tilde{\sigma}^2} $ and $ \kappa = \frac{\abs{m}}{  \tilde{\sigma}^2}$.
We have then:
\begin{equation}
X  \ \overset{\mathrm{law}}{=} \ 
\frac{2}{\sigma^2} V = \frac{2q}{\tilde{\sigma}^2} T \ V = \frac{1}{\abs{m}} \frac{  \frac{2\abs{m}}{\tilde{\sigma}^2}  }{1+ \frac{2\abs{m}}{\tilde{ \sigma}^2}} T \ V 
= \frac{1}{\abs{m}}  \  (1-q) T \ V
 \ \overset{\mathrm{law}}{\to} \  \frac{1}{\abs{m}} \InverseW
\end{equation}
where $\InverseW$ is an inverse-Wishart matrix with parameter $\kappa = \frac{\abs{m}}{\tilde{\sigma}^2}$. The last equation expresses that the rescaled matrix $(1-q) T \ V$ converges in law to the inverse-Wishart distribution, since its eigenvalue density converges to the inverse Marcenko-Pastur density as shown in \eqref{eq:IMPDistribution}. This is in perfect agreement with \eqref{eq:UinfinityInverseMP}, where the continuous-time and large-size limits are taken in the other order. \\

It is interesting to study the convergence of \eqref{eq:EvolSU} to the stationary solution. Let us define the deviation from stationarity as 
$\Delta \mathcal{S}(\omega,t) = \mathcal{S}(\omega,t) - \mathcal{S}_\infty(\omega)$. 
One can linearize the evolution equation \eqref{eq:EvolSU} around the stationary solution. One
finds that the solution of this linearized equation reads 
\bea
\label{eq:LinearizedSolutionSEvolution}
\Delta \mathcal{S}(\omega,t) = \frac{e^{m t}}{1+ \frac{\tilde\sigma^2 \omega}{2 m} (1-e^{m t})} \Delta \mathcal{S}\left(0,\frac{\omega e^{m t}}{1+ \frac{\tilde\sigma^2 \omega}{2 m} (1-e^{m t})}\right)
\eea
Since here $m<0$ we see that $\Delta \mathcal{S}(\omega,t)$ vanishes for any $\omega$ at large time, such that the convergence occurs. 
At finite time $t$ for large $\omega$, we see that $\Delta \mathcal{S}(\omega,t) = \mathcal{O}(\frac{1}{\omega})$, such that $\Delta \mathcal{S}$ is killed at finite time for large $\omega$ argument, regardless of the initial value. An exact solution of the full non-linear equation \eqref{eq:EvolSU} can also be obtained using the hodograph transform,
see Appendix \ref{app:ExactSolSandG}.

\subsubsection{Back to the discrete recursion} 

Let us give a short discussion of the discrete case, e.g. finite $\epsilon$.\\

\paragraph{Corrections to the stationary $\mathcal{S}$-transform.}

Although it is difficult to solve the recursion relation \eqref{eq:StransformRecursion} for finite $\epsilon$, we can easily obtain the next orders in $\epsilon$ of the stationary $S$ transform $\mathcal{S}_\infty (\omega)$, which show how the stationary distribution deviates from the inverse Wishart ensemble. With the same choice for $\mathcal{S}_\xi (\omega)$ in \eqref{Sxichoice}
we obtain upon expanding \eqref{scS}
\be
\label{eq:ExpansionStransformEpsilon}
\mathcal{S}_\infty (\omega) = - m - \frac{\tilde \sigma^2}{2} \omega + \epsilon \left( m^2+\frac{7}{4} m \tilde \sigma ^2 \omega +\frac{3 \tilde \sigma ^4
   \omega ^2}{4} \right) + \epsilon^2 
  \left( -m^3-\frac{31}{8} m^2 \tilde\sigma ^2 \omega -\frac{53}{12}
   m \tilde\sigma ^4 \omega ^2-\frac{25}{16} \tilde\sigma ^6 
   \omega ^3 \right) + \mathcal{O}(\epsilon^3) 
\ee

One can also expand \eqref{eq:StransformRecursion} in powers of $\omega$ at fixed $\epsilon$ for a
general noise and one obtains 
\bea \label{eq:moments1} 
\mathcal{S}_\infty(0)=
   \frac{\mathcal{S}_\xi(0)-1}{\epsilon} \quad , \quad \mathcal{S}_\infty'(0)=
   \frac{\mathcal{S}_\xi(0) \mathcal{S}_\xi'(0)}{\epsilon
   \mathcal{S}_\xi(0)+\epsilon} \quad , \quad \mathcal{S}_\infty''(0)=
   \frac{\mathcal{S}_\xi(0) \left(\mathcal{S}_\xi(0)
   (\mathcal{S}_\xi(0)+1)^2 \mathcal{S}_\xi''(0)+2
   \mathcal{S}_\xi'(0)^2\right)}{\epsilon
   (\mathcal{S}_\xi(0)+1)^2
   \left(\mathcal{S}_\xi(0)^2+\mathcal{S}_\xi(0)+1\right)}
\eea
In the case of the noise considered here in \eqref{Sxichoice} this leads to
\bea
\mathcal{S}_\infty(0)=  -\frac{m}{m \epsilon +1} \quad , \quad \mathcal{S}_\infty'(0)=
   -\frac{\tilde\sigma ^2}{(m \epsilon +1)^3 (m \epsilon
   +2)} \quad , \quad \mathcal{S}_\infty''(0) = \frac{2 \tilde\sigma ^4 \epsilon  (m
   \epsilon  (3 m \epsilon +10)+9)}{(m \epsilon +1)^5
   (m \epsilon +2)^2 (m \epsilon  (m \epsilon
   +3)+3)} 
   \eea

\paragraph{Stationary moments.}

The moments of the spectrum, denoted $\varphi_k = \lim\limits_{N\to \infty} \frac{1}{N} \sum\limits_{i=1}^N \lambda_i^k$ can be extracted from
the $\mathcal{S}$ transform. For instance the first moment is obtained from $\mathcal{S}(0) = \frac{1}{\varphi_1}$.
Denoting $\varphi_{k,\infty}$ the moments of $Z_\infty$ and $\varphi_{k,\xi}$ the moments of $\xi_n$ 
the first equation in \eqref{eq:moments1} gives the first moment 
\be \label{mom1} 
\varphi_{1,\infty} = \frac{\epsilon \varphi_{1,\xi}}{1- \varphi_{1,\xi}} = -\frac{m \epsilon +1}{m}
\ee
where the second equation is specified to the model \eqref{Sxichoice}. Similarly the second moment reads
\be \label{mom2} 
\varphi_{2,\infty} = 
\frac{\epsilon ^2 \left(\varphi_{2,\xi}- \varphi_{1,\xi}^4\right)}{(1-\varphi_{1,\xi})^3
   (1+\varphi_{1,\xi})} = \frac{m (m \epsilon +1)^2-\frac{\tilde\sigma ^2}{m \epsilon
   +2}}{m^3}
\ee
Higher moments can be obtained from the series expansion of the $\mathcal{S}$-transform, see Eq. \eqref{eq:InvertingSeriesMomentsSTransform}.
One can compare with the case $N=1$. In that case the stationary moments can be computed iteratively, by simply averaging the moments
of the recursion relation. The result coincides with \eqref{mom1} for the first moment (which can be understood from the fact that 
$\mathbb{E}[\xi_n] \propto I$). For the second moment however it reads $\varphi_{2,\infty} = 
\frac{\epsilon ^2 \varphi_{2,\xi}(1+ \varphi_{1,\xi})}{(1-\varphi_{2,\xi})
   (1-\varphi_{1,\xi})}$ which is markedly different from \eqref{mom2}. See more details on the moment computations in the scalar and matrix cases in App. \ref{app:Moments}. \\

\paragraph{Edges and stationary density corrections.}
We turn to the edges of the spectrum, that can be retrieved from the $\mathcal{S}$-transform as showed in App. \ref{app:FreeProba} by solving the system \eqref{eq:SystemEdgeStransform}. At order $\epsilon$ in the expansion \eqref{eq:ExpansionStransformEpsilon}, the two edges of the spectrum are found at positions: 
\be
\label{eq:EdgesFirstOrderEpsilon}
\lambda_\pm^{(1)} =
\lambda_\pm^{(0)}
+ \epsilon
\left( -1 \pm \frac{\tilde \sigma }{2 \sqrt{\tilde\sigma ^2-2 m}} \right) + \mathcal{O}(\epsilon^2)
\ee
where the order-zero terms $\lambda_\pm^{(0)}$ are the edges of the scaled inverse-Wishart stationary distribution given in Eq. \eqref{eq:DensityUinfini}. We note that the correction terms are negative, such that both edges are pushed to the left under the order-$\epsilon$ correction.

\begin{figure}[!t]
\centering
\includegraphics[width= 0.6 \textwidth]{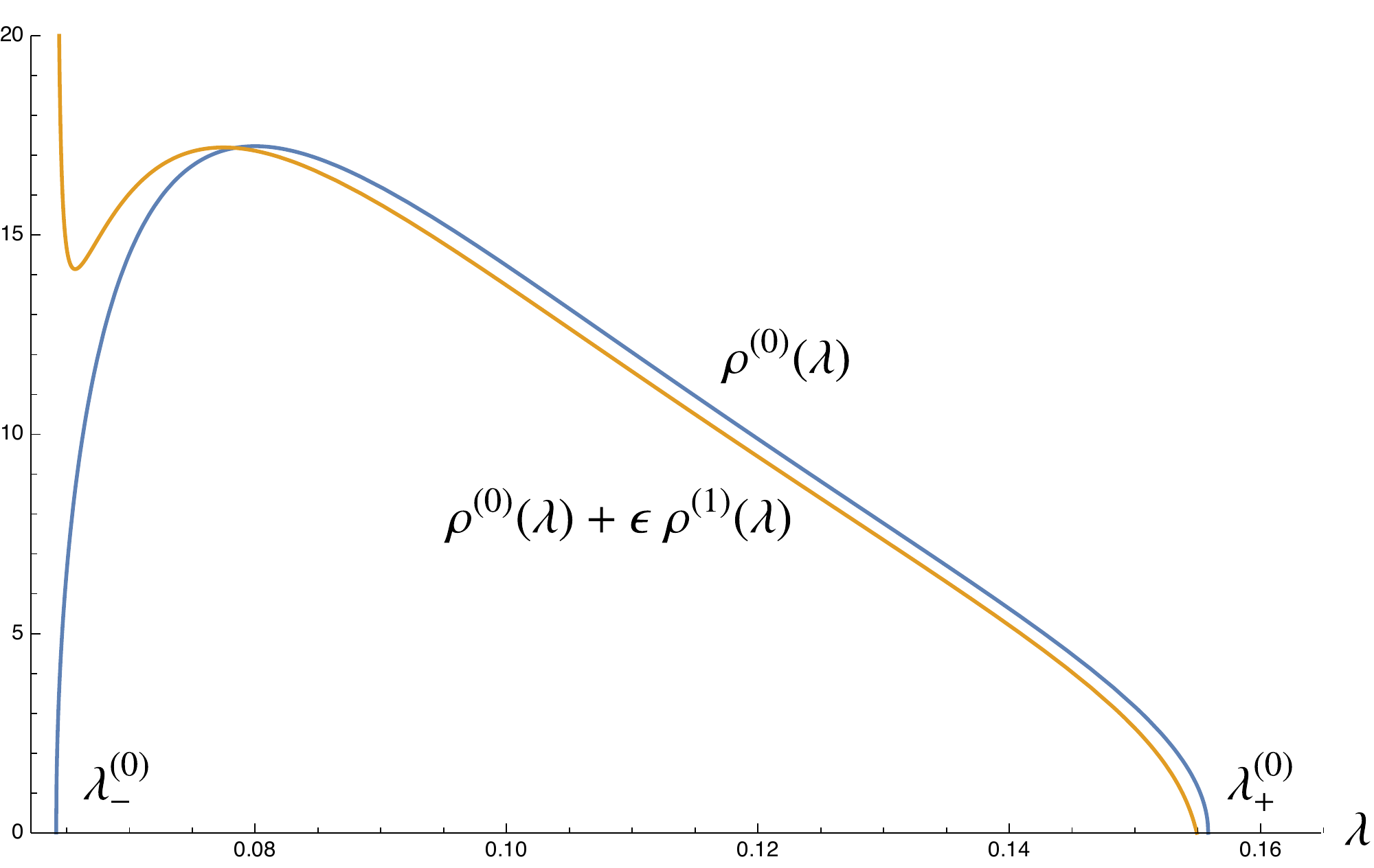}
\caption{This figure shows the plot of the stationary density at orders 0 and 1 in $\epsilon$, see Eq. \eqref{eq:DensityCorrections}, as the blue and orange lines respectively. Parameters were fixed as $m=-10, \tilde\sigma=1,\epsilon=2.10^{-3}$ in this plot. We see that the correction of the density at first order in $\epsilon$ induces a leftward translation consistent with the negative correction to both edges in Eq. \eqref{eq:EdgesFirstOrderEpsilon}. We also note that the expansion fails close to the left edge $\lambda^{(0)}_-$.} 
\label{fig:DensityCorrections}
\end{figure}

Finally, the stationary density can be obtained from the $\mathcal{S}$-transform \eqref{eq:ExpansionStransformEpsilon} by solving for $\mathfrak{g}$ and using the Stieltjes inversion formula or Sokhotski-Plemelj formula \eqref{eq:StieltjesInversion}. At first order in $\epsilon$, one obtains: 
\be 
\label{eq:DensityCorrections}
\rho(\lambda) =\rho^{(0)} (\lambda) + \epsilon
\frac{
6 + (11m - 9 \tilde \sigma^2) \lambda+2m (3m -\tilde \sigma^2) \lambda^2 + m^3 \lambda^3
}{2\pi \tilde\sigma^2 \lambda^3 \abs{m} \sqrt{(\lambda-\lambda_-^{(0)}) (\lambda_+^{(0)} -\lambda )}}
+ \mathcal{O}(\epsilon^2) 
\ee
where $\rho^{(0)}$ is the stationary distribution given in Eq. \eqref{eq:DensityUinfini}.
One can check that the integrated correction vanishes as required by normalization of $\rho$. A plot of the density at zeroth and first order in $\epsilon$ is showed in Fig. \ref{fig:DensityCorrections}.
 
\subsection{Stieltjes transform approach}

A complementary way to approach the large-$N$ problem is to study the evolution of the Stieltjes transform. 
We return here to the continuum limit.
Let us denote
${\sf g}$ the fluctuating Stieltjes transform defined as follows
\begin{equation}
{\sf g}(z,t) := \frac{1}{N} \sum_{i=1}^N \frac{1}{z - \lambda_i(t)}\; .
\end{equation}
Note that we recover the expected Stieltjes transform defined earlier through $\mathfrak{g}(z,t) = \mathbb{E}[ {\sf g}(z,t)]$. Note also that the Stieltjes transform is self-averaging such that the fluctuating transform ${\sf g}(z,t)$ converges almost surely to $\mathfrak{g}(z,t)$ for $N \to \infty$. We emphasize that we will treat the large-$N$ limit by dropping negligible terms, without attempt at mathematical rigor.

For a fixed argument $z$, the stochastic evolution of $\mathfrak{g}$ is, from the perturbative evolution of eigenvalues \eqref{eq:EvolutionLambdai}: 
\begin{eqnarray}
{\rm d}{\sf g}(z,t)  &=&  \frac{1}{N}  \sum_{i=1}^N  \frac{1}{(z - \lambda_i)^2} {\rm d}\lambda_i  + \frac{1}{2N}  \sum_{i=1}^N \frac{2}{(z-\lambda_i)^3} {\rm d}\lambda_i.{\rm d}\lambda_i   \\
&=& - \frac{\partial}{\partial z}  \frac{1}{N}  \sum_{i=1}^N \left( \frac{1}{z - \lambda_i} \left(1 + m \lambda_i + \sigma^2 \sum\limits_{\substack{1 \leqslant j \leqslant N \\j\neq i}}  \frac{\lambda_i \lambda_j }{\lambda_i - \lambda_j} \right) + \frac{\sigma^2 \lambda_i^2}{(z-\lambda_i)^2}  \right) {\rm d}t + \frac{1}{N} \sum_{i=1}^N \frac{\sqrt{2} \sigma \lambda_i}{(z-\lambda_i)^2} {\rm d}W_i 
\end{eqnarray}
up to $\mathcal{O}({\rm d}t^{3/2})$. With the scaling of $\sigma$ as $  \frac{\tilde{\sigma}}{\sqrt{N}}$, the noise term is negligible for large $N$, such that:
\begin{eqnarray}
{\rm d}{\sf g}(z,t) &=&
- \frac{\partial}{\partial z} \left(  
{\sf g}+ m ( z{\sf g} -1) +\frac{\tilde{\sigma}^2}{N^2 } \sum_{i} \left(
\frac{1}{2}\sum_{j \neq i} 
\frac{\lambda_i \lambda_j}{\lambda_i - \lambda_j}
\left(
\frac{1}{z - \lambda_i}
-
\frac{1}{z - \lambda_j}
\right)
  + \frac{\lambda_i^2}{(z-\lambda_i)^2} \right)
\right) + \mathcal{O}( \frac{1}{\sqrt{N}} )  \\
&=& - \frac{\partial}{\partial z} \left(  
{\sf g}+ m z{\sf g}  +
\frac{\tilde{\sigma}^2}{2}
(z {\sf g} -1)^2
+
\frac{\tilde{\sigma}^2}{2N^2}
\sum_{i}
\frac{\lambda_i^2}{(z-\lambda_i)^2}
\right) + \mathcal{O}( \frac{1}{\sqrt{N}} )
\end{eqnarray}
To go from the first to the second line, we have added and subtracted the term $i=j$ in the double sum, and removed the constant term $-m$ in the derivative. For large $N$ replacing ${\sf g}$ by the averaged Stieltjes transform and neglecting the last term which is of order $\frac{1}{N}$, the evolution of $\mathfrak{g}(z,t)$ is given by \cite{RMTBook}:
\begin{equation}
\label{eq:EvolutionStieltjes}
\frac{\partial}{\partial t} \mathfrak{g}(z,t)  = \frac{\partial}{\partial z} \left(  
- \mathfrak{g} + (\tilde{\sigma}^2 - m )z\mathfrak{g} - \frac{1}{2}  \tilde{\sigma}^2 (z\mathfrak{g})^2
\right)
\end{equation}
An exact solution of this non-linear partial differential equation can be obtained using the hodograph transform,
see Appendix \ref{app:ExactSolSandG}. It is furthermore shown in this Appendix that \eqref{eq:EvolutionStieltjes} is equivalent to the evolution \eqref{eq:EvolSU} of the $\mathcal{S}$-transform, as expected but not directly obvious. We also note that the stationary Stieltjes transform $\mathfrak{g}_{\rm stat}(z)$ is 
\be 
\label{eq:StationaryStieltjesTransform}
\mathfrak{g}_{\rm stat}(z)=|m| \ \mathfrak{g}_{\small{\InverseW}}(|m| z)
\ee
where 
\begin{equation}
\mathfrak{g}_{\small{\InverseW}}(z)=\frac{1}{z^{2}}\left[\left(1+\kappa \right) z- \kappa   - \kappa \sqrt{ (z - z_-) (z- z_+)  }\right]
\end{equation}
is the Stieltjes transform of $\InverseW$ following the inverse-Wishart law with coefficient $\kappa=\frac{\abs{m}}{\tilde{\sigma}^2}$, with $z_\pm = \frac{1}{\kappa} (1 + \kappa \pm \sqrt{2 \kappa +1})$.
Note that the branch has been chosen so that $\mathfrak{g}_W(z) \simeq 1/z$ at large $z$. This shows once again as in Eq. \eqref{eq:UinfinityInverseMP} that the stationary distribution for the Kesten evolution is the law of $\frac{\InverseW}{\abs{m}} $, with $\InverseW$ as above.

{Note that the evolution equation for the Stieltjes transform in Eq. \eqref{eq:EvolutionStieltjes} can be studied in the Hamilton-Jacobi framework recently introduced by Grela, Nowak and Tarnowsky in \cite{Nowak20}. In the notations of \cite{Nowak20}, where the Stieltjes transform $\mathfrak{g}$ is associated to a momentum variable $p$ and the variable $z$ is thought of as a position variable, the Hamiltonian related to our model is:
\be 
H_{HJ}(p,z) = p -\left(\tilde{\sigma}^{2}-m\right) z p+\frac{\tilde{\sigma}^{2}}{2} z^{2} p^{2}
\ee 
Our model and its Hamiltonian are an example of application of the Hamilton-Jacobi framework in large dynamical random matrix models, which we add to the list of examples given by the authors.
}

\subsection{Expected characteristic polynomial and Stieltjes transform for finite \texorpdfstring{$N$}{N}}

Here we study $\Pi(z,t) =  \prod_{i=1}^N (z - \lambda_i) $ the characteristic polynomial of the evolving matrix at time $t$. It is an 
interesting quantity because its average satisfies a closed equation for any finite $N$. This property has been also used to 
study the standard Dyson Brownian motion \cite{Blaizot10,Blaizot13,Blaizot14}.

Since it is linear in the $\lambda_i$'s its stochastic evolution is simply $d\Pi=- \sum_i \frac{\Pi}{z-\lambda_i} {\rm d}\lambda_i$. Using 
\eqref{eq:EvolutionLambdai} and the identities $\partial_z \Pi= \sum_i \frac{1}{z-\lambda_i} \Pi$ and 
$\partial^2_z \Pi= \sum_{i \neq j} \frac{1}{(z-\lambda_i)(z-\lambda_j)} \Pi$, one finds after some
manipulations the evolution of its expectation $\overline{\Pi}(z,t) := \mathbb{E}[\Pi(z,t)]$ as:
\be
\label{eq:EvolutionPibarre}
\partial_t  \; \overline{\Pi}(z,t) = N \left( m - (N-1) \frac{\sigma^2}{2} \right) \; \overline{\Pi}(z,t) + \left( -1 + z ( -m + (N-1) \sigma^2) \right) \partial_z \; \overline{\Pi} (z,t)  - \frac{\sigma^2}{2} z^2 \partial_z^2  \; \overline{\Pi}(z,t)
\ee
One can look for the stationary solution of this equation. By inspection we find that it is given by the degree $N$ polynomial
\be \label{Pistat} 
\lim_{t \to +\infty} \overline{\Pi}(z,t) = a_N \; z^N \; L_N^{-1-\frac{2 m}{\sigma^2}}(\frac{2}{\sigma^2 z}),  \quad \quad 
a_N= \binom{ N-\frac{2m}{\sigma^2}-1}{N}^{-1} 
\ee
It is interesting to note that the dynamics of $\overline{\Pi}(z,t)$ can be related to the same Morse
potential as studied in section \ref{sec:ContinuousTime}. Indeed one can perform a change of function and write 
\be
\overline{\Pi}(z=e^x,t) = f(x) \; \psi(x,t)  \quad , \quad f(x) = e^{(N-\frac{1}{2}- \frac{m}{\sigma^2} ) x + \frac{1}{\sigma^2} e^{-x}}
\ee
and one finds that $\psi(x,t)$ evolves according to {\it minus} the Schr\"odinger Hamiltonian in the Morse
potential, i.e. 
\be
\partial_t \psi 
= \frac{\sigma^2}{2} \left( - \partial_x^2 + V(x) - \epsilon_N \right) \psi  
\ee
where $V(x)$ is defined setting $\beta =2$ in \eqref{eq:QuantumPotential} and
$\epsilon_N = -\frac{(2 m +\sigma^2)^2}{4 \sigma^4}$ is the $N+1$th energy level of the Morse potential. We recall that the energy level and the eigenfunctions are given in \eqref{eq:EigenergiesEigenfunctionsbeta2}.
Because the initial condition for $\overline{\Pi}(z,t=0)$ is a polynomial in $z$ of degree $N$, the
initial wavefunction $\psi(x,t=0)$ belongs to the subspace spanned by the $N+1$ lowest levels of the
Morse potential. Since the time evolution is reversed, it converges to the highest energy accessible
which recovers \eqref{Pistat}.\\

It is interesting to note that the characteristic polynomial is linked to the fluctuating Stieltjes transform as:
\begin{equation}
{\sf g}(z,t) = \frac{1}{N} \partial_z \ln \Pi(z,t)
\end{equation}
which upon averaging over the noise leads to 
\begin{equation}
\label{eq:ggbarre}
\mathfrak{g}(z,t) =  \frac{1}{N} \partial_z \mathbb{E} [\ln \Pi(z,t)] \simeq \frac{1}{N} \partial_z \ln \; \overline{\Pi}(z,t) := \overline{\mathfrak{g}}(z,t)
\end{equation}
where the approximation is valid in the large $N$ limit
where the expectation value and the log can be exchanged. We introduce the notation $\overline{\mathfrak{g}}$ for simplicity. The evolution of the expected characteristic polynomial in Eq. \eqref{eq:EvolutionPibarre} gives, after some manipulations, an equation valid for any $N$ for the evolution of $\overline{\mathfrak{g}}$ defined in \eqref{eq:ggbarre}:
\begin{eqnarray}
\partial_t \overline{\mathfrak{g}} &=& 
\partial_z \left( \left( -1 + z ( -m + (N-1) \sigma^2) \right) \overline{\mathfrak{g}}     - \frac{\sigma^2N}{2} z^2 \overline{\mathfrak{g}}^2  -   \frac{\sigma^2}{2} z^2 \partial_z \overline{\mathfrak{g}}
\right)
\end{eqnarray}
Taking $N$ large such that $\overline{\mathfrak{g}} \simeq \mathfrak{g}$ with $ \sigma^2 = \tilde{\sigma}^2/  N$ and neglecting terms of order $1/N$ we recover by a different method the evolution for the Stieltjes transform, which was obtained in the previous section in Eq. \eqref{eq:EvolutionStieltjes}:
\begin{equation}
\frac{\partial}{\partial t} \mathfrak{g}(z,t)  = \frac{\partial}{\partial z} \left(  
- \mathfrak{g} + (\tilde{\sigma}^2 - m )z\mathfrak{g} - \frac{1}{2} \tilde{\sigma}^2 (z\mathfrak{g})^2
\right)
\end{equation}

\section{Links with Rider-Valk\'o and Grabsch-Texier}
\label{sec:LinksRVGT} 

Our contruction of Kesten matrices bears some similarities, but also some differences, with some matrix stochastic diffusions studied by Rider and Valk\'o on the one hand, and by Grabsch and Texier on the other. We discuss the precise relation between these models in the present section. 

\subsection{Rider-Valk\'o}
\label{sec:RD} 
 
Rider and Valk\'o define in \cite{RiderValko} the following geometric $N \times N$ drifted matrix Brownian motion $M$:
\begin{equation}
{\rm d}M_{t}=\left(\frac{1}{2} -  \mu\right) M_{t} {\rm d}t + M_{t} {\rm d}H_{t} , \quad M_{0}=I \quad \quad \quad \quad \text{(It\^o)} 
\end{equation}
where the elements of the Brownian matrix $H_t$ are $N^2$ independent real Brownian motions, in contrast with the Hermitian symmetry assumed in this work, with real ($\beta =1$) or complex ($\beta =2$) elements. 

Since $M$ is not symmetric, the authors study the law of $(MM^T)_s$. To this aim, they define the process $Q_t =  M_{-t}^{-1} \left(  \int_{-t}^\infty M_s M_s^T \mathrm{d}s \right) M_{-t}^{-T}$ for negative time $t$, and show that it obeys the It\^o SDE:
\begin{equation}
\label{eq:RiderValkoSDE}
{\rm d}Q_{t}=\left( (1 + \operatorname{tr} Q_{t} ) I +      (1-2 \mu) Q_{t} \right) {\rm d}t-{\rm d} H_{t} Q_{t}-Q_{t} {\rm d} H_{t}^{T}\quad \quad \quad \quad \text{(It\^o)} 
\end{equation}
The stationary distribution of this SDE, i.e.\ the law of $Q_0 = \int_{0}^\infty M_s M_s^T \mathrm{d}s$, is proved to be the following inverse-Wishart distribution:
\begin{equation}
\label{eq:RiderValkoInverseWishart}
\gamma_{2\mu}^{-1}(X) \ \propto \ (\operatorname{det} X)^{ - \mu -1 - \frac{N-1}{2}} \ e^{-\frac{1}{2} \operatorname{tr} X^{-1}}   
\end{equation}
As mentioned in the introduction, their work is an extension to the matrix realm of the Dufresne identity which gives the distribution of the exponential Brownian functional \eqref{eq:Uinfini} as an inverse-gamma law.

\subsection{Grabsch-Texier}

In \cite{GrabschTexier2016}, Grabsch and Texier study topological phase transitions occurring in a multichannel random wire modelled by a random-mass Dirac equation. They introduce the Riccati matrix $Z_t$ which follows the SDE:
\begin{equation}
{\rm d}Z_t= \left( -Z_t^{2} +k^2 I -\mu G Z_t-\mu Z_t G \right){\rm d}t  + \sigma_\beta (Z_t {\rm d}B_t +{\rm d}B_t   Z_t) \quad \quad \quad \quad \mathrm{(Stratonovich)}
\end{equation}
where both symmetry cases $\beta=1,2$ are considered, such that the matrix $Z$ is symmetric for $\beta = 1$ and Hermitian for $\beta=2$. $B_t$ is a matrix Brownian motion  with correlations given by $G$. In the case where $G=I$, $B_t$ is identical to the one used in this work, see definition below Eq. \eqref{eq:EvolutionU}, provided one sets $\sigma_\beta=\sqrt{\beta/2}$. We note here the time variable as $t$ in order to respect the conventions of our paper, but the evolution parameter is in fact the position $x$ along the longitudinal direction of the disordered wire. 

Notice that this stochastic equation is given with the Stratonovitch prescription, contrary to the It\^o prescription chosen for all SDEs in this work and in \cite{RiderValko}. We recall that the It\^o and Stratonovitch prescriptions are two different ways to interpret a SDE with multiplicative noise such as $Z_t {\rm d}B_t$.  $Z_t$ and ${\rm d}B_t$ are taken to be statistically independent under It\^o, whereas they are not independent under Stratonovitch, see \cite{VanKampenItoStrat} for a pedagogical introduction.

Using the Fokker-Planck operator acting on the entries of the matrix $Z$, Grabsch and Texier demonstrate that the matrix diffusion for $Z$ admits the following stationary distribution (for $\sigma_\beta=\sqrt{\beta/2}$):
\begin{equation}
 f(Z) \  \propto \  (\operatorname{det} Z)^{-\mu-1-\beta\frac{N-1}{2}   }  \  e^{-\frac{1}{2} \operatorname{tr}\left\{G^{-1}\left(Z+k^{2} Z^{-1}\right)\right\} }
\end{equation}
Let us take the special case $G=I$. After the change of variables $\{Z,k^2\} \to \{ \epsilon Z, \epsilon   \}$ the SDE becomes in the $\epsilon \to 0$ limit:
\begin{equation}
\label{eq:LimitSDE}
{\rm d}Z_t= \left(   I - 2 \mu Z_t \right){\rm d}t + \sigma_\beta( Z_t {\rm d}B_t +{\rm d}B_t   Z_t ) \quad \quad \quad \quad \mathrm{(Stratonovich)}
\end{equation}
which admits as a consequence the following stationary distribution, restoring the parameter $\sigma_\beta$ for future comparisons:
\begin{equation}
\label{eq:StatDistLimit}
 \tilde{f}(Z) \ \propto \ (\operatorname{det} Z)^{
 - \frac{\beta}{2\sigma_\beta^2}\mu -1 - \beta \frac{N-1}{2}}
 \  e^{ -\frac{\beta}{4\sigma_\beta^2} \operatorname{tr} Z^{-1}}
 \end{equation}

In a recent work \cite{GrabschTexier2020}, Grabsch and Texier encountered the SDE \eqref{eq:LimitSDE} in a different problem: the evolution of the (symmetrized) Wigner-Smith time delay matrix $\tilde{Q}$ of a multi-channel disordered wire, where the size $L$ of the disordered region plays the role of the time parameter in the SDE. More precisely, the scattering process is defined by a Schr\"odinger equation with random potential coupling the $N$ channels. Their work generalizes a problem which is known to coincide in the case $N=1$ with the Brownian exponential functional that we presented in equation \eqref{eq:Uinfini}. In the general $N$ case, they show that the symmetrized Wigner-Smith matrix $\tilde{Q}$ can be written as:
\be
\label{eq:QTildeDef}
\tilde Q_t = X_t \left( \int_0^t \mathrm{d}s X_s^{-1} \left(X_s^{ \dagger}\right)^{-1} \right) X_t^\dagger
\ee
where $X_t$ follows the following Stratonovitch flow:
\be 
{\rm d}X_t = (-\mu + \sigma_\beta {\rm d}B_t) X_t\quad \quad \quad \quad \mathrm{(Stratonovich)}
\ee 
such that $\tilde Q_t$ satisfies the SDE \eqref{eq:LimitSDE} where $Z_t$ is replaced by $\tilde Q_t$. They rewrite \eqref{eq:QTildeDef} by defining $\Lambda^\dagger_s = X_t X_{t-s}^{-1}$, which satisfies ${\rm d}\lambda_s = (-\mu + \sigma_\beta {\rm d}\tilde B_s)\Lambda_s$ in Stratonovitch prescription, where ${\rm d}\tilde B_s = {\rm d}B_{t-s}$. Using a reordering $s \to t -s$ in \eqref{eq:QTildeDef}, similar to the one explained for the scalar case in Footnote \ref{footnoteReordering}, they obtain:
\be 
\label{eq:QTildeIntegraleLambdacroixLambda}
\tilde Q_t  \overset{\mathrm{law}}{=} 
\int_{0}^t \Lambda^\dagger_s \Lambda_s \mathrm{d}s
\ee
As a result they show that both $\tilde Q_t $ and the r.h.s. of Eq. \eqref{eq:QTildeIntegraleLambdacroixLambda} are distributed as \eqref{eq:StatDistLimit} in the limit of large time.

As pointed out by Grabsch and Texier, this is very close to the results of Rider and Valk\'o in the real case ($\beta =1$) as can be seen by comparing the stationary distributions \eqref{eq:RiderValkoInverseWishart} and \eqref{eq:StatDistLimit} with $\sigma_\beta = \sqrt{\beta/2}$. Fixing $\beta= 1$, it is worth noting that the two SDEs \eqref{eq:RiderValkoSDE} and \eqref{eq:LimitSDE} differ by a drift term $ ( Q + I \operatorname{tr} Q  ) {\rm d}t$ which can be shown to be exactly the drift obtained by transforming a Stratonovich SDE with noise term $Z {\rm d}B + {\rm d}B Z$ into an It\^o SDE. 
The only difference is the absence of structure in the noise matrix ${\rm d}H_t$ used by Rider and Valk\'o. Grabsch and Texier show in \cite{GrabschTexier2020} that this is transparent at the level of the distribution. Indeed, the non-symmetric structure  
can be dealt with by gauging out the non-Hermitian part  of ${\rm d}H_t$ in a way that leaves the law of $\int_{0}^\infty M_s M_s^T \mathrm{d}s$ unaltered.  
 
Since the expression of $\tilde{Q}$ as a matrix Brownian functional and the stationary distribution for SDE \eqref{eq:LimitSDE} were both established also in the complex Hermitian case ($\beta =2$) by Grabsch and Texier, their work is in fact an extension of the matrix Dufresne identity to the complex case.

\subsection{Connections with the present work}
\label{sec:LinksRVGTConnectionsPresentWork}

The works by Rider-Valk\'o and Grabsch-Texier are closely interconnected since they study the same matrix diffusion, as we have seen from the SDEs or from the Brownian functional expression. This latter point of view is naturally related to the Kesten recursion as was shown for the scalar case in \ref{sec:Models}. However, instead of defining the matrix Kesten problem from the Brownian functional, we have chosen in this work to study the recursion \eqref{eq:defmodel}. We showed that, in the continuous limit, this recursion yields the SDE \eqref{eq:EvolutionU}:
\begin{equation}
\label{eq:ItoSDEIVC}
{\rm d}U_t = \left( I +m U_t\right) {\rm d}t +\sigma \sqrt{U_t} {\rm d}B_{t} \sqrt{U_t}
 \quad \quad \quad \quad \text{(It\^o)} 
 \end{equation}
This is close to the diffusion of Rider-Valk\'o and Grabsch-Texier, as the drift terms are essentially the same, but there is a crucial difference in the noise term, which is $\sqrt{U} {\rm d}B \sqrt{U}$ instead of $U{\rm d}B + {\rm d}B U$. 

Remarkably, although these two matrix processes are \emph{different}, the eigenvalue processes are actually \emph{identical} provided one compares Eq. \eqref{eq:ItoSDEIVC} in It\^o prescription and Eq. \eqref{eq:LimitSDE} in Stratonovitch prescription, with some correspondence between the parameters. This is showed in Appendix \ref{app:PerturbationGT} where we study the joint evolution of the eigenvalues in the Grabsch-Texier model (which was not studied in \cite{GrabschTexier2016,GrabschTexier2020}). We find that we can identify the eigenvalues of $U_t$ and those of $Z_t$ if we set $\sigma_\beta = \frac{\sigma}{2}$ and $\mu = -\frac{m}{2} + \frac{\sigma^2}{4}(N+2-\beta)$. Inserting these values in the stationary distribution \eqref{eq:StatDistLimit} reproduces exactly our result \eqref{eq:StationaryMatrixDist} for $\beta=1,2$.

The evolution processes of the eigenvectors, however, have no obvious reason to be related. Since the stationary matrix measure is isotropic in both cases, the two models converge to the identical inverse-Wishart distribution.

We see that there is a subtle interplay between the form of the noise and the prescription used, such that terms of the Stratonovitch-It\^o drift combine with the perturbative eigenvalue drift to recover the same eigenvalue flow.

As the last note on relations to other works, let us finally mention that our Fokker-Planck equation \eqref{eq:FPEigenval} appears in a very recent work of Ossipov \cite{Ossipov} related to the Wigner-Smith time-delay matrix evolution in a multichannel disordered wire. Indeed, see Eq. (8) of this work, where the interaction term $\frac{\tau_i^2}{\tau_i - \tau_k}$ is equal to the interaction term of Eq. \eqref{eq:FPEigenval} plus a term proportional to $\tau_i$.
In our notations, the work in Ref. \cite{Ossipov} is restricted to the case $m=0$ and $\sigma=1/\sqrt{2 N}$, with null initial condition. In this special case and for large $N$, the author obtains the solution of the Stieltjes transform evolution equation, by a mapping to a Burgers equation. We extend this result by solving Eq. \eqref{eq:EvolutionStieltjes} with arbitrary parameters $m, \tilde\sigma$ and arbitrary initial condition in Appendix \ref{app:ExactSolSandG}. We stress that the finite $N$ evolution equation solved in \cite{Ossipov} lacks a noise term, such that it does not qualify as a solution of the finite $N$ diffusion problem.

\subsection{Stochastic generalization : the matrix Bougerol identity}

Bougerol's identity is a probabilistic result related to Dufresne's identity, which gives the law of the same Brownian exponential functional of Eq. \eqref{eq:Uinfini} where one replaces the integral by a stochastic integral with respect to an independent Brownian motion $w_t^{\nu}$ (with drift $\nu$). It states that the law of
\begin{equation}
\label{eq:Bougerol}
V_\infty \overset{\mathrm{law}}{=} \int_{0}^{\infty} e^{\gamma t+  W_{t}} \mathrm{d} w_t^{(\nu)}
\end{equation}
is characterized by the density
\begin{equation}
\label{eq:MeasureBougerol}
P(V_\infty) \propto \frac{e^{2 \nu \arctan (V_\infty)}}{\left(1+{V_\infty}^{2}\right)^{\frac{1}{2}-\gamma}} \; .
\end{equation}
Note that the heavy tail at $V_\infty$ has the same power-law exponent as $U_\infty$ in the Dufresne case, see Eq. \eqref{eq:1DInverseGamma}. This result has recently been extended to the matrix case by Assiotis in \cite{AssiotisBougerol}, where it was showed that the matrix analogue of \eqref{eq:Bougerol} is distributed according to a Hua-Pickrell measure, the natural matrix extension of the generalized Cauchy measure of the scalar case \eqref{eq:MeasureBougerol}. The authors of the present paper believe that a generalized Kesten recursion 
\begin{equation}
    Z_{n+1} = \sqrt{\nu \epsilon  I + \sqrt{\epsilon} C_n  + Z_n } \ \xi_n \ \sqrt{\nu \epsilon I + \sqrt{\epsilon} C_n  + Z_n }
\end{equation}
where $C_n$ is a standard GOE or GUE, would allow to explore this stochastic generalization. We conjecture that the Hua-Pickrell measure is the stationary distribution of the generalized Kesten recursion in the limit $\epsilon \to 0$ and leave the study for a future work.
 
\section{Conclusion}

In this paper we have proposed and studied an extension of the celebrated Kesten recursion and random variable $1+ z_1+ z_1 z_2 + \dots$ 
to $N \times N$ positive defined real ($\beta=1$) and Hermitian ($\beta=2$) matrices, the scalar case being recovered for $N=1$. By studying the evolution of the matrix eigenvalues
we have shown that the salient feature of the Kesten variable, i.e. the spontaneous generation of a heavy tail distribution, 
persists in the case of matrices for any $N$, while at large $N$ the effects of the correlations and of the spectral rigidity, familiar in random matrix theory, also become important.

We have studied in details the continuum limit of the Kesten recursion, which is related to the exponential functional of the Brownian motion. In the matrix case, we have derived the Fokker-Planck equation which describes the flow of the eigenvalues and obtained the stationary distribution limit. It shows that for the continuum matrix Kesten recursion the stationary matrix distribution is given by the inverse Wishart ensemble for any $N$, which generalizes the classical results of Dufresne and of two of the present authors for $N=1$. By relating the Fokker-Planck equation to an imaginary time Schr\"odinger problem, we have shown that the evolution of the eigenvalues, under a simple transformation, maps onto the quantum evolution of $N$ fermions in a Morse potential. Although both cases are presumably integrable, in the Hermitian case $\beta=2$ the fermions are non-interacting, leading to a simple solution for the dynamics of the original eigenvalues.
We showed that the above mentioned stationary state for the Kesten recursion can be retrieved from the ground state of the fermion system. 

Since it is an interesting problem in its own right, we have explored in details the physics of the ground state of $N$ non-interacting fermions in the Morse potential. In principle this system can be realized in cold atom experiments, where the fermions are trapped in potentials of tunable shape.
Although the Morse potential has only a finite number of bound states, and can accommodate only a finite number of fermions, 
one can tune its parameters so that this number is large. 
In the large $N$ limit the Fermi gas develops two edges and we have computed the scaling forms of the density and correlations
in the bulk and near the edges. These can be related, via some (a priori non obvious) transformation, to the Marcenko-Pastur distribution and to the
correlations in the Wishart-Laguerre ensemble. If we specify the parameters of the Morse potential to be relevant for the matrix Kesten problem, we find two distinct large $N$
scaling limits, depending on whether the variance of the noise is scaled as $1/N$ or remains $\mathcal{O}(1)$. In the first case (weak noise) the two spectral edges are ``soft" and the correlations at these edges are described by the Airy kernel. In the second case (strong noise) the right edge is pushed to infinity. The
largest eigenvalues (which correspond to the fermions at the far right)
are described by the Bessel determinantal point process, i.e. the physics of the ``hard edge" in random matrix theory. 
Remarkably, this provides a precise description of the many body correlations of the heavy tail of the matrix Kesten recursion. Note that these results for non-interacting fermions were obtained from the case $\beta=2$. In the case $\beta=1$, the fermions have in addition a Sutherland type interaction, and we have obtained the explicit form of their ground-state wavefunction.

Finally, we have shown that the matrix Kesten recursion proposed in the present work, although distinct from those studied by Rider and Valko and
by Grabsch and Texier, share the same distribution of eigenvalues in its continuum limit (modulo some subtleties with It\^o and Stratonovich prescriptions
that we elucidated). It is thus rewarding that a universal behavior seems to emerge in that problem. 

The present work leaves open several questions and suggests some interesting extensions. First, concerning the discrete matrix recursion
we have been able to derive exact results only in the large $N$ limit, using free probability theory. A self-consistent equation for 
the $\mathcal{S}$-transform of the Kesten matrix was obtained in the stationary limit. We have not been able to solve this equation, except in the
case of the continuum limit. It would thus be nice to study this equation further, and also to extend the analysis to finite $N$.
In particular, a problem which remains open is whether the remarkably simple condition $\mathbb{E}(z^\nu) = 1$, which determines the heavy tail exponent 
$\nu$ for the scalar case $N=1$, has any analog in the matrix case.

Let us finish by recalling our initial motivation for studying the present model, which is to explore possible matrix (non-commuting)
generalizations of the famous directed polymer problem (which is related to the KPZ stochastic growth equation). The directed polymer
problem amounts to study sums over a set of paths of products of scalar weights along these paths. The idea is to replace these weights by random matrices.
Such objects already appeared, e.g. in Chalker-Coddington type of models for localization in $d=2$ \cite{Cardy10}. In the case of 
positive definite matrix weights however, it has only been studied very recently \cite{OConnell2019}. Note also an operator generalization, which also recently appeared in the context of stochastic quantum spin chains \cite{Krajenbrink2020}.
We hope that the present study will also stimulate progress in these directions.

\acknowledgements

We thank A. Krajenbrink, M. Potters, G. Baverez and A. Lafay for useful discussions as well as S. N. Majumdar and C. Texier for pointing out connected references. We are grateful to a referee for bringing the Bougerol identity and its matrix generalization to our attention. P.L.D. thanks the LPTMS Orsay for hospitality while this work was completed and acknowledges support from ANR grant ANR-17-CE30-0027-01 RaMaTraF.

\begin{appendix}

\section{Perturbative evolution for eigenvalues}
\label{app:PT}

Let ${\rm d}H_t$ a matrix with entries $\sqrt{{\rm d}t} \ \xi_{i,j}$ with $\xi_{i,j} \sim \mathcal{N}(0,1)$ {\sc {iid}}. Letting $\{ \ket{v_i} \}$ an orthonormal basis of vectors, let us define $A_{i,j}$ as $\sqrt{{\rm d}t} A_{i,j} = \bra{v_i} {\rm d}H_t \ket{v_j}$, which is a Gaussian variable as a linear combination of independent Gaussian variables. Furthermore $A_{i,j}$ and $A_{i',j'}$ with a different pair of indices are independent, indeed we have:
\begin{eqnarray}
 && \mathbb{E} [A_{i,j} ] =  0  \\
  && \mathbb{E} [A_{i,j}^2 ] =  \bra{v_i}\ket{v_{i}} \bra{v_j}\ket{v_{j}} = 1 \\
   && \mathbb{E} [A_{i,j} A_{i',j'}]  = \bra{v_i}\ket{v_{i'}} \bra{v_j}\ket{v_{j'}} = \delta_{i,i'} \delta_{j,j'}
\end{eqnarray}
Using these notations, let us compute the perturbation theory for eigenvalue evolution in the Matrix Kesten model.
 
\tocless\subsection{Symmetric case \texorpdfstring{$\beta =1$}{beta=1}}
In the symmetric case, the noise matrix is ${\rm d}B_t = \frac{ {\rm d}H_t + {\rm d}H_t^T}{\sqrt{2}}$. Let us truncate the matrix evolution of the model studied in this paper and keep only the noise term:
\begin{equation}
{\rm d}U_t = 
\sigma \sqrt{U_t} {\rm d}B_t  \sqrt{U_t}
=
\sigma \sqrt{U_t} \left( \frac{{\rm d}H_t  + {\rm d}H_t^T}{\sqrt{2} } \right)  \sqrt{U_t}
\end{equation}
The evolution of eigenvalue $\lambda_{i}$, with corresponding eigenvectors $\{ \ket{v_{i,t}} \}$, can be expressed perturbatively \cite{Tao2012} as:
\begin{equation}
\lambda_{i,t+{\rm d}t} = \lambda_{i,t} + \sigma\bra{v_{i,t}} \sqrt{U_t} \left(\frac{{\rm d}H_t  +  {\rm d}H_t^T}{\sqrt{2} }  \right) \sqrt{U_t} \ket{v_{i,t}} + 
\sum\limits_{\substack{1 \leqslant j \leqslant N \\j\neq i}}
\frac{\abs{\bra{v_{j,t}} \sigma \sqrt{U_t} ( \frac{{\rm d}H_t  + {\rm d}H_t^T}{\sqrt{2} }  ) \sqrt{U_t}\ket{v_{i,t}} }^2
}{\lambda_{i,t} - \lambda_{j,t}}
\end{equation}
\begin{equation}
\lambda_{i,t+{\rm d}t} = \lambda_{i,t} + \sigma \lambda_{i,t}  \bra{v_{i,t}}\frac{{\rm d}H_t  + {\rm d}H_t^T}{\sqrt{2} }   \ket{v_{i,t}} + 
\sigma^2\sum\limits_{\substack{1 \leqslant j \leqslant N \\j\neq i}}
 \frac{ \lambda_{i,t}  \lambda_{j,t}  
\abs{\bra{v_{j,t}} \frac{{\rm d}H_t  + {\rm d}H_t^T}{\sqrt{2} }  \ket{v_{i,t}} }^2
 }{\lambda_{i,t} - \lambda_{j,t}}
\end{equation}
From the independence relations of the $A_{i,j}$ variables, let us define independent $\eta_{i,j}$ variables following a $\mathcal{N}(0,1)$ law: 
\begin{equation}
\bra{v_{i,t}} \frac{{\rm d}H_t  + {\rm d}H_t^T}{\sqrt{2} }   \ket{v_{j,t}}   =\sqrt{{\rm d}t}( \frac{ A_{i,j} + A_{j,i} }{\sqrt{2} } ) =  \sqrt{{\rm d}t} \left\{
    \begin{array}{ll}
    \sqrt{2}  \ \eta_{i,i} & \mbox{if } i =j \\
     \eta_{i,j}   & \mbox{if } i \neq j  
    \end{array}
\right.
\end{equation}
The eigenvalue $\lambda_{i}$ then evolves according to:
\begin{equation}
\lambda_{i,t+{\rm d}t} = \lambda_{i,t} + \sigma \lambda_{i,t} \sqrt{2 {\rm d}t}  \ \eta_{i,i} + 
\sigma^2\sum\limits_{\substack{1 \leqslant j \leqslant N \\j\neq i}}
\frac{ \lambda_{i,t}  \lambda_{j,t}  
 \ \eta_{i,j}^2
 }{\lambda_{i,t} - \lambda_{j,t}} {\rm d}t
\end{equation}
\begin{equation}
d \lambda_{i,t} =  \sigma \lambda_{i,t} \sqrt{2 {\rm d}t}  \ \eta_{i,i} + 
\sigma^2\sum\limits_{\substack{1 \leqslant j \leqslant N \\j\neq i}}
\frac{ \lambda_{i,t}  \lambda_{j,t}  
 \
 }{\lambda_{i,t} - \lambda_{j,t}} {\rm d}t
 +
\sigma^2\sum\limits_{\substack{1 \leqslant j \leqslant N \\j\neq i}}
 \frac{ \lambda_{i,t}  \lambda_{j,t}  
 \ (\eta_{i,j}^2 -1)
 }{\lambda_{i,t} - \lambda_{j,t}} {\rm d}t
\end{equation}
Following \cite{Tao2012}, the last term can be discarded in the stochastic equation as it has zero mean and order ${\rm d}t^3$ in the third moment. Finally, defining $N$ independents Brownian motions $(B_i)_{1\leqslant i \leqslant N}$, the evolution of the eigenvalues of $U$ is:
\begin{equation}
\label{eq:evModelASymmetric}
 d \lambda_i =  \sqrt{2} \sigma \lambda_i  \ {\rm d}W_i + 
\sigma^2\sum\limits_{\substack{1 \leqslant j \leqslant N \\j\neq i}}
\frac{ \lambda_{i}  \lambda_{j}  
 \
 }{\lambda_{i} - \lambda_{j}} {\rm d}t
 \quad \quad \quad \quad 
 \text{(It\^o)}
 \end{equation}

\tocless\subsection{Hermitian case \texorpdfstring{$\beta =2$}{beta=2}}

The previous computation extends to the Hermitian case by taking ${\rm d}B_t = \frac{{\rm d}H_t + {\rm d}H_t^T + i ( {\rm d}\tilde{H}_t - {\rm d}\tilde{H}_t^T)}{2}$ where ${\rm d}H_t$ and ${\rm d}\tilde{H}_t$ are independent samples of the same distribution. The perturbative evolution of eigenvalues can be computed as above. Indeed, defining $\eta_{i,j}$ and  $\tilde{\eta}_{i,j}$ independent sets of standard gaussian variables:
\begin{equation}
\abs{\bra{v_{j,t}}  \frac{{\rm d}H_t + {\rm d}H_t^T + i ( d \tilde{H}_t - {\rm d}\tilde{H}_t^T)}{2}   \ket{v_{i,t}} }^2
= \frac{1}{2} {\rm d}t (\eta_{i,j}^2 + \tilde{\eta}_{i,j}^2) =  {\rm d}t  + \mathcal{O} ( {\rm d}t ( \eta_{i,j}^2 + \tilde{\eta}_{i,j}^2 -2 ) )
\end{equation}

We conclude that the evolution of the eigenvalues of $U$ is, in the Hermitian case:
\begin{equation}
\label{eq:evModelAHermitian}
 d \lambda_i =   \lambda_i  \ {\rm d}W_i + 
\sigma^2\sum\limits_{\substack{1 \leqslant j \leqslant N \\j\neq i}}
\frac{ \lambda_{i}  \lambda_{j}  
 \
 }{\lambda_{i} - \lambda_{j}} {\rm d}t
  \quad \quad \quad \quad 
 \text{(It\^o)}
 \end{equation} 
 
We conclude with the full evolution of the main text, by adding the drift that was left aside in this section. The perturbative evolution of eigenvalues of $U_t$ evolving according to \eqref{eq:EvolutionU} is given, in symmetric and Hermitian cases, by:
\begin{equation}
\label{eq:PTmodelsqUdBsqU}
{\rm d}\lambda_i =
\left(1+m \lambda_{i}\right) {\rm d} t+
  \sqrt{\frac{2}{\beta}} \sigma \lambda_i  \ {\rm d}W_i + 
 \sigma^2 \sum\limits_{\substack{1 \leqslant j \leqslant N \\j\neq i}}
\frac{ \lambda_{i}  \lambda_{j}  
 \
 }{\lambda_{i} - \lambda_{j}} {\rm d}t
  \quad \quad \quad  \quad 
 \text{(It\^o)}
\end{equation}

\tocless\subsection{Eigenvalue evolution of the Grabsch-Texier model}
\label{app:PerturbationGT}
We derive the perturbative evolution of the set of eigenvalues of the matrix $Z_t$ evolving as the Grabsch-Texier model:
\begin{equation} 
\label{eq:AppGTStratoSDE}
{\rm d}Z_t = \left(   I - 2 \mu Z_t \right){\rm d}t  + \sigma_\beta ( Z_t {\rm d}B_t + {\rm d}B_t   Z_t) \quad \quad \quad \quad \mathrm{(Stratonovich)}
\end{equation}
where the matrix is real symmetric ($\beta =1$) or complex Hermitian ($\beta=2$), and the noise matrix is structured as in Eq. \eqref{eq:DefinitionBn}:
\be 
\label{eq:StructureNoiseGT}
{\rm d}B_t =  
\frac{{\rm d}H_t + {\rm d}H_t^T}{\sqrt{2\beta}} + i \frac{{\rm d}\tilde H_t - {\rm d}\tilde H_t^T}{2} \delta_{\beta=2} 
\ee
where $H_t$ and $\tilde H_t$ are, as above, both filled with $N^2$ independent standard real Brownian motions. This reproduces the
noise correlations given in Eqs. (14-15) of \cite{GrabschTexier2020} by setting $\sigma_\beta$ to $ \sqrt{\frac{\beta}{2}}$ times a constant.

In order to derive the evolution of the eigenvalues, it is now convenient to take the matrix SDE \eqref{eq:AppGTStratoSDE} to the It\^o prescription, which gives: 
\begin{equation} 
\label{eq:AppGTItoSDE}
dZ_t = \left(   I - 2 \mu Z_t \right){\rm d}t + \sigma_\beta^2
\left(
\operatorname{tr}(Z_t) I + (N+4-2\beta) Z_t \right){\rm d}t 
 + \sigma_\beta (Z_t {\rm d}B_t +{\rm d}B_t   Z_t) \quad \quad \quad \quad \text{(It\^o)} 
\end{equation}
This is obtained as follows. First one recalls that in Stratonovitch prescription, the noise part of \eqref{eq:AppGTStratoSDE} has the form $\sigma_\beta(Z_{t+\frac{{\rm d}t}{2}}{\rm d}B_t +{\rm d}B_tZ_{t+\frac{{\rm d}t}{2}})$. Then, expanding (at order ${\rm d}t^{1/2}$) $Z_{t+\frac{{\rm d}t}{2}} = Z_t + \frac{\sigma_\beta}{2} (Z_t {\rm d}B_t +{\rm d}B_t Z_t)$ and using the noise structure given in \eqref{eq:StructureNoiseGT} leads to \eqref{eq:AppGTItoSDE}.

Let us truncate the evolution as above and keep the noise term only $dZ_t = \sigma_\beta (Z_t {\rm d}B_t + {\rm d}B_t Z_t)$. The perturbative evolution of eigenvalue $\lambda_i$ is:
\bea
\lambda_{i,t+{\rm d}t} &=& \lambda_{i,t} + \sigma_\beta  \bra{v_{i,t}} Z_t {\rm d}B_t + {\rm d}B_t Z_t \ket{v_{i,t}} + \sigma_\beta^2
\sum\limits_{\substack{1 \leqslant j \leqslant N \\j\neq i}}
\frac{\abs{\bra{v_{j,t}} Z_t {\rm d}B_t + {\rm d}B_t Z_t\ket{v_{i,t}} }^2
}{\lambda_{i,t} - \lambda_{j,t}} \\
&=& \lambda_{i,t} + \sigma_\beta  \lambda_{i,t} 
\bra{v_{i,t}} 2 {\rm d}B_t \ket{v_{i,t}}+  \sigma_\beta^2
\sum\limits_{\substack{1 \leqslant j \leqslant N \\j\neq i}}
\frac{ (  \lambda_{i,t}+  \lambda_{j,t})^2 \abs{ \bra{v_{j,t}}{\rm d}B_t\ket{v_{i,t} }  }^2
}{\lambda_{i,t} - \lambda_{j,t}}
\eea 
From the noise structure \eqref{eq:StructureNoiseGT}, we obtain with $2N^2$ independent $\eta_{i,j}$ and $\tilde \eta_{i,j} \sim \mathcal{N}(0,1)$ variables:
\be
\lambda_{i,t+{\rm d}t} = \lambda_{i,t} + 2 \sqrt{\frac{2}{\beta}}  \sigma_\beta  \lambda_{i,t} \sqrt{{\rm d}t} \eta_{i,i} + \frac{\sigma_\beta^2}{2\beta} \sum\limits_{\substack{1 \leqslant j \leqslant N \\j\neq i}}
\frac{(\lambda_{i,t} + \lambda_{j,t})^2}{\lambda_{i,t} - \lambda_{j,t}}
\left((\eta_{i,j}+\eta_{j,i})^2 + (\tilde\eta_{i,j}- \tilde\eta_{j,i})^2 \delta_{\beta=2} \right)
{\rm d}t
\ee

Adding finally the drift terms, the perturbative evolution of the eigenvalues $\lambda$ evolving under the It\^o SDE \eqref{eq:AppGTItoSDE} (or equivalently the Stratonovitch SDE \eqref{eq:AppGTStratoSDE}) is:
\be 
{\rm d}\lambda_i = (1 - 2 \mu \lambda_i) {\rm d}t + 
\sigma_\beta^2
\left(
 \sum\limits_{1\leqslant j \leqslant N} \lambda_j  + (N+4-2\beta) \lambda_i \right){\rm d}t 
+   2  \sqrt{\frac{2}{\beta}} \sigma_\beta \lambda_i {\rm d}W_i + \sigma_\beta^2 \sum\limits_{\substack{1 \leqslant j \leqslant N \\j\neq i}}
\frac{(\lambda_i + \lambda_j)^2}{\lambda_i - \lambda_j} {\rm d}t
\ee 
Reordering terms, we obtain:
\be 
{\rm d} \lambda_i = (1 +(\sigma_\beta^2(N+2-\beta)- \mu) 2\lambda_i) {\rm d}t 
+ 2  \sqrt{\frac{2}{\beta}} \sigma_\beta \lambda_i {\rm d}W_i + 4 \sigma_\beta^2 \sum\limits_{\substack{1 \leqslant j \leqslant N \\j\neq i}}
\frac{\lambda_i \lambda_j}{\lambda_i - \lambda_j} {\rm d}t
 \quad   \quad    \quad  
 \text{(It\^o)}
\ee 

We recover the evolution obtained for our model in Eq. \eqref{eq:PTmodelsqUdBsqU} by fixing $\sigma_\beta =\frac{\sigma}{2}$ and $\frac{\sigma^2}{2}(N+2-\beta)-  2 \mu= m$.\\\\

\section{Free probability}
\label{app:FreeProba}

\tocless\subsection{RMT transforms and main results}

Free probability theory is the study of free random variables, where freeness is a generalization of independence to non-commuting variables. We will not detail the definition of freeness further than stating that two large matrices are free if their eigenbasis are randomly rotated with respect to one another. For detailed introductions to the field and its application to random matrices, see \cite{RMTBook,MingoSpeicher2017,TulinoVerdu,Novak14,BunBouchaudPotters2017}.  In this Appendix section, we detail the main RMT and free probability tools that we use in this paper. Let us consider a large $N \times N$ matrix $M$, which has a one-cut spectrum with a density of eigenvalues $\rho = \frac{1}{N} \sum_{\lambda \in \mathrm{Sp}(M)} \delta_\lambda$ such that:
\begin{equation}
\mathrm{supp}(\rho)=[a,b] \quad \quad \int_a^b \rho(u) \mathrm{d}u = 1
\end{equation}
A standard RMT object is the Stieltjes transform, defined on $\mathbb{C}\setminus[a,b]$:
\begin{equation}
\mathfrak{g}(z)=\int_a^b \frac{\rho(u)}{z-u} \mathrm{d} u
\end{equation}
We note that the density can be retrieved from the Stieltjes transform as:
\begin{equation}
\label{eq:StieltjesInversion}
\rho(u) = \frac{1}{\pi} \lim_{\eta \to 0^+} \mathrm{Im}\left( \mathfrak{g}(u - i \eta) \right)
\end{equation}
We also give the following scaling and inverting properties of $\mathfrak{g}$ (if $M$ is invertible):
\begin{eqnarray}
\mathfrak{g}_{a M}(z)&=&\frac{1}{a} \mathfrak{g}_M\left(\frac{z}{a}\right) \\
1&=&z \mathfrak{g}_{M}(z)+\frac{1}{z} \mathfrak{g}_{M^{-1}}\left(\frac{1}{z}\right)
\end{eqnarray}
The behaviour of $\mathfrak{g}$ at infinity is clear from the definition, as well as the general series expansion in terms of the moments $\varphi_k$ of $M$, with $\varphi_k = \frac{1}{N} \operatorname{tr}( M^k)= \int_a^b u^k \rho(u) \, \mathrm{d}u $ for $k\geqslant 1$:
\begin{equation}
\mathfrak{g}(z) \sim_{\abs{z} \to \infty} \frac{1}{z} \quad \quad \quad 
\mathfrak{g}(z)  =_{\abs{z} \to \infty}  \frac{1}{z} +\sum\limits_{k=1}^\infty \frac{\varphi_k }{z^{k+1} }
\end{equation}
The moment generating function, or $\mathcal{T}$-transform, is then:
\begin{equation}
\mathcal{T}(z) = z \, \mathfrak{g}(z) - 1 =\int_a^b \frac{u \, \rho(u)}{z-u} \mathrm{d} u   = \sum\limits_{k=1}^\infty \frac{\varphi_k }{z^{k} }
\end{equation}
and has the following scaling property:
\begin{eqnarray}
\mathcal{T}_{a M} (z)  = \mathcal{T}_{M} (\frac{z}{a}) 
\end{eqnarray}
The $\mathcal{B}$ (or Blue) transform is defined as the functional inverse of the Stieltjes transform $\mathcal{B}(\mathfrak{g}(z)) = z$ for all $z \text{ in } \mathbb{C}\setminus[a,b]$, and the $\mathcal{R}$-transform is defined as:
\begin{equation}
\mathcal{R}(\omega ) = \mathcal{B}(\omega) - \frac{1}{\omega}
\end{equation}
From the scaling property of $\mathfrak{g}$, we obtain scaling properties for $\mathcal{B}$ and $\mathcal{R}$:
\begin{eqnarray}
\mathcal{B}_{a M }(\omega)&=&a \mathcal{B}_{M}(a \omega) \\
\mathcal{R}_{a M}(\omega)&=&a \mathcal{R}_{M}(a \omega)
\end{eqnarray}
It can be shown that the $\mathcal{R}$-transform admits a Taylor expansion as $\omega \to 0$:
\begin{equation}
\mathcal{R}(\omega) = \sum\limits_{k=1}^\infty \kappa_k \omega^{k-1}
\end{equation}
where the coefficients $\kappa_k(M)$ are the free cumulants of $M$, which can be expressed in terms of the moments $\varphi_k$ in a different way from cumulants in standard probability theory. The definition of free cumulants and their combinatorial interpretation in terms of non-crossing partitions can be found in \cite{SpeicherCombinatorics,MingoSpeicher2017,RMTBook}. An important result from free probability theory is the additivity of free cumulants for the sum of two free matrices $A$ and $B$, i.e. for all $k$:
\begin{equation}
\kappa_k(A+B) = \kappa_k(A) + \kappa_k(B)
\end{equation}
Additivity then extends to the $\mathcal{R}$-transfoms of free matrices $A$ and $B$:
\begin{equation}
\mathcal{R}_{A+B}(\omega)=\mathcal{R}_{A}(\omega)+\mathcal{R}_{B}(\omega)
\end{equation}
This is usually referred to as Voiculescu's free addition. A similar result holds for free multiplication, that we now turn to. In this regard, we define the $\mathcal{S}$-transform as:
\begin{equation}
\mathcal{S}(\omega) = \frac{\omega +1 }{\omega \mathcal{T}^{-1}(\omega) }
\end{equation}
where $\mathcal{T}^{-1}$ is the functional inverse of the $\mathcal{T}$-transform. From the scaling property of $\mathcal{T}$, we obtain:
\begin{equation}
\mathcal{S}_{a M} (\omega)  = \frac{1}{a}  \mathcal{S}_{ M} (\omega)  
\end{equation}
We state the following relations between $\mathcal{S}$ and $\mathcal{R}$ transforms:
\begin{eqnarray}
\label{eq:RelationSR}
\mathcal{S}(\omega) &=& \frac{1}{\mathcal{R}( \omega \mathcal{S}(\omega) )  }\\
\mathcal{R}(\omega) &=& \frac{1}{\mathcal{S}( \omega \mathcal{R}(\omega) ) }
\end{eqnarray}
The main free probability result that we use in this paper is the following: for two free self-adjoint matrices $A$ and $B$, the self-adjoint product $M= A^{1/2} B A^{1/2}$ verifies the following factorization property on the $\mathcal{S}$ transforms \cite{Voiculescu91,RMTBook}:
\begin{equation}
\label{eq:FreeMultiplication}
\mathcal{S}_{M} (\omega) = \mathcal{S}_A(\omega) \times  \mathcal{S}_B(\omega)
\end{equation}
This is the free multiplication, or free multiplicative convolution, theorem, expressed here for the self-adjoint product of free random matrices.

Note that the $\mathcal{S}$-transform can be expressed as a function of the moments $\varphi_k$ 
as 
\be
\mathcal{S} (\omega) =\frac{\omega+1}{\omega}\left(\sum_{k=1}^{\infty} \varphi_k \omega^{k}\right)^{<-1>}
\ee
where subscript ${<-1>}$ denotes the operation of taking the inverse with respect to composition of formal power series.
To invert that relation and obtain the moments from the expansion of $\mathcal{S} (\omega)$ in powers of $\omega$ one can
write, denoting $z=\mathcal{T}^{-1}(\omega)$:
\bea
\label{eq:InvertingSeriesMomentsSTransform}
\frac{1}{z}  = \frac{\omega \mathcal{S}(\omega)}{\omega + 1} = \text{series in } \omega
\eea
One then inverts this series, and obtains $\omega$ as a series in $1/z$ whose coefficients are the moments, i.e. 
$\omega = \sum_{k \geqslant 1} \frac{\varphi_k}{z^k}=\mathcal{T}(z)$. 

It is interesting to note that the knowledge of the $\mathcal{S}$-transform allows to find the position of the edges
$\lambda_e$ of the spectrum.
In generic cases these edges are determined by the roots $\mathfrak{g}_e$ of the equation
\be
\mathcal{B}'(\mathfrak{g}_e) = 0 \quad , \quad \lambda_e = \mathcal{B}(\mathfrak{g}_e) 
\ee
From their definitions, one sees that the relation between the $\mathcal{B}$ and $\mathcal{S}$ transforms is obtained by eliminating $\omega$ in the following system:
\bea
&& \mathcal{B}(\mathfrak{g}) = \frac{\omega+1}{\omega \mathcal{S}(\omega)} \\
&& \omega = \mathfrak{g} \mathcal{B}(\mathfrak{g}) - 1
\eea
As a consequence we find that the position of the edges are the solutions of the following system:
\be 
\label{eq:SystemEdgeStransform}
\begin{array}{ll}
        \mathcal{S}(\omega_e) + \omega_e (\omega_e+1) \mathcal{S}'(\omega_e) = 0  \\[5pt]
        \lambda_e = \frac{\omega_e+1}{\omega_e \; \mathcal{S}(\omega_e)}
\end{array}
\ee
One can check for instance that for $\mathcal{S}(\omega)= 1 - \frac{\omega}{2 \kappa}$ one recovers as solutions
the two edges of the inverse-Wishart ensemble given in \eqref{eq:IMPDistribution}. At order $\epsilon$ in the expansion \eqref{eq:ExpansionStransformEpsilon}, one recovers the edges given in \eqref{eq:EdgesFirstOrderEpsilon}.

\tocless\subsection{\texorpdfstring{$\mathcal{S}$}{S}-transform of \texorpdfstring{$\xi$}{xi}}
We obtain the $\mathcal{S}$-transform of the noise matrix $\xi=(1+m \, \epsilon) I +\tilde{\sigma} \, \sqrt{\epsilon} \frac{B}{\sqrt{N}} $ considered in the text, where we assume $1+m \epsilon >0$. Here $\frac{B}{\sqrt{N}}$ is a symmetric Wigner matrix with semi-circle eigenvalue density with support $[-2,2]$. Its Stieltjes transform is $\mathfrak{g}(z) = \frac{z}{2} \left(1- \sqrt{1-\frac{4}{z^2}}\right)$ and its $\mathcal{R}$-transform is: $\mathcal{R}_{\frac{B}{\sqrt{N}} } ( \omega) = \omega $. By free sum of a scaled identity and a scaled Wigner matrix, the $\mathcal{R}$-transform of $\xi$ is:
\begin{equation}
\mathcal{R}_\xi (\omega) = 1+ m \epsilon + \tilde{\sigma}^2 \epsilon \omega
\end{equation}
We deduce from \eqref{eq:RelationSR} the $\mathcal{S}$-transform of the noise matrix $\xi$:
\begin{equation}
\label{eq:StransformSigma}
\mathcal{S}_\xi (\omega) =  \frac{-1-m \epsilon + \sqrt{(1+m\epsilon)^2 +4  \tilde{\sigma}^2 \epsilon \omega}}{2  \tilde{\sigma}^2 \epsilon \omega} 
\end{equation}
where the correct square-root branch is chosen here such that $\mathcal{S}$ has a regular behaviour as $\omega \to 0$. In this region, we have then $\mathcal{S}(\omega) = \frac{1}{1 + \epsilon m} - \frac{\epsilon \sigma^2}{(1+\epsilon m)^3}\omega + \mathcal{O}(\omega^2)$.\\\\

\section{Wishart and inverse-Wishart distributions}
\label{app:WishartInverseWishart}

\tocless\subsection{The Wishart ensemble}

The Wishart ensemble consists of $N \times N$ matrices $M = X^\dagger X$ where $X$ is a $T \times N$ matrix ($T\geqslant N$ if $T$ and $N$ are integers, $T> N-1$ in the general case) with independent entries following a real centered gaussian law of variance $C^{-1}$ if $\beta =1$ (such that $X^\dagger = X^T$), or complex centered gaussian  law of variance $C^{-1}$ if $\beta =2$. By construction, $M$ is a symmetric (resp. Hermitian) positive semidefinite matrix with real (resp. complex) entries if $\beta =1$ (resp. $\beta = 2$).
The matrix distribution of $X$ is $P(X) \propto e^{-C \frac{\beta}{2} \operatorname{tr} (X^\dagger X) }$, and taking into account the Jacobian of the transformation $X \to M$ we obtain the Wishart matrix distribution of $M$:
\begin{equation}
\mathcal{P}_{\mathrm{w}}(M) =  Q_{T, N, C, \beta}  \ 
\operatorname{det}(M)^{ \frac{\beta}{2} (T-N+1)-1} 
\ e^{- C \frac{\beta}{2} \operatorname{tr}M} 
\end{equation}
where the normalization constant $Q_{T, N, C, \beta}  $ is: 
\begin{equation}
Q_{T, N, C, \beta} =\left(\frac{C \beta}{2}  \right) ^{\frac{\beta}{2} N T}    \frac{\pi^{- \frac{\beta}{2} \frac{N(N-1)}{2}  }}{    \prod_{j=1}^N \Gamma \left( \frac{\beta}{2}( T - N +j) \right)} = 
\frac{\left(C \beta / 2  \right) ^{\frac{\beta}{2} N T}  }{\Gamma_{N,\beta}( \frac{\beta}{2} T)}
\end{equation}
where the generalized multivariate Gamma function is $\Gamma_{N,\beta}(a) = \pi^{\frac{\beta N(N-1)}{4}} \prod_{j=1}^N \Gamma \left( a- \beta \frac{j-1}{2} \right)$.
The joint eigenvalue distribution of $M$ on $(\mathbb{R}^+)^N$ follows:
\begin{equation}
P_{\mathrm{w}}(\vec{\lambda} ) =K_{T, N, C, \beta}  \ \left|\Delta(\vec{\lambda})\right|^{\beta}  \ \prod_{i=1}^{N} \lambda_{i}^{\frac{\beta}{2}(T-N+1)-1} 
e^{-C \frac{\beta}{2} \sum_{i=1}^N \lambda_{i}} 
\end{equation}
where the normalization constant $K_{T, N, C, \beta}  $ is:
\begin{equation}
K_{T, N, C, \beta}  =\left(\frac{C \beta}{2}  \right) ^{\frac{\beta}{2} N T} \ \prod_{j=1}^{N} \frac{\Gamma\left(1+\frac{\beta}{2}\right)}{\Gamma\left(1+\frac{\beta}{2} j\right) \Gamma\left(\frac{\beta}{2} ( T-N+j)\right)}
= \frac{1}{N!}\left( \frac{C \beta}{2}  \right) ^{\frac{\beta}{2} N T }
\frac{ \Gamma\left(\frac{\beta}{2}\right)^N \pi^{\frac{\beta}{2}N(N-1) }}{\Gamma_{N,\beta}\left( \frac{\beta}{2}N\right) \Gamma_{N, \beta} \left( \frac{\beta}{2}T \right) }
\end{equation}

In the limit of large $N$ and $T$ with $\frac{N}{T} = q$ fixed, and with the usual convention $C  = 1$, the Empirical Spectral Distribution (ESD) of the rescaled matrix $\frac{1}{T} M$ converges to the Marcenko-Pastur distribution:
\begin{equation}
\rho_{\frac{1}{T}  M}(\lambda) = \frac{1}{N} \sum_{i=1}^N  \delta(\lambda- \lambda_i)  \to  \rho_{\mathrm{MP}}(\lambda)=\frac{\sqrt{4 \lambda q-(\lambda+q-1)^{2}}}{2 q \pi \lambda}, \quad  \lambda \in\left[(1-\sqrt{q})^2, (1+\sqrt{q})^2\right]
\end{equation}

This limit distribution is characterized by the following Stieltjes, $\mathcal{R}$ and $\mathcal{S}$ transforms:
\begin{equation}
\mathfrak{g}_{\mathrm{MP}}(z)=\frac{(z+q-1)-\sqrt{  z-(1- \sqrt{q})^2 } \sqrt{ z-(1+ \sqrt{q})^2 }}{2 q z} 
\quad \quad \quad
\mathcal{R}_\mathrm{MP}(\omega) =  \frac{1}{1- q \omega}
\quad \quad \quad
\mathcal{S}_\mathrm{MP}(\omega) =  \frac{1}{1+q \omega}
\end{equation}

\tocless\subsection{The inverse-Wishart ensemble}
 
The inverse-Wishart distribution is the distribution of $M=W^{-1}$, where $W$ follows the Wishart distribution, with the parameters defined above. Taking into account the Jacobian of the $M \to M^{-1}$ transform, we obtain the inverse-Wishart matrix probability density, with the same normalization constant $Q_{T, N, C, \beta}  $:
 \begin{equation}
\mathcal{P}_{\mathrm{iw}}\left( M \right) = Q_{T, N, C, \beta}   \
\operatorname{det}(M)^{ -\frac{\beta}{2} (T + N -1 ) -1 }
 e^{- C \frac{\beta}{2} \operatorname{tr} M^{-1}}
 \end{equation} 
The corresponding joint eigenvalue distribution is, with the same normalization constant  $K_{T, N, C, \beta}  $:
\begin{equation}
P_{\mathrm{iw}}(\vec{\lambda} ) =K_{T, N, C, \beta}  \ \left|\Delta(\vec{\lambda})\right|^{\beta}  \ \prod_{i=1}^{N} \lambda_{i}^{-\frac{\beta}{2} (T + N -1 ) -1 }
e^{-C \frac{\beta}{2} \sum_{i=1}^N \frac{1}{\lambda_{i}}  } 
\end{equation}

In the limit of large $N$ and $T$ with $\frac{N}{T} = q$ fixed, and with the usual convention $C= 1$, the ESD of the rescaled matrix $(1-q)T \times M$ then converges to the inverse Marcenko-Pastur distribution:
\begin{equation}
\label{eq:IMPDistribution}
\rho_{(1-q)T \times M} (\lambda) = \frac{1}{N} \sum_{i=1}^N  \delta(\lambda- \lambda_i)  \quad \longrightarrow \quad 
\rho_{\mathrm{IMP}}(\lambda)= \frac{\kappa}{\pi \lambda^{2}} \sqrt{\left(\lambda_{+}-\lambda \right)\left(\lambda-\lambda_{-}\right)}, \quad  \lambda \in\left[\lambda_{-}, \lambda_{+}\right]
\end{equation}
where $q = \frac{1}{2 \kappa +1 }$ and $\lambda_\pm=\frac{1}{\kappa}[\kappa+1 \pm \sqrt{2 \kappa+1}]$. In addition to the rescaling by $T$, which is obvious from the previous subsection, we insert an additional factor $1-q$ such that the mean of the limit spectrum is 1. In the main text, we denote the rescaled matrix by $\InverseW = (1-q)T\times M$. This limit distribution is characterized by the following Stieltjes, $\mathcal{R}$ and $\mathcal{S}$ transforms:
\begin{equation}
\mathfrak{g}_\mathrm{IMP}(z)=\frac{z(\kappa+1)-\kappa-\kappa \sqrt{z-\lambda_{-}} \sqrt{z-\lambda_{+}}}{z^{2}}
\quad \quad \quad
\mathcal{R}_\mathrm{IMP}(\omega) =  \frac{\kappa-\sqrt{\kappa(\kappa-2 \omega)}}{\omega}
\quad \quad \quad
\mathcal{S}_\mathrm{IMP}(\omega) =  1- \frac{\omega}{2 \kappa}
\end{equation} 

From the joint eigenvalue density of the inverse-Wishart distribution, we can obtain the marginal density for the highest eigenvalue $\lambda_1$ in the limit where it is very large, with a factor $N$ for the choice of the eigenvalue pushed to large values:
\begin{eqnarray}
P(\lambda_1) &\isEquivTo{\lambda_1 \to \infty}&  N \ K_{T, N, C, \beta}  \ \lambda_1^{- \frac{\beta}{2} ( T-N +1) -1}  \int 
\left|\Delta(\lambda_2, \cdots, \lambda_N)\right|^{\beta}  \ \prod_{j=2}^{N} \lambda_{j}^{-\frac{\beta}{2} (T + N -1 ) -1 }
e^{-C \frac{\beta}{2} \sum_{i=2}^N \frac{1}{\lambda_{i}}  } \prod_{j=2}^N \mathrm{d}\lambda_j 
\\
&=& 
\frac{N \ K_{T, N, C, \beta} }{K_{T+1, N-1, C, \beta} }   \lambda_1^{- \frac{\beta}{2} ( T-N +1) -1}     
=
\frac{ N \left( C \beta/2  \right) ^{\frac{\beta}{2}(T-N+1)} 
\Gamma(1+\frac{\beta}{2}) \ \Gamma( \frac{\beta}{2} ( T+1) )
\
 \lambda_1^{- \frac{\beta}{2} ( T-N +1) -1}  
}{
\Gamma(1+ \frac{\beta}{2}N) \ \Gamma( \frac{\beta}{2} ( T-N+ 1) ) \ \Gamma( \frac{\beta}{2} ( T-N+ 2)) 
}
\end{eqnarray}
Inserting in this equation the values for $C,T$ given in \eqref{eq:parameters} one obtains the equation \eqref{eq:QueueMarginalLargestEval} given in the text. For some results on the distribution of the smallest eigenvalue of the Wishart matrix, see \cite{Edelman88}.

We note that recent mathematical interest in the Wishart and inverse-Wishart ensembles has unveiled connections with the combinatorics of Hurwitz numbers \cite{Cunden18,Gissoni20} and an explicit ergodic decomposition \cite{Assiotis19}.
\\\\

\section{Morse potential}
\label{app:MorsePresentation}

We review the single-particle quantum system in a Morse potential \cite{Morse29,Dong07}. We are looking to solve the following eigenproblem, where $g>0$ and $x_0$ are two constants:
\begin{equation}
\begin{array}{l}\left(-\frac{\partial^{2}}{\partial x^{2}}+V(x)\right) \psi_{k}(x)=\varepsilon_{k} \psi_{k}(x) \\ V(x)=g^{2}\left(e^{-2\left(x-x_0\right)}-2 e^{-\left(x-x_0\right)}\right)\end{array}
\end{equation}
Let us take $z = 2 g e^{-(x-x_0)}$ and define $\phi (z) =  z^{g-k-\frac{1}{2}} e^{-\frac{1}{2} z} f(z)$. Inserting this in the differential equation gives:
\begin{equation}
\left(-\frac{\partial^{2}}{\partial x^{2}}+V(x)\right) \phi(2 g e^{-(x-x_0)})= 
-\left(g - k -\frac{1}{2}\right)^2 \phi(z) 
-  z^{g-k+\frac{1}{2}}   e^{-z/2} \left[  z f''(z) + (2 g -2 k -z ) f'(z) + k f(z)\right]
\end{equation}
The eigenproblem is solved if $f$ solves the generalized Laguerre equation $0 = z f''(z) + (2 g -2 k -z ) f'(z) + k f(z)$ such that regularity conditions impose that $f = L^{(2g-2k- 1)}_{k}$ is a generalized Laguerre polynomial.
The $k$-th eigenfunction is then: 
\begin{equation}
\psi_{k}(x)=N_{k} \left(2 g e^{-(x-x_0)}\right)^{g-k-\frac{1}{2}} e^{-  g e^{-(x-x_0)}} L_{k}^{(2 g-2 k-1)}\left(2 g e^{-(x-x_0)} \right)
\end{equation}
where $N_{k}=\left[\frac{k !(2 g-2 k-1)}{\Gamma(2 g-k)}\right]^{\frac{1}{2}}$ and $k < g - \frac{1}{2}$. This is thus a state in the discrete spectrum, with corresponding eigenvalue:
\begin{equation}
\varepsilon_k = -\left(g - k -\frac{1}{2}\right)^2
\end{equation}
The maximum number of fermions that can be stacked in the discrete spectrum of the Morse potential is the largest integer $N$ such that the last level exists i.e. $N -1  <   g - \frac{1}{2} \iff N <g + \frac{1}{2}  $. We denote this maximum occupation number by a strict integer part $N= \lfloor \left(g + \frac{1}{2} \right)^- \rfloor$.

In addition to the discrete spectrum of bound states at negative energies $\epsilon_k$, there is a continuum of
positive energy eigenstates $\epsilon=p^2 >0$. One can check that the associated eigenfunctions are of the form
$\psi_p(x) \sim e^{(x-x_0)/2} W_{g, i p}(2 g e^{-(x-x_0)})$ in terms of the Whittaker $W$ function, such that $\psi_p$ is a real function decreasing fast for
$x<0$ and oscillating for $x>0$. 

Putting all together we can write the Green's function as follows \cite{Zhang10}:
\bea
\nonumber
&& G(x,x',t)= \sum_{k=0}^{\lfloor \left(g - \frac{1}{2} \right)^- \rfloor} \psi_k(x) \psi_k(x') e^{- (g - k - \frac{1}{2})^2 t} \\
&& + \int_0^{+\infty} \frac{dp}{2 g \pi^2} p \sinh(2 \pi p) |\Gamma(i p - g + \frac{1}{2})|^2   e^{\frac{x+x'}{2} - x_0} W_{g, i p}(2 g e^{-(x-x_0)}) W_{g, i p}(2 g e^{-(x'-x_0)}) e^{- p^2 t} 
\eea
We point to \cite{Duru83} for an expression of this Green's function in path integral form.

\section{Exact solutions for the evolution of \texorpdfstring{$\mathcal{S}$}{S} and Stieltjes transforms}
\label{app:ExactSolSandG}

\tocless\subsection{\texorpdfstring{$\mathcal{S}$}{S}-transform} 

In this Appendix we study the evolution equation \eqref{eq:EvolSU} for the $\mathcal{S}$ transform, first to linear order
near the stationary solution and next in an exact manner using an hodograph transformation.

Recalling the stationary solution $\mathcal{S}_{\text{stat}}(\omega) = -m - \frac{\tilde \sigma^2}{2} \omega$ and writing $\mathcal{S}(\omega,t) = \mathcal{S}_{\text{stat}}(\omega) + \Delta \mathcal{S}(\omega,t) $, we have the following equation for $\Delta \mathcal{S}$ :
\begin{equation}
\frac{\partial}{\partial t } \Delta \mathcal{S}(\omega,t)= - \left( \Delta \mathcal{S}(\omega,t) + \mathcal{S}_{\text{stat}}(\omega)  \right) \frac{\partial}{\partial \omega} \left( \omega \Delta \mathcal{S}(\omega,t) \right)
\end{equation}
Expanding to linear order in $\Delta \mathcal{S}$, one obtains a linear differential equation whose solution is given in \eqref{eq:LinearizedSolutionSEvolution}.

An exact solution of the non-linear equation \eqref{eq:EvolSU} can be obtained using the hodograph transform, we point to \cite{PDEBook1} for a general presentation of the technique. The result is given in an implicit form
for $\Delta \mathcal{S}(\omega,t)$ as
\bea \label{soluexacte}
\frac{\tilde \sigma (\omega - \Omega(2 \omega \Delta \mathcal{S}(\omega,t))) }{\sqrt{ 2 \omega \Delta \mathcal{S}(\omega,t) + (\frac{m}{\tilde \sigma})^2}
- (\tilde \sigma \omega + \frac{m}{\tilde \sigma}) \frac{\tilde \sigma \Omega(2 \omega \Delta \mathcal{S}(\omega,t)) + \frac{m}{\tilde \sigma}}{
\sqrt{ 2 \omega \Delta \mathcal{S}(\omega,t) + (\frac{m}{\tilde \sigma})^2}} } = \tanh \left( 
t\frac{\tilde\sigma}{2} \sqrt{ 2 \omega \Delta \mathcal{S}(\omega,t) + (\frac{m}{\tilde \sigma})^2} \right)
\eea
where the function $\Omega(u)$ is obtained by inverting
\be \label{Omega} 
\omega = \Omega(u) \quad \Leftrightarrow \quad u = 2 \omega \Delta \mathcal{S}_0(\omega)
\ee 
This result is obtained as follows. Let us rewrite the equation \eqref{eq:EvolSU} 
\begin{equation} \label{ew} 
\omega \mathcal{S} \partial_\omega \mathcal{S} + \partial_t \mathcal{S}  + \mathcal{S}^2 + \tilde \sigma^2 \omega \mathcal{S} + m \mathcal{S} = 0
\end{equation}
We now perform the hodograph transformation, that is we take $\mathcal{S}$ and $\omega$ as variables and consider $t$ as
a function $t(\omega,\mathcal{S})$. The rules for the partial derivatives are such that $\partial_\mathcal{S} t = \frac{1}{\partial_t \mathcal{S}}$
and $\partial_\omega t = - \frac{\partial_\omega \mathcal{S}}{\partial_t \mathcal{S}}$.
The equation \eqref{ew} leads to
\be \label{lin} 
\mathcal{S} \omega \partial_\omega t - \mathcal{S} (\mathcal{S} + \tilde \sigma^2 \omega + m) \partial_\mathcal{S} t = 1
\ee
which is a first-order linear differential equation for the function $t(\omega,\mathcal{S})$.
Let us denote $u(\omega,\mathcal{S})$ a solution of the homogeneous equation
\be \label{aga} 
u(\omega,\mathcal{S}) = \tilde \sigma^2 \omega^2 + 2 \omega (\mathcal{S}+ m) 
\ee
A general method in first-order partial differential equations, see \cite{PDEBook2}, consists in choosing now $\omega$ and $u$ as independent variables,
i.e. we look for a solution of \eqref{lin} in the form
\be
t(\omega,\mathcal{S}) = \tilde t ( \omega, u(\omega,\mathcal{S}) ) 
\ee
Using that $u$ is a solution of \eqref{lin} setting the r.h.s. to zero, we obtain the following equation for the function
$\tilde t(\omega,u)$
\be
\mathcal{S}(\omega,u) \omega \partial_\omega \tilde t = 1   \quad  \Rightarrow \quad \partial_\omega \tilde t = \frac{2}{u - \tilde \sigma^2 \omega^2 - 2 m \omega}
\ee
where $\mathcal{S}(\omega,u)= \frac{1}{2 \omega} (u - \tilde \sigma^2 \omega^2) - m$ is obtained from inverting $u(\omega,\mathcal{S})$. The general solution of this equation is
\bea
\tilde t(\omega,u) = \frac{2}{\tilde \sigma \sqrt{ u + (\frac{m}{\tilde \sigma})^2}} {\rm arctanh} \left( \frac{\tilde \sigma \omega + \frac{m}{\tilde \sigma}}{\sqrt{u + (\frac{m}{\tilde \sigma})^2}} \right) + \phi(u) 
\eea
Replacing now $u = u(\omega,\mathcal{S})=\tilde \sigma^2 \omega^2 + 2 \omega (\mathcal{S}+ m)  $ we finally obtain the result
announced above in \eqref{soluexacte}. The function $\phi$ is determined from the initial condition $\Delta \mathcal{S}_0(\omega) = \Delta \mathcal{S}(\omega,t=0)$
\be
\phi(u) = - \frac{2}{\tilde \sigma \sqrt{ u + (\frac{m}{\tilde \sigma})^2}} {\rm arctanh} \left( \frac{\tilde \sigma \Omega(u) + \frac{m}{\tilde \sigma}}{\sqrt{ u + (\frac{m}{\tilde \sigma})^2}} \right)  
\ee
where the function $\Omega(u)$ is obtained by inverting \eqref{Omega}.
Putting all together we find
\bea
&& t =  
\frac{2}{\tilde \sigma \sqrt{ 2 \omega \Delta \mathcal{S}(\omega,t) + (\frac{m}{\tilde \sigma})^2}}
{\rm arctanh} \left( 
\frac{\tilde \sigma (\omega - \Omega(2 \omega \Delta \mathcal{S}(\omega,t))) }{\sqrt{ 2 \omega \Delta \mathcal{S}(\omega,t) + (\frac{m}{\tilde \sigma})^2}
- (\tilde \sigma \omega + \frac{m}{\tilde \sigma}) \frac{\tilde \sigma \Omega(2 \omega \Delta \mathcal{S}(\omega,t)) + \frac{m}{\tilde \sigma}}{
\sqrt{ 2 \omega \Delta \mathcal{S}(\omega,t) + (\frac{m}{\tilde \sigma})^2}}} \right)
\eea
which is finally equivalent to the result given in \eqref{soluexacte}.

\tocless\subsection{Stieltjes transform} 

The same method can be applied to the Stieltjes transform. Starting from Equation \eqref{eq:EvolutionStieltjes} for $\mathfrak{g}(z,t)$,
we write a partial differential equation on the function $t(z,\mathfrak{g})$ using $\partial_\mathfrak{g} t=\frac{1}{\partial_t \mathfrak{g}}$ and $\partial_z t = - \frac{\partial_z \mathfrak{g}}{\partial_t \mathfrak{g}}$. One
obtains the linear equation
\begin{equation} \label{lin2} 
\left(1+(m-\tilde \sigma^2) z + \tilde \sigma^2 z^2 \mathfrak{g} \right) \partial_z t -  \left( (m-\tilde \sigma^2)\mathfrak{g} + \tilde \sigma^2 z \mathfrak{g}^2 \right) \partial_\mathfrak{g} t = 1
\end{equation}
One homogeneous solution is found as :
\begin{equation} \label{udef} 
u(z,\mathfrak{g}) = \mathfrak{g} + (m-\tilde \sigma^2) z \mathfrak{g} + \frac{\tilde \sigma^2}{2} z^2 \mathfrak{g}^2
\end{equation}
Inverting $u(z,g)$ gives 
\begin{equation}
\mathfrak{g}(z,u) = \frac{ - 1 -  (m- \tilde \sigma^2 )z  - z  \sqrt{ (1/z + (m- \tilde \sigma^2 ))^2 + 2 \tilde \sigma^2  u }  }{\tilde \sigma^2 z^2}
\end{equation}
where the branch is chosen in order to satisfy $\mathfrak{g}(z)\sim \frac{1}{z}$ for $z \to \infty$, with $m<0$. 
Changing variables as $t(z,g) \to t(z,u)$ in \eqref{lin2} we obtain
\begin{equation}
\left(1+(m-\tilde \sigma^2) z + \tilde \sigma^2 z^2 \mathfrak{g}(z,u) \right) \partial_z t = 1
\end{equation}
thus :
\begin{equation}
\partial_z t = \frac{- 1 }{ z\sqrt{ (1/z +  (m- \tilde \sigma^2 ) )^2 + 2 \tilde \sigma^2  u } } 
\end{equation}
This integrates as :
\be
t(z,u) =\frac{- 1}{\sqrt{2  \tilde \sigma^2 u + ( m - \tilde \sigma^2 )^2}} 
{\rm arcsinh} \left(
\frac{
(2\tilde\sigma^2 u+(m-\tilde\sigma^2)^2)z +m-\tilde\sigma^2}{\tilde\sigma \sqrt{2 u} }
\right)
+ \phi(u)
\ee
where $\phi$ is an arbitrary function of $u$ that can be determined from the initial condition at time $t=0$. 

Inserting the definition of $u=u(z,\mathfrak{g})$ from \eqref{udef} we obtain the 
exact solution for the Stieltjes transform $\mathfrak{g} = \mathfrak{g}(z,t)$ in the implicit form by solving
\be
t =  \frac{- 1}{\sqrt{  2 \tilde \sigma^2 \mathfrak{g} + (\tilde \sigma^2 z \mathfrak{g} + (m-\tilde \sigma^2) ) ^2}}  {\rm arcsinh} \left(
\frac{
( 2 \tilde \sigma^2 \mathfrak{g} + (\tilde \sigma^2 z \mathfrak{g} + (m-\tilde \sigma^2) ) ^2)z +m-\tilde\sigma^2}{ \sqrt{
2 \tilde \sigma^2  (1+  (m-\tilde \sigma^2)  z )\mathfrak{g} + \tilde \sigma^4 z^2 \mathfrak{g}^2 
} }
\right)
+\phi(u(z,\mathfrak{g}))
\ee
It is possible to rewrite it using the initial condition $\mathfrak{g}(z,t=0)=\mathfrak{g}_0(z)$. Let us define the function $Z(u)$ such that 
\be
\label{eq:DefinitionZofu}
u(z,\mathfrak{g}_0(z)) = \mathfrak{g}_0(z) + (m-\tilde \sigma^2) z \mathfrak{g}_0(z) + \frac{\tilde \sigma^2}{2} z^2 \mathfrak{g}_0(z)^2 = u \quad \Leftrightarrow \quad z=Z(u) 
\ee
Then the solution is
\begin{equation}
\label{eq:ResSolutionG}
\mathfrak{g}(z,t) = 
\frac{ - 1 -  (m- \tilde \sigma^2 )z  - z \sqrt{ (1/z + (m- \tilde \sigma^2 ))^2 
+ 2 \tilde \sigma^2 U(z,t) }  }{\tilde \sigma^2 z^2}
\end{equation}
where $U(z,t)$ is obtained by inverting
\bea \label{eq:ResInverting}
&& U(z,t)= u \quad \Leftrightarrow \quad \\ 
&& \nonumber z= \frac{1}{A(u)} \bigg( 
-\tilde \sigma  \sqrt{2u} \sqrt{1+ \frac{\left( A(u) Z(u) +m-\tilde \sigma ^2\right)^2}{2 \tilde \sigma ^2 u}}
   \sinh \left(t \sqrt{A(u)}\right)
    +\left( A(u) Z(u) +m-\tilde \sigma ^2\right)
   \cosh \left(t \sqrt{A(u)}\right)-m+\tilde \sigma ^2 \bigg)
\eea
where we denote $A(u)= \left(m-\tilde \sigma ^2\right)^2+2\tilde \sigma ^2 u $ and we recall that $Z(u)$ is defined in \eqref{eq:DefinitionZofu} from the initial condition $\mathfrak{g}_0$. One can check that the second equation in \eqref{eq:ResInverting} gives $z=Z(u)$ for $t=0$, which is equivalent to $U(z,0)=u(z,\mathfrak{g}_0(z))$ which, plugged in \eqref{eq:ResSolutionG}, gives back $\mathfrak{g}(z,t)=\mathfrak{g}_0(z)$. In the large time limit, the r.h.s. in the second equation diverges with corresponds to $z$ going to $\infty$, hence we recover the stationary value $U(z,t\to \infty)=m-\frac{\tilde\sigma^2}{2}$ which in turn yields the stationary Stieltjes transform \eqref{eq:StationaryStieltjesTransform}.
   
This solution reproduces the one obtained by Ossipov in Eq. (18) of \cite{Ossipov} in the particular case
$m=0$, $\tilde \sigma=1/\sqrt{2}$ and $r=t/2$ plays the role of the time variable, with
his $z_0=z_0(z,t)$ related to our $u=U(z,t)$ by $1/z_0^2 = u + \frac{1}{4}$.
Note that in \cite{Ossipov} the solution is only given for the initial condition $\mathfrak{g}_0(z)=1/z$ while we have given here the solution for a general initial condition. In the limit $t\to\infty$ (equivalently $r\gg 1$), a scaling form was found in \cite{Ossipov} in the case $m=0$. Note however that a well-defined stationary solution does not exist then, since $m=0$ is the critical case.\\

\tocless\subsection{Time evolution equivalence of \texorpdfstring{$\mathcal{S}$}{S} and Stieltjes transforms}
\label{app:EquivSStieltjes}

Let us show that the evolution equations obtained for the $\mathcal{S}$-transform \eqref{eq:EvolSU} and for the Stieltjes transform \eqref{eq:EvolutionStieltjes} are indeed equivalent. For ease of notation, let us drop the dependence on $t$ for all functions and denote $f = \mathcal{T}^{-1}$ the functional inverse of the $\mathcal{T}$-transform of $U_t$ such that $\mathcal{S}(\omega)=  \frac{\omega +1 }{\omega f(\omega) }$. Then, we obtain from \eqref{eq:EvolSU}:
\begin{equation}
 - \frac{\omega + 1}{\omega f^2} \frac{\partial f}{\partial t} =\frac{\partial}{\partial t} \mathcal{S} = - \frac{\omega +1 }{\omega f(\omega) }\left( m + \tilde{\sigma}^2 \omega + \frac{\omega +1 }{\omega f(\omega) } + \omega \frac{\omega f(\omega) - (\omega +1 ) ( f(\omega) + \omega f'(\omega)) }{ \omega^2 f^2(\omega) } \right)
\end{equation}
Simplifying, this leads to:
\begin{equation}
\frac{\partial}{\partial t} f = 1 +  f(\omega)   \left( m +  \tilde{\sigma}^2 \omega \right) - \frac{  (\omega +1 )   f'(\omega) }{  f(\omega) } 
\end{equation}
Let us deduce the evolution of $\mathcal{T} = f^{-1}$, between two consecutive instants $t_2 = t_1 + {\rm d}t$. We know that :
\begin{equation}
\mathcal{T}_{t_2}^{-1}( \omega) - \mathcal{T}_{t_1}^{-1}(\omega) = {\rm d}t \left(
1 +  f(\omega)   \left( m +  \tilde{\sigma}^2 \omega \right) - \frac{  (\omega +1 )   f'(\omega) }{  f(\omega) } 
\right)
\end{equation}
Evaluating at $\omega = \mathcal{T}_{t_2}(z)= \mathcal{T}_{t_1}(z) + {\rm d}t \; \frac{\partial  \mathcal{T} }{\partial t} $:
\begin{equation}
z - \left( z + {\rm d}t \; \frac{\partial  \mathcal{T} }{\partial t} \times \left(\mathcal{T}^{-1}\right)' \left(\mathcal{T}(z) \right) \right) = {\rm d}t \left(
1 +  z  \left( m +  \tilde{\sigma}^2 \mathcal{T}(z) \right) - \frac{  (\mathcal{T}(z) +1 )   (\mathcal{T}^{-1})'(\mathcal{T}(z)) }{  z} 
\right)
\end{equation}
Using $ (\mathcal{T}^{-1})'(\mathcal{T}(z)) = \frac{1}{\mathcal{T}'(z)} $, we obtain:
\begin{equation}
\frac{\partial}{\partial t}\mathcal{T}= \frac{\mathcal{T}(z) +1 }{z} - \mathcal{T}'(z) \left( 1+z (m+\tilde{\sigma}^2 \mathcal{T}(z) ) \right)
\end{equation}
From the definition of $\mathcal{T}$ as $\mathcal{T}(z) = z \mathfrak{g}(z) -1$, we finally obtain the evolution of $\mathfrak{g}$ as in Eq. \eqref{eq:EvolutionStieltjes}:
\begin{equation}
\frac{\partial}{\partial t} \mathfrak{g} =\frac{\partial}{\partial z}\left[-\mathfrak{g}+\left(\tilde{\sigma}^{2}-m\right) z \mathfrak{g}-\frac{1}{2} \tilde{\sigma}^{2} z^{2} \mathfrak{g}^{2}\right]
\end{equation}\\

\section{Moments}
\label{app:Moments}

In this Appendix, we compute the evolution of the first moments in the discrete and continuous versions of the scalar and matrix Kesten models. For clarity, we recall the relatively standard results for the scalar case and in the matrix case we show that there are considerations of invariant theory, which are interesting as they constrain the stationary measure.

\tocless\subsection{Discrete scalar case}

The scalar Kesten recursion, with the regularization parameter $\epsilon$, is $Z_{n} = \xi_{n}\left(\epsilon+Z_{n-1}\right)$ with the initial condition $Z_0=0$. Denoting $\varphi_{k,n} = \mathbb{E}[Z_n^k]$ the $k$th moment of the scalar random variable $Z_n$ and $\varphi_{k,\xi}$ the $k$th moment of $\xi$, we have the following recursion for the first moment:
\be 
\label{eq:FirstMomentDiscreteScalar}
\varphi_{1,n} = \varphi_{1,\xi} (\epsilon +\varphi_{1,n-1}) \quad \Rightarrow  \quad  \varphi_{1,n}  = \epsilon (  \varphi_{1,\xi} + \varphi_{1,\xi}^2 + \cdots + \varphi_{1,\xi}^n  )= \epsilon \varphi_{1,\xi} \frac{\varphi_{1,\xi}^n-1}{\varphi_{1,\xi}-1} \ \xrightarrow[n\to \infty]{} \  \frac{\epsilon \varphi_{1,\xi}}{ 1- \varphi_{1,\xi} }
\ee 
If we choose, as in Eq. \eqref{choicexi} of the main text, $\xi = 1+m \epsilon + \sqrt{2}\sigma \sqrt{\epsilon}X$ with $X \sim \mathcal{N}(0,1)$, we obtain 
\be 
\label{eq:ResultDiscrete}
\varphi_{1,n=+\infty}= - \frac{m \epsilon+1}{m} \ee
The limit $n=+\infty$ is finite only for $\varphi_{1,\xi}<1$, i.e. $m<0$. The recursion for the second moment is $\varphi_{2,n} = \varphi_{2,\xi} (\epsilon^2 + 2 \epsilon \varphi_{1,n-1} + \varphi_{2,n-1} )$ and we obtain the second moment for every step $n$ as:
\be
\label{eq:SecondMomentDiscreteScalar}
\varphi_{2,n} =
\epsilon ^2 \varphi_{2,\xi} \frac{  \left(\varphi_{1,\xi}-1\right)
    \left(\varphi_{1,\xi}+\varphi_{2,\xi}\right)\varphi_{2,\xi}^n-\varphi_{1,\xi} \left(2 \left(\varphi_{2,\xi}-1\right)
   \varphi_{1,\xi}^n+\varphi_{1,\xi}-\varphi_{2,\xi}+1\right)+\varphi_{2,\xi} }{\left(1- \varphi_{1,\xi}\right)
   \left(1- \varphi_{2,\xi}\right) \left(\varphi_{2,\xi}-\varphi_{1,\xi}\right)} 
   \xrightarrow[n\to \infty]{}   
\frac{\epsilon^2 \varphi_{2,\xi}(1+\varphi_{1,\xi})}{(1-\varphi_{1,\xi}) ( 1- \varphi_{2,\xi}) }  
\ee
Injecting $\xi = 1+m \epsilon + \sqrt{2}\sigma \sqrt{\epsilon}X$, we obtain
\be
\label{eq:ResultDiscreteSecondMoment}
\varphi_{2,n=+\infty} =
\frac{(2+m\epsilon) (1+\epsilon (2\sigma^2 +m (2+ m \epsilon))) }{m(2\sigma^2+m(2+m \epsilon) ) }
\ee
The limit $n=+\infty$ is finite only for $\varphi_{2,\xi}<1$, i.e. $2\sigma^2+m(2+m \epsilon)<0$.
In this work we are interested in random variables $\xi_n$ which are restricted to be positive. The Eqs \eqref{eq:FirstMomentDiscreteScalar} and \eqref{eq:SecondMomentDiscreteScalar} are still valid. The Gaussian choice for $\xi_n$ violates this positivity condition, but only with 
exponentially small probability (as $\exp(-1/\epsilon)$) for small $\epsilon$. Hence we expect that the explicit formula
\eqref{eq:ResultDiscrete} and \eqref{eq:ResultDiscreteSecondMoment} are useful in that case within an expansion in powers of $\epsilon$ at small $\epsilon$. The same applies below in the matrix case.

\tocless\subsection{Continuous scalar case}

In the continuous limit, the scalar Kesten evolution is the It\^o stochastic equation \eqref{eq:StochasticEvolution1D}: $ {\rm d}U_t = \left( 1+ m \; U_t \right) {\rm d}t + \sqrt{2}\sigma \; U_t {\rm d}W_t $, and we choose here for simplicity $U_0=0$. The evolution of the moments $\varphi_{k,t}=\mathbb{E}[U_t^k]$ can be computed recursively from this stochastic evolution. The first moment is found as:
\be 
\partial_t \varphi_{1,t} = (1 + m \varphi_{1,t} )  \quad \Rightarrow \quad \varphi_{1,t} = - \frac{1-e^{m t}}{m} \ \xrightarrow[t \to \infty]{}  \ - \frac{1}{m}
\ee 
where the large time coincides as expected with the $\epsilon \to 0$ limit of the discrete result \eqref{eq:ResultDiscrete}. The evolution of the squared process $U_t^2$ is obtained from the It\^o rules as $ {\rm d}U_t^2 = 2U_t\left( 1+ (m+\sigma^2) \; U_t \right) {\rm d}t + 2\sqrt{2}\sigma \; U_t^2 {\rm d}W_t$. Taking the expectation yields the evolution equation of $\varphi_{2,t}$:
\be 
\label{eq:Phi2ContinuousScalar}
\partial_t \varphi_{2,t} = 2(\varphi_{1,t}+(m+\sigma^2)\varphi_{2,t})
\quad \Rightarrow \quad 
\varphi_{2,t} = 
\frac{m \left(e^{2 t \left(m+\sigma ^2\right)}-2 e^{m t}+1\right)+2 \sigma ^2 \left(1-e^{m
   t}\right)}{m \left(m+\sigma ^2\right) \left(m+2 \sigma ^2\right)}
   \ \xrightarrow[t \to \infty]{}  \ \frac{1}{m(m+\sigma^2)}
\ee 
where the large time limit coincides as expected with the $\epsilon \to 0$ limit of the discrete result \eqref{eq:ResultDiscreteSecondMoment}.
Note that the stationary variable $U_\infty$ exists for $m < \sigma^2$, but that the first and second moments are finite at $t=+\infty$ only for $m<0$
and $m< - \sigma^2$ respectively. The above time dependent solutions however are valid in all cases, and show how the moments
grow with time in the cases where their limits do not exist.

\tocless\subsection{Discrete matrix case}

The $N\times N$ matrix Kesten recursion $Z_{n}= \sqrt{\epsilon I + Z_{n-1}} \xi_n\sqrt{\epsilon I + Z_{n-1}}$ makes it possible to obtain the moments $\varphi_{k,n}^{(N)}=\frac{1}{N}\operatorname{Tr} \mathbb{E}[Z_n^k]$ from the distribution of $\xi$ and the initial condition that we fix to $Z_0=0$. We denote $\varphi_{k,\xi}^{(N)}=\frac{1}{N}\operatorname{Tr} \mathbb{E}[\xi^k]$ and assume that the distribution of $\xi$ is invariant under rotations, i.e. conjugation by an element of the orthogonal (resp. unitary) group for $\beta=1$ (resp. 2). This assumption imposes that $\mathbb{E}[\xi] = \varphi_{1,\xi} I $ and we thus have a similar recursion for the first moment as in the scalar case:
\be
\label{eq:FirstMomentGeneralN}
\mathbb{E}[Z_n]=(\epsilon I + \mathbb{E}[Z_{n-1})] \varphi_{1,\xi} \quad  \Rightarrow  \quad 
\mathbb{E}[Z_n] =  \varphi_{1,n}^{(N)} I \quad \text{with} \quad
\varphi_{1,n}^{(N)} =\epsilon \varphi_{1,\xi}\frac{\varphi_{1,\xi}^n-1}{\varphi_{1,\xi}-1} \ \xrightarrow[n \to \infty]{}  \  \frac{\epsilon \varphi_{1,\xi}}{1-\varphi_{1,\xi}}
\ee
With $\xi$ defined as in Eq. \eqref{eq:structuresigma} we obtain 
$\varphi_{1,\xi}^{(N)}=1+m\epsilon$ such that $\varphi_{1,n=+\infty}^{(N)}$ coincides with the large-$N$ result \eqref{mom1}. Let us turn to the second moment. The isotropy condition on the $\xi$ distribution imposes that there exist four coefficients $a,b,c,d$ such that for any symmetric (resp. Hermitian if $\beta=2$) matrix $M$ \cite{Dolgachev03}:
\be 
\label{eq:InvariantTheoryConstraint}
\mathbb{E}[\xi M \xi] = a M + \frac{ b }{N} \operatorname{Tr}(M) I \quad , \quad \mathbb{E} [(\operatorname{Tr}M \xi )^2 ] =  N c \operatorname{Tr}(M^2) + d (\operatorname{Tr}M )^2 
\ee 
Note that $\varphi_{2,\xi} = a + b$ and $\mathbb{E} [( \frac{1}{N} \operatorname{Tr} \xi)^2 ]= c  + d $. These two sets
of parameters are not independent however, indeed
choosing the matrix $M$ to be $M_{ij} = \delta_{i1} \delta_{j1}$, one obtains the relation
\be
\label{eq:FinalRelationObtainedCongrats}
\mathbb{E} \xi_{11}^2 = a + \frac{b}{N} = N c + d
\ee

We can then write the expectation of the matrix $Z_n^2$ as:
\bea
\mathbb{E}[Z_n^2] &&= \mathbb{E}[ \sqrt{\epsilon I + Z_{n-1}} \;  \xi_n \; (\epsilon I + Z_{n-1}) \; \xi_n \; \sqrt{\epsilon I + Z_{n-1}} ]  \\ 
 &&= a \mathbb{E}\left[(\epsilon I + Z_{n-1})^2\right]+
\frac{ b }{N}  \mathbb{E}\left[ (\epsilon I + Z_{n-1}) \operatorname{Tr}(\epsilon I + Z_{n-1})\right]
\eea
Taking the trace and dividing by $N$ yields:
\be 
\label{eq:EqPhi2arbitraryN}
\varphi_{2,n}^{(N)}
= a  (\epsilon^2 + 2 \epsilon \varphi_{1,n-1}^{(N)} + \varphi_{2,n-1}^{(N)})  + b \left(  \epsilon^2 + 2  \epsilon \varphi_{1,n-1}^{(N)} + \frac{1}{N^2}\mathbb{E}\left[(\operatorname{Tr}Z_{n-1})^2\right]  \right)
\ee 
The correlations of $\xi$ in Eq. \eqref{eq:InvariantTheoryConstraint} give furthermore: 
\bea
\label{eq:EqETRZ2Discrete}
&& \mathbb{E}\left[ (\operatorname{Tr}Z_{n})^2 \right]= \mathbb{E}\left[(\operatorname{Tr}(\xi(\epsilon I + Z_{n-1}))^2\right] =
N  c \; \mathbb{E}\left[ \operatorname{Tr} (  (\epsilon I + Z_{n-1})^2) \right] + 
d \; \mathbb{E}\left[ ( \operatorname{Tr} (  (\epsilon I + Z_{n-1}) )^2 \right]  \\
&& =
N^2 c \; ( \epsilon^2 + 2 \epsilon \varphi_{1,n-1}^{(N)} + \varphi_{2,n-1}^{(N)} )+
N^2 d  \; ( \epsilon^2 + 2  \epsilon  \varphi_{1,n-1}^{(N)} + \frac{1}{N^2} \mathbb{E}\left[ (\operatorname{Tr}Z_{n-1})^2 \right] ) 
\eea 
Equations \eqref{eq:EqPhi2arbitraryN} and \eqref{eq:EqETRZ2Discrete} form a coupled recursion system which is fully solvable. For simplicity, we simply give the stationary values $\varphi_{2,n=+\infty}^{(N)}$ and $ \mathbb{E}\left[ (\operatorname{Tr}Z_{\infty})^2 \right]$:  
\be 
\label{eq:Phi2InftyDiscrete}
\varphi_{2,n=+\infty}^{(N)} =
\epsilon ^2   \frac{ (\varphi_{1,\xi}+1) (a+ b + b c-ad  ) }{(\varphi_{1,\xi}-1) (a+d+  bc-ad-1)}
\ee
\be 
\label{eq:ETRZ2InftyDiscrete}
 \mathbb{E}\left[ (\operatorname{Tr}Z_{\infty})^2 \right]  = 
N^2 \epsilon^2 \frac{ (\varphi_{1,\xi}+1) ( c+ d+  b c-a d)}{(\varphi_{1,\xi}-1) (a+d+ bc-ad-1)}
\ee 
We check that fixing $N=1$, $a= d=\varphi_{2,\xi}$ and $b=c=0$ recovers the scalar result \eqref{eq:SecondMomentDiscreteScalar}. In the large-$N$ limit and for an arbitrary matrix $M$, the random variable $\frac{1}{N}\operatorname{Tr}(M \xi)$ is self-averaging. The vanishing of its variance yields $d=\varphi_{1,\xi}^2$ and $c=o(\frac{1}{N})$. Recalling that $\varphi_{2,\xi}=a+b$ such that $b$ must remain $\mathcal{O}(1)$, relation \eqref{eq:FinalRelationObtainedCongrats} then yields that $a=\varphi_{1,\xi}^2$ in this limit. With these substitutions, we verify that $\varphi_{2,n=+\infty}^{(N)}$ of Eq. \eqref{eq:Phi2InftyDiscrete} matches the general formula for the second moment obtained for large $N$ from free probability tools in Eq. \eqref{mom2}. 

With $\xi$ defined as in Eq. \eqref{eq:structuresigma} we obtain $a=(1+m\epsilon)^2 +\sigma^2 \epsilon \delta_{\beta=1}, b= N \sigma^2 \epsilon ,   c = \sigma^2 \epsilon (1 +\delta_{\beta=1})/N , d = (1 + m \epsilon)^2 $ such that:
\be 
\label{eq:Phi2InftyDiscreteXiEpsilonStructure}
\varphi_{2,n=+\infty}^{(N)} =
\epsilon^2 + \frac{2 \epsilon}{m}
+\frac{(2+m \epsilon) (m ( 2+m \epsilon) -N \sigma^2)}{m \left(m (m
   \epsilon +2)-\sigma ^2\right) \left(\sigma ^2 (1+ \delta_{\beta=1}) +m (m \epsilon +2)\right)}
\ee
\be
\label{eq:ETRZ2InftyDiscreteXiEpsilonStructure}
 \mathbb{E}\left[ (\operatorname{Tr}Z_{\infty})^2 \right]  = \frac{N (m \epsilon +2) \left(- \sigma ^2 (1+\delta_{\beta=1}) +N \left(-\sigma ^4 (1+ \delta_{\beta=1}) 
   \epsilon +  (m \sigma  \epsilon +\sigma )^2 \delta_{\beta=1} +m (m \epsilon +1)^2 (m \epsilon
   +2)\right)\right)}{m \left(m (m \epsilon +2)-\sigma ^2\right) \left(\sigma ^2(1+\delta_{\beta=1} )
   +m (m \epsilon +2)\right)}
\ee 
We check that the large $N$ limit of $\varphi_{2,n=+\infty}^{(N)}$, while fixing $\sigma=\tilde\sigma/\sqrt{N}$, coincides for $\beta=1,2$ with the r.h.s. of \eqref{mom2}. The corrections are $\mathcal{O}(1/N^\beta)$ for $\beta=1,2$.
 We also find in this limit that
\be
\label{eq:DiffMoments2Discrete}
\mathbb{E}\left[ (\operatorname{Tr}Z_{\infty})^2 \right] -  (\operatorname{Tr} \mathbb{E}\left[ Z_{\infty} \right])^2  = \frac{(1+\delta_{\beta=1})
\tilde \sigma^2 (\tilde \sigma^2 - m (2 + \epsilon m))}{m^4 (2 + \epsilon m)^2} + \mathcal{O}(\frac{1}{N}) 
\ee
where each term of the l.h.s. are $\mathcal{O}(N^2)$ but the difference is $\mathcal{O}(1)$.

\tocless\subsection{Continuous matrix case}

Let us consider the It\^o matrix stochastic differential equation \eqref{eq:EvolutionU}: $ {\rm d}U_t = \left( I +m U_t\right) {\rm d}t +\sigma \sqrt{U_t} d B_{t} \sqrt{U_t}$, with $U_0=0$, and study the evolution of moments $\varphi_{k,t}^{(N)}= \frac{1}{N}\operatorname{Tr}\mathbb{E}[U_t^k]$. We have the following for the process $\mathbb{E}[U_t]$:
\be 
\partial_t \mathbb{E}[U_t] = (I + m \mathbb{E}[U_t] )  \quad \Rightarrow  \quad
\mathbb{E}[U_t] = - \frac{1-e^{m t}}{m}  I \quad  \Rightarrow  \quad
\varphi_{1,t}^{(N)} = - \frac{1-e^{m t}}{m}  \to_{t \to \infty} - \frac{1}{m}
\ee 
where $\varphi_{1,t=+\infty}^{(N)}$ coincides with the $\epsilon \to 0$ limit of the discrete result \eqref{eq:FirstMomentGeneralN}.
For the second moment, the evolution of the squared matrix is, according to It\^o rules:
\be
{\rm d}U_t^2 = 2 U_t (I + m U_t) {\rm d}t + \sigma^2 \sqrt{U_t} (\operatorname{Tr}(U_t) I + U_t \delta_{\beta=1}) \sqrt{U_t} +  \sigma ( U_t \sqrt{U_t} {\rm d}B_t \sqrt{U_t} + \sqrt{U_t} {\rm d}B_t \sqrt{U_t} U_t)
\ee
where we used that $\mathbb{E} [{\rm d}B_{t} U_t {\rm d}B_t ]= \mathbb{E} [ \operatorname{Tr}(U_t) I + U_t \delta_{\beta=1}]$, as seen from the following correlations:
\be
\label{eq:correlationsBt}
\mathbb{E}[ ({\rm d}B_t)_{i,j} ({\rm d}B_t)_{k,l}^*] = {\rm d}t  \; C_{i,j,k,l} \ ,
\quad \quad  \quad 
C_{i,j,k,l} = 
\left\{
    \begin{array}{ll}
     \delta_{i,k}\delta_{j,l} + \delta_{i,l}\delta_{j,k}     \quad \quad  &\beta =1 \\
     \delta_{i,k}\delta_{j,l}  &\beta =2
    \end{array}
\right.
\ee  
Taking the expectation value one finds
\be 
\label{eq:EvolutionPhi2NContinuous}
\partial_t \varphi_{2,t}^{(N)}  =2 \varphi_{1,t}^{(N)}  + (2 m+\sigma^2 \delta_{\beta=1}) \varphi_{2,t}^{(N)}  + \frac{\sigma^2}{N} \mathbb{E}[(\operatorname{Tr}U_t)^2]
\ee 
The It\^o rules and the correlations \eqref{eq:correlationsBt} also give the evolution of $\mathbb{E}[(\operatorname{Tr}U_t)^2]$ as:
\be 
d\mathbb{E}[(\operatorname{Tr}U_t)^2]
= \mathbb{E} \left[  2 \operatorname{Tr}(U_t)(N+m \operatorname{Tr}U_t){\rm d}t + \sigma^2 (1+\delta_{\beta=1}) \operatorname{Tr}(U_t^2) {\rm d}t  \right]
\ee 
such that
\be 
\label{eq:EvolutionETRU2Continuous}
\partial_t  \mathbb{E}[(\operatorname{Tr}U_t)^2]
= 2N^2 \varphi_{1,t}^{(N)} + \sigma^2 N (1+\delta_{\beta=1}) \varphi_{2,t}^{(N)} + 2m \mathbb{E}[(\operatorname{Tr}U_t)^2]
\ee 
Equations \eqref{eq:EvolutionPhi2NContinuous} and \eqref{eq:EvolutionETRU2Continuous} form a closed first-order differential system. For the sake of simplicity, we only give the stationary solutions $\varphi_{2,t=+\infty}^{(N)}$ and $\mathbb{E}[(\operatorname{Tr}U_\infty)^2]$: 
\be 
\varphi_{2,t=+\infty}^{(N)} = \frac{2 \left(2 m-N \sigma ^2\right)}{m \left(2 m-\sigma ^2\right) \left( 2 m+\sigma ^2 (1+\delta_{\beta=1} ) \right)}
\ee  
\be 
\mathbb{E}[(\operatorname{Tr}U_\infty)^2] = \frac{2 N \left(2 m N+\delta_{\beta=1}\sigma ^2  (N-1) - \sigma ^2\right)}{m
   \left(2 m-\sigma ^2\right) \left( 2 m+\sigma ^2 (1+\delta_{\beta=1} ) \right)}
\ee  
We see that $\varphi_{2,t=+\infty}^{(N)}$ and $\mathbb{E}[(\operatorname{Tr}U_\infty]^2)$ coincide as expected with the $\epsilon \to 0$ limit of the discrete results \eqref{eq:Phi2InftyDiscrete} and \eqref{eq:ETRZ2InftyDiscrete}. In the limit of large $N$ with $\sigma =\tilde \sigma/\sqrt{N}$, we find that
\be
\mathbb{E}\left[ (\operatorname{Tr}U_{\infty})^2 \right] -  (\operatorname{Tr} \mathbb{E}\left[ U_{\infty} \right])^2  = \frac{(1+\delta_{\beta=1})
\tilde \sigma^2 (\tilde \sigma^2 - 2m )}{4m^4} + \mathcal{O}(\frac{1}{N}) 
\ee
which coincides with the limit $\epsilon\to 0$ of the discrete result \eqref{eq:DiffMoments2Discrete}.
We recover the scalar continuous result \eqref{eq:Phi2ContinuousScalar} by injecting $N=1$ and $\beta=1$ (or equivalently by also rescaling $\sigma\to \sqrt{2}\sigma$ in the complex case $\beta=2$). Finally, we stress that $\varphi_{1,t=+\infty}^{(N)}$ and $\varphi_{2,t=+\infty}^{(N)}$ can be verified to agree with the first and second moments of the matrix inverse-Wishart distribution obtained as the stationary distribution of the continuous model in Eq. \eqref{eq:StationaryMatrixDist}.

\end{appendix}


\begin{thebibliography}{9}

\bibitem{BouchaudGeorges} J. P. Bouchaud, A. Georges, {\it Anomalous diffusion in disordered media: statistical mechanisms, models and physical applications}. Physics reports {\bf195}(4-5), 127-293 (1989).

\bibitem{DerridaSpohn} B. Derrida, H. Spohn, {\it Polymers on disordered trees, spin glasses, and traveling waves}. Journal of Statistical Physics {\bf51}(5-6), 817-840 (1988).

\bibitem{SouillardPRL1984} F. Delyon, B. Simon, B. Souillard, {\it From Power-Localized
to Extended States in a Class of One-Dimensional Disordered
Systems}. Phys. Rev. Lett. {\bf 52}, 2187 (1984).

\bibitem{Progodin1980} V. N Prigodin, {\it One-dimensional disordered system in an electric
field}. Zh. Eksp. Teor. Fiz. {\bf 79}, 2338 (1980) [Sov. Phys. JETP {\bf 52},
1185 (1980)].

\bibitem{GeorgesAspect2018} C. Crosnier de Bellaistre, C. Trefzger, A. Aspect, A. Georges, L. Sanchez-Palencia,
   {\it Expansion of a quantum wave packet in a one-dimensional disordered potential in the presence
of a uniform bias force}. Phys. Rev. A {\bf 97}, 013613 (2018).

\bibitem{Dorogovtsev} S. N. Dorogovtsev, \& J. F. Mendes, {\it Evolution of networks: From biological nets to the Internet and WWW}. OUP Oxford (2013). 

\bibitem{ABBM} B. Alessandro, C. Beatrice, G. Bertotti, A. Montorsi, {\it Domain-wall dynamics and Barkhausen effect in metallic
ferromagnetic materials I. Theory}. J. Appl. Phys. {\bf 68}, 2901-2907 (1990).

\bibitem{Bak} P. Bak, C.Tang, K. Wiesenfeld, {\it Self-organized criticality: an explanation of 1/f noise}. Phys. Rev. Lett. {\bf 59}(4), 381 (1987).

\bibitem{Bouchaud2001} J. P. Bouchaud, {\it Power laws in economics and finance: some ideas from physics}. Quantitative Finance {\bf 1}, 105-112 (2001).

\bibitem{Gabaix} X. Gabaix, {\it Power laws in economics and finance}. Annu. Rev. Econ. {\bf 1}, 255-294 (2009).

\bibitem{BouchaudMezard} J. P. Bouchaud, M. M\'ezard, {\it Wealth condensation in a simple model of economy}. Physica A {\bf 282}(3-4), 536-545 (2000).

\bibitem{Axtell} R. L. Axtell. {\it Zipf Distribution of U.S. Firm Sizes}. Science {\bf 293}, 1818-1820 (2001).

\bibitem{GabaixZipf} X. Gabaix, {\it Zipf's law for cities: an explanation}. The Quarterly journal of economics {\bf114}(3), 739-767 (1999).

\bibitem{SornetteZipf} A. I. Saichev, Y. Malevergne, and D. Sornette. {\it Theory
of Zipf's law and beyond}. SSBM {\bf 632} (2009).

\bibitem{Kesten1973} H. Kesten, {\it Random difference equations and Renewal theory for products of random matrices}. Acta Math. {\bf131}, 207-248 (1973).

\bibitem{Solomon1975} F. Solomon, {\it Random walks in a random environment}. Ann. Probab. 1-31 (1975).

\bibitem{Kesten1975} H. Kesten, M. V. Kozlov, F. Spitzer {\it A limit law for random walk in a random environment}. Compos. Math. {\bf30}(2), 145-168 (1975).

\bibitem{BouchaudPLD1990} J. P. Bouchaud, A. Comtet, A. Georges, P. Le Doussal, {\it Classical diffusion of a particle in a one-dimensional random force field}. Annals of Physics {\bf201}(2), 285-341 (1990).

\bibitem{MehtaBook} M. L. Mehta, {\it Random matrices}. Elsevier (2004).

\bibitem{Forrester} P. J. Forrester, {\it Log-Gases and Random Matrices}. London Math. Soc. monographs, Princeton Univ. Press (2010).

\bibitem{RMTBook} M. Potters, J. P. Bouchaud, {\it A First Course in Random Matrix Theory for Physicists, Engineers and Data Scientists}. Cambridge University Press (2020).

\bibitem{AllezBouchaud2012} R. Allez, J. P. Bouchaud, {\it Eigenvector dynamics: General theory and some applications}. Phys. Rev. E {\bf86}(4), 046202 (2012).

\bibitem{BunBouchaudPotters2017} J. Bun, J. P. Bouchaud, M. Potters, {\it Cleaning large correlation matrices: Tools from Random Matrix Theory}. Phys. Rep. {\bf666}, 1-109 (2017).

\bibitem{AllezBouchaud2013} R. Allez, J. P. Bouchaud, S. N. Majumdar, P. Vivo, {\it Invariant $\beta$-Wishart ensembles, crossover densities and asymptotic corrections to the Marcenko-Pastur law}. J. Phys. A Math. Theor. {\bf46}(1), 015001 (2012).

\bibitem{DeanLeDoussal19} D. S. Dean, P. Le Doussal, S. N. Majumdar, G. Schehr, {\it Noninteracting fermions in a trap and random matrix theory}. J. Phys. A Math. Theor. \textbf{52}(14), 144006 (2019).

\bibitem{LACTLeDoussal17} B. Lacroix-A-Chez-Toine, P. Le Doussal, S. N. Majumdar, G. Schehr, {\it Statistics of fermions in a d-dimensional box near a hard wall}. EPL \textbf{120}(1), 10006 (2017).

\bibitem{LACTLeDoussal18} B.Lacroix-A-Chez-Toine, P. Le Doussal, S. N. Majumdar, G. Schehr, {\it Non-interacting fermions in hard-edge potentials}. J. Stat. Mech. {\bf2018}(12), 123103 (2018).

\bibitem{NadalMaj09} C. Nadal, S. N. Majumdar, {\it Nonintersecting Brownian interfaces and Wishart random matrices}. Phys. Rev. E {\bf79}(6), 061117 (2009).

\bibitem{CundenMezzOconnell18} F. D. Cunden, F. Mezzadri, N. O'Connell, {\it Free fermions and the classical compact groups}. J. Stat. Phys. {\bf171}(5), 768-801 (2018).

\bibitem{Kesten1984} H. Kesten, F. Spitzer, {\it Convergence in distribution of products of random matrices}. Z. Wahrscheinlichkeit {\bf67}(4), 363-386 (1984).

\bibitem{Calan} C. de Calan, J. M. Luck, T. M. Nieuwenhuizen, D. Petritis, {\it On the distribution of a random variable occurring in 1D disordered systems}. J Phys A {\bf18}, 501 (1985).

\bibitem{goldie1991} C. M. Goldie, {\it Implicit renewal theory and tails of solutions of random equations}. Ann. Appl. Probab. {\bf1}(1), 126-166 (1991).

\bibitem{Buraczewski2013} D. Buraczewski, E. Damek, T. Mikosch, J. Zienkiewicz, {\it Large deviations for solutions to stochastic recursion equations under Kesten's condition}. Ann. Probab. {\bf41}(4), 2755-2790 (2013).

\bibitem{Buraczewski2015} D. Buraczewski, E. Damek, T. Przebinda, {\it On the rate of convergence in the Kesten renewal theorem}. Electronic Journal of Probability {\bf20} (2015).

\bibitem{Sinai1982} Y. G. Sinai, {\it The limiting behavior of a one-dimensional random walk in a random medium}. Theory Probab. its Appl. {\bf27}(2), 256-268 (1982).

\bibitem{DerridaPomeau1982} B. Derrida, Y. Pomeau, {\it Classical diffusion on a random chain}. Phys. Rev. Lett. {\bf48}(9), 627 (1982).

\bibitem{Derrida1983a} B. Derrida, {\it Velocity and diffusion constant of a periodic one-dimensional hopping model}. J. Stat. Phys. {\bf31}(3), 433-450 (1983).

\bibitem{Kesten1986} H. Kesten, {\it The limit distribution of Sinai's random walk in random environment}. Phys. A Stat. Mech. its Appl. {\bf138}(1-2), 299-309 (1986).

\bibitem{BouchaudPLD1987} J. P. Bouchaud, A. Comtet, A. Georges, P. Le Doussal, {\it The relaxation-time spectrum of diffusion in a one-dimensional random medium: an exactly solvable case}. EPL {\bf3}(6), 653 (1987).

\bibitem{PLD1989} P. Le Doussal, {\it First-passage time for random walks in random environments}. Phys. Rev. Lett. {\bf62}(26), 3097 (1989).

\bibitem{ComtetDean} A. Comtet, D. S. Dean, {\it Exact results on Sinai's diffusion}. J. Phys. A Math. Gen. {\bf31}(43), 8595-8605 (1998).

\bibitem{MonthusPLD1999} P. Le Doussal, C. Monthus, D. S. Fisher,
{\it Random walkers in one-dimensional random environments:
exact renormalization group analysis}. Phys. Rev. E {\bf 59}(5), 4795 (1999).

\bibitem{Derrida1983} B. Derrida, H. J. Hilhorst, {\it Singular behaviour of certain infinite products of random 2$\times$2 matrices}. J. Phys. A Math. Gen. {\bf16}(12), 2641-2654 (1983).

\bibitem{Texier1997} A. Comtet, C. Texier, {\it On the distribution of the Wigner time delay in one-dimensional disordered systems}. J. Phys. A {\bf30}(23), 8017 (1997).

\bibitem{Texier1999} C. Texier, A. Comtet, {\it Universality of the Wigner time delay distribution for one-dimensional random potentials}. PRL {\bf82}(21), 4220 (1999).

\bibitem{Steiner1999} M. Steiner, Y. Chen, M. Fabrizio, A. O. Gogolin, {\it Statistical properties of a localization-delocalization transition in one dimension}. PRB {\bf59}(23), 14848 (1999).

\bibitem{Dufresne1990} D. Dufresne, {\it The distribution of a perpetuity, with applications to risk theory and pension funding}. Scand. Actuar. J. {\bf1990}(1), 39-79 (1990).

\bibitem{Paulson1972} A. S. Paulson, V. R. R. Uppuluri, {\it Limit laws of a sequence determined by a random difference equation governing a one-compartment system}. Math. Biosci. {\bf13}(3-4), 325-333 (1972).

\bibitem{Feldman1973} L. Cavalli-Sforza, M.W. Feldman, {\it Models for cultural inheritance, I. Group mean and within group variation}. Theor. Popul. Biol. {\bf4}(1), 42-55 (1973).

\bibitem{Dufresne2001} D. Dufresne, {\it On the integral of geometric Brownian motion}. Adv. Appl. Probab. {\bf33}, 223-241 (2001).

\bibitem{MatYor1} H. Matsumoto, M. Yor, {\it Exponential functionals of Brownian motion, I: Probability laws at fixed time}. Probab. Surv. {\bf2}, 312-347 (2005).

\bibitem{MatYor2} H. Matsumoto, M. Yor, {\it Exponential functionals of Brownian motion, II: Some related diffusion processes}. Probab. Surv. {\bf2}, 348-384 (2005).

\bibitem{Comtet1998} A. Comtet, C. Monthus, M. Yor, {\it Exponential functionals of Brownian motion and disordered systems}. J. Appl. Proba {\bf35}, 255 (1998). 

\bibitem{Comtet2005} A. Comtet, J. Desbois, C. Texier, {\it Functionals of Brownian motion, localization and metric graphs}. J. Phys. A {\bf38}(37), R341 (2005).

\bibitem{RiderValko} B. Rider, B. Valk\'o, {\it Matrix Dufresne identities}. Int. Math. Res. Not. {\bf 2016}(1), 174-218 (2016).

\bibitem{OConnell2019} N. O'Connell, {\it Interacting diffusions on positive definite matrices}. arXiv preprint 1910.03389 (2019).

\bibitem{GrabschTexier2016} A. Grabsch, C. Texier, {\it Topological phase transitions in the 1D multichannel Dirac equation with random mass and a random matrix model}. EPL {\bf116}(1), 17004 (2016).

\bibitem{GrabschTexier2020} A. Grabsch, C. Texier, {\it Wigner-Smith matrix, exponential functional of the matrix Brownian motion and matrix Dufresne identity}. J. Phys. A Math. Theor. {\bf53}, 425003 (2020).

\bibitem{Ossipov} A. Ossipov, {\it Scattering Approach to Anderson Localization}. Phys. Rev. Lett. {\bf121}(7), 076601 (2018).

\bibitem{TexierMajumdar13} C. Texier, S. N. Majumdar, {\it Wigner time-delay distribution in chaotic cavities and freezing transition}. Phys. Rev. Lett. {\bf110}(25), 250602 (2013).

\bibitem{FyodorovSommers97} Y. V. Fyodorov, H. J. Sommers, {\it Statistics of resonance poles, phase shifts and time delays in quantum chaotic scattering: Random matrix approach for systems with broken time-reversal invariance}. J. Math. Phys. {\bf38}(4), 1918-1981 (1997).

\bibitem{Texier2016} C. Texier, {\it Wigner time delay and related concepts: Application to transport in coherent conductors}. Physica E {\bf82}, 16-33 (2016); see updated Arxiv version arXiv:1507.00075v6.

\bibitem{DysonBM65} F. Dyson, {\it On the Brownian-motion model for the eigenvalues of a random matrix}. Nuovo Cim. {\bf38}(2), 1047-1053 (1965).

\bibitem{Brouwer97} P. W. Brouwer, K. M. Frahm, C. W. Beenakker, {\it Quantum mechanical time-delay matrix in chaotic scattering}. Phys. Rev. Lett. {\bf78}(25), 4737 (1997).

\bibitem{Brouwer99} P. W. Brouwer, K. M. Frahm, C. W. Beenakker, {\it Distribution of the quantum mechanical time-delay matrix for a chaotic cavity}. Waves in Random Media {\bf9}(2), 91-104 (1999).
 
\bibitem{Cunden16} F. D. Cunden, F. Mezzadri, N. Simm, P. Vivo, {\it Large-N expansion for the time-delay matrix of ballistic chaotic cavities}. J. Math. Phys. 57 (2016).

\bibitem{Gautie1} T. Gauti\'e, P. Le Doussal, S. N. Majumdar, G. Schehr, {\it Non-crossing Brownian paths and Dyson Brownian motion under a moving boundary}. J. Stat. Phys. {\bf 177}, 752-805 (2019).

\bibitem{Morse29} P. M. Morse, {\it Diatomic molecules according to the wave mechanics. II. Vibrational levels}. Phys. Rev. {\bf34}(1), 57-64 (1929).

\bibitem{SutherlandBook} B. Sutherland, {\it Beautiful models: 70 years of exactly solved quantum many-body problems}. World Scientific Publishing Company (2004).

\bibitem{Calogero69} F. Calogero, {\it Solution of a three-body problem in one dimension}. J. Math. Phys.f {\bf 10}(12), 2191-2196 (1969).

\bibitem{yor} H. Geman, M. Yor. {\it Bessel processes, Asian options, and perpetuities}. Mathematical finance {\bf3}, 349-375 (1993).

\bibitem{Zhang10} P. Zhang, {\it Morse Potential, Contour Integrals, and Asian Options}. arXiv preprint 1010.3820 (2010).

\bibitem{Gautie2} T. Gauti\'e, N. R. Smith, {\it Constrained non-crossing Brownian motions, fermions and the Ferrari-Spohn distribution}. arXiv preprint 2011.12995 (2020).

\bibitem{Johansson05} K. Johansson, {\it Random matrices and determinantal processes}. Lecture Notes of the Les Houches Summer School 2005, arXiv preprint math-ph/0510038 (2005).

\bibitem{Borodin11} A. Borodin, {\it Determinantal point processes. The Oxford Han{\rm d}Book of Random Matrix Theory}.
G. Akemann, J. Baik, P. Di Francesco (Eds.), Oxford University Press, Oxford (2011).

\bibitem{TracyWidom94} C. Tracy, H. Widom, {\it Level spacing distributions and the Bessel kernel}. Commun. Math. Phys. {\bf161}(2), 289-309 (1994).

\bibitem{Rider09} J. A. Ramirez, B. Rider, {\it Diffusion at the random matrix hard edge}. Comm. Math. Phys. {\bf288}(3), 887-906 (2009).

\bibitem{Voiculescu85} D. Voiculescu, {\it Symmetries of some reduced free product C*-algebras}. Springer (1985).

\bibitem{Voiculescu91} D. Voiculescu, {\it Limit laws for random matrices and free products}. Invent. Math. {\bf104}(1), 201-220 (1991).

\bibitem{Voiculescu95} D. Voiculescu, {\it Free probability theory: random matrices and von Neumann Algebras}. Proceedings of the International Congress of Mathematicians, 227-242 (1995).

\bibitem{MingoSpeicher2017} J. Mingo, R. Speicher, {\it Free Probability and Random Matrices}. Fields Monograph Series (2017).

\bibitem{TulinoVerdu} A. M. Tulino, S. Verd\'u {\it Random Matrix Theory and Wireless Communications}. Now Publishers Inc. (2004).

\bibitem{Novak14} J. Novak, {\it Three lectures on free probability. In Random Matrix Theory, Interacting Particle Systems, and Integrable Systems}. Cambridge University Press (2014).

\bibitem{Nowak20} J. Grela, M. A. Nowak, W. Tarnowski, {\it Eikonal formulation of large dynamical random matrix models}. arXiv preprint 2010.01690 (2020).

\bibitem{Blaizot10} J. P. Blaizot, M. A. Nowak, {\it Universal shocks in random matrix theory}. Phys. Rev. E {\bf82}(5), 051115 (2010).

\bibitem{Blaizot13} J. P. Blaizot, M. A. Nowak, P. Warchol, {\it Universal shocks in the Wishart random-matrix ensemble}. Phys. Rev. E {\bf87}(5), 052134 (2013).

\bibitem{Blaizot14} J. P. Blaizot, M. A. Nowak, P. Warchol, {\it Universal shocks in the Wishart random-matrix ensemble. II. Nontrivial initial conditions}. Phys. Rev. E {\bf89}(4), 042130 (2014).

\bibitem{VanKampenItoStrat} N. G. Van Kampen, {\it It\^o versus Stratonovich}. J. Stat. Phys. {\bf24}(1), 175-187 (1981).
 
\bibitem{AssiotisBougerol} T. Assiotis, {\it A matrix Bougerol identity and the Hua-Pickrell measures}. ECP {\bf23} (2018).

\bibitem{Cardy10} J. Cardy, {\it Quantum network models and classical localization problems}. Int. J. Mod. Phys. B {\bf24}(12n13), 1989-2014 (2010).

\bibitem{Krajenbrink2020} T. Jin, A. Krajenbrink, D. Bernard, {\it From stochastic spin chains to quantum Kardar-Parisi-Zhang dynamics}. Phys. Rev. Lett. {\bf 125}, 040603 (2020).

\bibitem{Tao2012} T. Tao, {\it Topics in Random Matrix Theory}. Am. Math. Soc. (2012).

\bibitem{SpeicherCombinatorics} R. Speicher, {\it Free probability theory and non-crossing partitions}. S\'em. Lothar. Combin. {\bf 39}, B39c-38 (1997).

\bibitem{Edelman88} A. Edelman, {\it Eigenvalues and Condition Numbers of Random Matrices}. SIAM J. Matrix Anal. Appl. {\bf9}, 543-560 (1988).

\bibitem{Cunden18} F. D. Cunden, A. Dahlqvist, N. O'Connell, {\it Integer moments of complex Wishart matrices and Hurwitz numbers}. Ann. IHP D (2021).

\bibitem{Gissoni20} M. Gisonni, T. Grava, G. Ruzza, {\it Laguerre Ensemble: Correlators, Hurwitz Numbers and Hodge Integrals}. Ann. Henri Poincar\'e {\bf21}(10), 3285-3339 (2020).

\bibitem{Assiotis19} T. Assiotis, {\it Ergodic Decomposition for Inverse Wishart Measures on Infinite Positive-Definite Matrices}. SIGMA {\bf15} (2019).

\bibitem{Dong07} S. H. Dong, {\it Factorization methods in quantum mechanics}. Springer Netherlands (2007).

\bibitem{Duru83} I. H. Duru, {\it Morse-potential Green's function with path integrals}. Phys. Rev. D {\bf28}(10), 2689 (1983).

\bibitem{PDEBook1} A. D. Polyanin, V. F. Zaitsev, {\it Han{\rm d}Book of nonlinear partial differential equations}. CRC press (2004).

\bibitem{PDEBook2} A. D. Polyanin, V. F. Zaitsev, A. Moussiaux, {\it Han{\rm d}Book of first-order partial differential equations}. CRC Press (2001).

\bibitem{Dolgachev03} I. Dolgachev, {\it Lectures on invariant theory}. Cambridge University Press (2003).


\end{thebibliography}
\end{document}